\documentclass[a4paper, traditabstract, longauth]{aa} % for the abstract without structuration 

\def\setsymbol#1#2{\expandafter\def\csname #1\endcsname{#2}}
\def\getsymbol#1{\csname #1\endcsname}

\def\Planck{{\it Planck\/}}

\newbox\tablebox    \newdimen\tablewidth
\def\leaderfil{\leaders\hbox to 5pt{\hss.\hss}\hfil}
\def\tablenote#1 #2\par{\begingroup \parindent=0.8em
    \abovedisplayshortskip=0pt\belowdisplayshortskip=0pt
    \noindent
    $$\hss\vbox{\hsize\tablewidth \hangindent=\parindent \hangafter=1 \noindent
    \hbox to \parindent{\sup{\rm #1}\hss}\strut#2\strut\par}\hss$$
    \endgroup}

\def\L2{\ifmmode L_2\else $L_2$\fi}

\def\DeltaT{\ifmmode \Delta T\else $\Delta T$\fi}
\def\deltat{\ifmmode \Delta t\else $\Delta t$\fi}
\def\fknee{\ifmmode f_{\rm knee}\else $f_{\rm knee}$\fi}
\def\Fmax{\ifmmode F_{\rm max}\else $F_{\rm max}$\fi}
\def\solar{\ifmmode{\rm M}_{\mathord\odot}\else${\rm M}_{\mathord\odot}$\fi}

\def\inv{\ifmmode^{-1}\else$^{-1}$\fi}
\def\mo{\ifmmode^{-1}\else$^{-1}$\fi}
\def\sup#1{\ifmmode ^{\rm #1}\else $^{\rm #1}$\fi}
\def\expo#1{\ifmmode \times 10^{#1}\else $\times 10^{#1}$\fi}
\def\,{\thinspace}
\def\lsim{\mathrel{\raise .4ex\hbox{\rlap{$<$}\lower 1.2ex\hbox{$\sim$}}}}
\def\gsim{\mathrel{\raise .4ex\hbox{\rlap{$>$}\lower 1.2ex\hbox{$\sim$}}}}

\def\simprop{\mathrel{\raise .4ex\hbox{\rlap{$\propto$}\lower 1.2ex\hbox{$\sim$}}}}
\def\deg{\ifmmode^\circ\else$^\circ$\fi}
\def\pdeg{\ifmmode $\setbox0=\hbox{$^{\circ}$}\rlap{\hskip.11\wd0 .}$^{\circ}
          \else \setbox0=\hbox{$^{\circ}$}\rlap{\hskip.11\wd0 .}$^{\circ}$\fi}
\def\arcs{\ifmmode {^{\scriptstyle\prime\prime}}
          \else $^{\scriptstyle\prime\prime}$\fi}
\def\arcm{\ifmmode {^{\scriptstyle\prime}}
          \else $^{\scriptstyle\prime}$\fi}
\newdimen\sa  \newdimen\sb
\def\parcs{\sa=.07em \sb=.03em
     \ifmmode \hbox{\rlap{.}}^{\scriptstyle\prime\kern -\sb\prime}\hbox{\kern -\sa}
     \else \rlap{.}$^{\scriptstyle\prime\kern -\sb\prime}$\kern -\sa\fi}
\def\parcm{\sa=.08em \sb=.03em
     \ifmmode \hbox{\rlap{.}\kern\sa}^{\scriptstyle\prime}\hbox{\kern-\sb}
     \else \rlap{.}\kern\sa$^{\scriptstyle\prime}$\kern-\sb\fi}
\def\ra[#1 #2 #3.#4]{#1\sup{h}#2\sup{m}#3\sup{s}\llap.#4}
\def\dec[#1 #2 #3.#4]{#1\deg#2\arcm#3\arcs\llap.#4}
\def\deco[#1 #2 #3]{#1\deg#2\arcm#3\arcs}
\def\rra[#1 #2]{#1\sup{h}#2\sup{m}}

\def\dots{\relax\ifmmode \ldots\else $\ldots$\fi}
\def\WHzsr{\ifmmode $W\,Hz\mo\,sr\mo$\else W\,Hz\mo\,sr\mo\fi}
\def\mHz{\ifmmode $\,mHz$\else \,mHz\fi}
\def\GHz{\ifmmode $\,GHz$\else \,GHz\fi}
\def\mKs{\ifmmode $\,mK\,s$^{1/2}\else \,mK\,s$^{1/2}$\fi}
\def\muKs{\ifmmode \,\mu$K\,s$^{1/2}\else \,$\mu$K\,s$^{1/2}$\fi}
\def\muKRJs{\ifmmode \,\mu$K$_{\rm RJ}$\,s$^{1/2}\else \,$\mu$K$_{\rm RJ}$\,s$^{1/2}$\fi}
\def\muKHz{\ifmmode \,\mu$K\,Hz$^{-1/2}\else \,$\mu$K\,Hz$^{-1/2}$\fi}
\def\MJysr{\ifmmode \,$MJy\,sr\mo$\else \,MJy\,sr\mo\fi}
\def\MJysrmK{\ifmmode \,$MJy\,sr\mo$\,mK$_{\rm CMB}\mo\else \,MJy\,sr\mo\,mK$_{\rm CMB}\mo$\fi}
\def\microns{\ifmmode \,\mu$m$\else \,$\mu$m\fi}

\def\muK{\ifmmode \,\mu$K$\else \,$\mu$\hbox{K}\fi}
\def\microK{\ifmmode \,\mu$K$\else \,$\mu$\hbox{K}\fi}
\def\muW{\ifmmode \,\mu$W$\else \,$\mu$\hbox{W}\fi}
\def\kms{\ifmmode $\,km\,s$^{-1}\else \,km\,s$^{-1}$\fi}
\def\kmsMpc{\ifmmode $\,\kms\,Mpc\mo$\else \,\kms\,Mpc\mo\fi}
\setsymbol{LFI:center:frequency:70GHz:units}{70.3\,GHz}
\setsymbol{LFI:center:frequency:44GHz:units}{44.1\,GHz}
\setsymbol{LFI:center:frequency:30GHz:units}{28.5\,GHz}

\setsymbol{LFI:center:frequency:70GHz}{70.3}
\setsymbol{LFI:center:frequency:44GHz}{44.1}
\setsymbol{LFI:center:frequency:30GHz}{28.5}

\setsymbol{LFI:center:frequency:LFI18:Rad:M:units}{71.7\GHz}
\setsymbol{LFI:center:frequency:LFI19:Rad:M:units}{67.5\GHz}
\setsymbol{LFI:center:frequency:LFI20:Rad:M:units}{69.2\GHz}
\setsymbol{LFI:center:frequency:LFI21:Rad:M:units}{70.4\GHz}
\setsymbol{LFI:center:frequency:LFI22:Rad:M:units}{71.5\GHz}
\setsymbol{LFI:center:frequency:LFI23:Rad:M:units}{70.8\GHz}
\setsymbol{LFI:center:frequency:LFI24:Rad:M:units}{44.4\GHz}
\setsymbol{LFI:center:frequency:LFI25:Rad:M:units}{44.0\GHz}
\setsymbol{LFI:center:frequency:LFI26:Rad:M:units}{43.9\GHz}
\setsymbol{LFI:center:frequency:LFI27:Rad:M:units}{28.3\GHz}
\setsymbol{LFI:center:frequency:LFI28:Rad:M:units}{28.8\GHz}
\setsymbol{LFI:center:frequency:LFI18:Rad:S:units}{70.1\GHz}
\setsymbol{LFI:center:frequency:LFI19:Rad:S:units}{69.6\GHz}
\setsymbol{LFI:center:frequency:LFI20:Rad:S:units}{69.5\GHz}
\setsymbol{LFI:center:frequency:LFI21:Rad:S:units}{69.5\GHz}
\setsymbol{LFI:center:frequency:LFI22:Rad:S:units}{72.8\GHz}
\setsymbol{LFI:center:frequency:LFI23:Rad:S:units}{71.3\GHz}
\setsymbol{LFI:center:frequency:LFI24:Rad:S:units}{44.1\GHz}
\setsymbol{LFI:center:frequency:LFI25:Rad:S:units}{44.1\GHz}
\setsymbol{LFI:center:frequency:LFI26:Rad:S:units}{44.1\GHz}
\setsymbol{LFI:center:frequency:LFI27:Rad:S:units}{28.5\GHz}
\setsymbol{LFI:center:frequency:LFI28:Rad:S:units}{28.2\GHz}

\setsymbol{LFI:center:frequency:LFI18:Rad:M}{71.7}
\setsymbol{LFI:center:frequency:LFI19:Rad:M}{67.5}
\setsymbol{LFI:center:frequency:LFI20:Rad:M}{69.2}
\setsymbol{LFI:center:frequency:LFI21:Rad:M}{70.4}
\setsymbol{LFI:center:frequency:LFI22:Rad:M}{71.5}
\setsymbol{LFI:center:frequency:LFI23:Rad:M}{70.8}
\setsymbol{LFI:center:frequency:LFI24:Rad:M}{44.4}
\setsymbol{LFI:center:frequency:LFI25:Rad:M}{44.0}
\setsymbol{LFI:center:frequency:LFI26:Rad:M}{43.9}
\setsymbol{LFI:center:frequency:LFI27:Rad:M}{28.3}
\setsymbol{LFI:center:frequency:LFI28:Rad:M}{28.8}
\setsymbol{LFI:center:frequency:LFI18:Rad:S}{70.1}
\setsymbol{LFI:center:frequency:LFI19:Rad:S}{69.6}
\setsymbol{LFI:center:frequency:LFI20:Rad:S}{69.5}
\setsymbol{LFI:center:frequency:LFI21:Rad:S}{69.5}
\setsymbol{LFI:center:frequency:LFI22:Rad:S}{72.8}
\setsymbol{LFI:center:frequency:LFI23:Rad:S}{71.3}
\setsymbol{LFI:center:frequency:LFI24:Rad:S}{44.1}
\setsymbol{LFI:center:frequency:LFI25:Rad:S}{44.1}
\setsymbol{LFI:center:frequency:LFI26:Rad:S}{44.1}
\setsymbol{LFI:center:frequency:LFI27:Rad:S}{28.5}
\setsymbol{LFI:center:frequency:LFI28:Rad:S}{28.2}

\setsymbol{LFI:white:noise:sensitivity:70GHz:units}{134.7\muKs}
\setsymbol{LFI:white:noise:sensitivity:44GHz:units}{164.7\muKs}
\setsymbol{LFI:white:noise:sensitivity:30GHz:units}{143.4\muKs}

\setsymbol{LFI:white:noise:sensitivity:70GHz}{134.7}
\setsymbol{LFI:white:noise:sensitivity:44GHz}{164.7}
\setsymbol{LFI:white:noise:sensitivity:30GHz}{143.4}

\setsymbol{LFI:white:noise:sensitivity:LFI18:Rad:M:units}{512.0\muKs}
\setsymbol{LFI:white:noise:sensitivity:LFI19:Rad:M:units}{581.4\muKs}
\setsymbol{LFI:white:noise:sensitivity:LFI20:Rad:M:units}{590.8\muKs}
\setsymbol{LFI:white:noise:sensitivity:LFI21:Rad:M:units}{455.2\muKs}
\setsymbol{LFI:white:noise:sensitivity:LFI22:Rad:M:units}{492.0\muKs}
\setsymbol{LFI:white:noise:sensitivity:LFI23:Rad:M:units}{507.7\muKs}
\setsymbol{LFI:white:noise:sensitivity:LFI24:Rad:M:units}{462.2\muKs}
\setsymbol{LFI:white:noise:sensitivity:LFI25:Rad:M:units}{413.6\muKs}
\setsymbol{LFI:white:noise:sensitivity:LFI26:Rad:M:units}{478.6\muKs}
\setsymbol{LFI:white:noise:sensitivity:LFI27:Rad:M:units}{277.7\muKs}
\setsymbol{LFI:white:noise:sensitivity:LFI28:Rad:M:units}{312.3\muKs}
\setsymbol{LFI:white:noise:sensitivity:LFI18:Rad:S:units}{465.7\muKs}
\setsymbol{LFI:white:noise:sensitivity:LFI19:Rad:S:units}{555.6\muKs}
\setsymbol{LFI:white:noise:sensitivity:LFI20:Rad:S:units}{623.2\muKs}
\setsymbol{LFI:white:noise:sensitivity:LFI21:Rad:S:units}{564.1\muKs}
\setsymbol{LFI:white:noise:sensitivity:LFI22:Rad:S:units}{534.4\muKs}
\setsymbol{LFI:white:noise:sensitivity:LFI23:Rad:S:units}{542.4\muKs}
\setsymbol{LFI:white:noise:sensitivity:LFI24:Rad:S:units}{399.2\muKs}
\setsymbol{LFI:white:noise:sensitivity:LFI25:Rad:S:units}{392.6\muKs}
\setsymbol{LFI:white:noise:sensitivity:LFI26:Rad:S:units}{418.6\muKs}
\setsymbol{LFI:white:noise:sensitivity:LFI27:Rad:S:units}{302.9\muKs}
\setsymbol{LFI:white:noise:sensitivity:LFI28:Rad:S:units}{285.3\muKs}

\setsymbol{LFI:white:noise:sensitivity:LFI18:Rad:M}{512.0}
\setsymbol{LFI:white:noise:sensitivity:LFI19:Rad:M}{581.4}
\setsymbol{LFI:white:noise:sensitivity:LFI20:Rad:M}{590.8}
\setsymbol{LFI:white:noise:sensitivity:LFI21:Rad:M}{455.2}
\setsymbol{LFI:white:noise:sensitivity:LFI22:Rad:M}{492.0}
\setsymbol{LFI:white:noise:sensitivity:LFI23:Rad:M}{507.7}
\setsymbol{LFI:white:noise:sensitivity:LFI24:Rad:M}{462.2}
\setsymbol{LFI:white:noise:sensitivity:LFI25:Rad:M}{413.6}
\setsymbol{LFI:white:noise:sensitivity:LFI26:Rad:M}{478.6}
\setsymbol{LFI:white:noise:sensitivity:LFI27:Rad:M}{277.7}
\setsymbol{LFI:white:noise:sensitivity:LFI28:Rad:M}{312.3}
\setsymbol{LFI:white:noise:sensitivity:LFI18:Rad:S}{465.7}
\setsymbol{LFI:white:noise:sensitivity:LFI19:Rad:S}{555.6}
\setsymbol{LFI:white:noise:sensitivity:LFI20:Rad:S}{623.2}
\setsymbol{LFI:white:noise:sensitivity:LFI21:Rad:S}{564.1}
\setsymbol{LFI:white:noise:sensitivity:LFI22:Rad:S}{534.4}
\setsymbol{LFI:white:noise:sensitivity:LFI23:Rad:S}{542.4}
\setsymbol{LFI:white:noise:sensitivity:LFI24:Rad:S}{399.2}
\setsymbol{LFI:white:noise:sensitivity:LFI25:Rad:S}{392.6}
\setsymbol{LFI:white:noise:sensitivity:LFI26:Rad:S}{418.6}
\setsymbol{LFI:white:noise:sensitivity:LFI27:Rad:S}{302.9}
\setsymbol{LFI:white:noise:sensitivity:LFI28:Rad:S}{285.3}

\setsymbol{LFI:knee:frequency:70GHz:units}{29.5\mHz}
\setsymbol{LFI:knee:frequency:44GHz:units}{56.2\mHz}
\setsymbol{LFI:knee:frequency:30GHz:units}{113.7\mHz}

\setsymbol{LFI:knee:frequency:70GHz}{29.5}
\setsymbol{LFI:knee:frequency:44GHz}{56.2}
\setsymbol{LFI:knee:frequency:30GHz}{113.7}

\setsymbol{LFI:knee:frequency:LFI18:Rad:M:units}{16.3\mHz}
\setsymbol{LFI:knee:frequency:LFI19:Rad:M:units}{15.1\mHz}
\setsymbol{LFI:knee:frequency:LFI20:Rad:M:units}{18.7\mHz}
\setsymbol{LFI:knee:frequency:LFI21:Rad:M:units}{37.2\mHz}
\setsymbol{LFI:knee:frequency:LFI22:Rad:M:units}{12.7\mHz}
\setsymbol{LFI:knee:frequency:LFI23:Rad:M:units}{34.6\mHz}
\setsymbol{LFI:knee:frequency:LFI24:Rad:M:units}{46.2\mHz}
\setsymbol{LFI:knee:frequency:LFI25:Rad:M:units}{24.9\mHz}
\setsymbol{LFI:knee:frequency:LFI26:Rad:M:units}{67.6\mHz}
\setsymbol{LFI:knee:frequency:LFI27:Rad:M:units}{187.4\mHz}
\setsymbol{LFI:knee:frequency:LFI28:Rad:M:units}{122.2\mHz}
\setsymbol{LFI:knee:frequency:LFI18:Rad:S:units}{17.7\mHz}
\setsymbol{LFI:knee:frequency:LFI19:Rad:S:units}{22.0\mHz}
\setsymbol{LFI:knee:frequency:LFI20:Rad:S:units}{8.7\mHz}
\setsymbol{LFI:knee:frequency:LFI21:Rad:S:units}{25.9\mHz}
\setsymbol{LFI:knee:frequency:LFI22:Rad:S:units}{15.8\mHz}
\setsymbol{LFI:knee:frequency:LFI23:Rad:S:units}{129.8\mHz}
\setsymbol{LFI:knee:frequency:LFI24:Rad:S:units}{100.9\mHz}
\setsymbol{LFI:knee:frequency:LFI25:Rad:S:units}{38.9\mHz}
\setsymbol{LFI:knee:frequency:LFI26:Rad:S:units}{58.9\mHz}
\setsymbol{LFI:knee:frequency:LFI27:Rad:S:units}{104.4\mHz}
\setsymbol{LFI:knee:frequency:LFI28:Rad:S:units}{40.7\mHz}

\setsymbol{LFI:knee:frequency:LFI18:Rad:M}{16.3}
\setsymbol{LFI:knee:frequency:LFI19:Rad:M}{15.1}
\setsymbol{LFI:knee:frequency:LFI20:Rad:M}{18.7}
\setsymbol{LFI:knee:frequency:LFI21:Rad:M}{37.2}
\setsymbol{LFI:knee:frequency:LFI22:Rad:M}{12.7}
\setsymbol{LFI:knee:frequency:LFI23:Rad:M}{34.6}
\setsymbol{LFI:knee:frequency:LFI24:Rad:M}{46.2}
\setsymbol{LFI:knee:frequency:LFI25:Rad:M}{24.9}
\setsymbol{LFI:knee:frequency:LFI26:Rad:M}{67.6}
\setsymbol{LFI:knee:frequency:LFI27:Rad:M}{187.4}
\setsymbol{LFI:knee:frequency:LFI28:Rad:M}{122.2}
\setsymbol{LFI:knee:frequency:LFI18:Rad:S}{17.7}
\setsymbol{LFI:knee:frequency:LFI19:Rad:S}{22.0}
\setsymbol{LFI:knee:frequency:LFI20:Rad:S}{8.7}
\setsymbol{LFI:knee:frequency:LFI21:Rad:S}{25.9}
\setsymbol{LFI:knee:frequency:LFI22:Rad:S}{15.8}
\setsymbol{LFI:knee:frequency:LFI23:Rad:S}{129.8}
\setsymbol{LFI:knee:frequency:LFI24:Rad:S}{100.9}
\setsymbol{LFI:knee:frequency:LFI25:Rad:S}{38.9}
\setsymbol{LFI:knee:frequency:LFI26:Rad:S}{58.9}
\setsymbol{LFI:knee:frequency:LFI27:Rad:S}{104.4}
\setsymbol{LFI:knee:frequency:LFI28:Rad:S}{40.7}

\setsymbol{LFI:slope:70GHz:units}{$-1.03$\mHz}
\setsymbol{LFI:slope:44GHz:units}{$-0.89$\mHz}
\setsymbol{LFI:slope:30GHz:units}{$-0.87$\mHz}

\setsymbol{LFI:slope:70GHz}{$-1.03$}
\setsymbol{LFI:slope:44GHz}{$-0.89$}
\setsymbol{LFI:slope:30GHz}{$-0.87$}

\setsymbol{LFI:slope:LFI18:Rad:M:units}{$-1.04$\mHz}
\setsymbol{LFI:slope:LFI19:Rad:M:units}{$-1.09$\mHz}
\setsymbol{LFI:slope:LFI20:Rad:M:units}{$-0.69$\mHz}
\setsymbol{LFI:slope:LFI21:Rad:M:units}{$-1.56$\mHz}
\setsymbol{LFI:slope:LFI22:Rad:M:units}{$-1.01$\mHz}
\setsymbol{LFI:slope:LFI23:Rad:M:units}{$-0.96$\mHz}
\setsymbol{LFI:slope:LFI24:Rad:M:units}{$-0.83$\mHz}
\setsymbol{LFI:slope:LFI25:Rad:M:units}{$-0.91$\mHz}
\setsymbol{LFI:slope:LFI26:Rad:M:units}{$-0.95$\mHz}
\setsymbol{LFI:slope:LFI27:Rad:M:units}{$-0.87$\mHz}
\setsymbol{LFI:slope:LFI28:Rad:M:units}{$-0.88$\mHz}
\setsymbol{LFI:slope:LFI18:Rad:S:units}{$-1.15$\mHz}
\setsymbol{LFI:slope:LFI19:Rad:S:units}{$-1.00$\mHz}
\setsymbol{LFI:slope:LFI20:Rad:S:units}{$-0.95$\mHz}
\setsymbol{LFI:slope:LFI21:Rad:S:units}{$-0.92$\mHz}
\setsymbol{LFI:slope:LFI22:Rad:S:units}{$-1.01$\mHz}
\setsymbol{LFI:slope:LFI23:Rad:S:units}{$-0.95$\mHz}
\setsymbol{LFI:slope:LFI24:Rad:S:units}{$-0.73$\mHz}
\setsymbol{LFI:slope:LFI25:Rad:S:units}{$-1.16$\mHz}
\setsymbol{LFI:slope:LFI26:Rad:S:units}{$-0.79$\mHz}
\setsymbol{LFI:slope:LFI27:Rad:S:units}{$-0.82$\mHz}
\setsymbol{LFI:slope:LFI28:Rad:S:units}{$-0.91$\mHz}

\setsymbol{LFI:slope:LFI18:Rad:M}{$-1.04$}
\setsymbol{LFI:slope:LFI19:Rad:M}{$-1.09$}
\setsymbol{LFI:slope:LFI20:Rad:M}{$-0.69$}
\setsymbol{LFI:slope:LFI21:Rad:M}{$-1.56$}
\setsymbol{LFI:slope:LFI22:Rad:M}{$-1.01$}
\setsymbol{LFI:slope:LFI23:Rad:M}{$-0.96$}
\setsymbol{LFI:slope:LFI24:Rad:M}{$-0.83$}
\setsymbol{LFI:slope:LFI25:Rad:M}{$-0.91$}
\setsymbol{LFI:slope:LFI26:Rad:M}{$-0.95$}
\setsymbol{LFI:slope:LFI27:Rad:M}{$-0.87$}
\setsymbol{LFI:slope:LFI28:Rad:M}{$-0.88$}
\setsymbol{LFI:slope:LFI18:Rad:S}{$-1.15$}
\setsymbol{LFI:slope:LFI19:Rad:S}{$-1.00$}
\setsymbol{LFI:slope:LFI20:Rad:S}{$-0.95$}
\setsymbol{LFI:slope:LFI21:Rad:S}{$-0.92$}
\setsymbol{LFI:slope:LFI22:Rad:S}{$-1.01$}
\setsymbol{LFI:slope:LFI23:Rad:S}{$-0.95$}
\setsymbol{LFI:slope:LFI24:Rad:S}{$-0.73$}
\setsymbol{LFI:slope:LFI25:Rad:S}{$-1.16$}
\setsymbol{LFI:slope:LFI26:Rad:S}{$-0.79$}
\setsymbol{LFI:slope:LFI27:Rad:S}{$-0.82$}
\setsymbol{LFI:slope:LFI28:Rad:S}{$-0.91$}

\setsymbol{LFI:FWHM:70GHz:units}{13\parcm01}
\setsymbol{LFI:FWHM:44GHz:units}{27\parcm92}
\setsymbol{LFI:FWHM:30GHz:units}{32\parcm65}

\setsymbol{LFI:FWHM:70GHz}{13.01}
\setsymbol{LFI:FWHM:44GHz}{27.92}
\setsymbol{LFI:FWHM:30GHz}{32.65}

\setsymbol{LFI:FWHM:LFI18:units}{13\parcm39}
\setsymbol{LFI:FWHM:LFI19:units}{13\parcm01}
\setsymbol{LFI:FWHM:LFI20:units}{12\parcm75}
\setsymbol{LFI:FWHM:LFI21:units}{12\parcm74}
\setsymbol{LFI:FWHM:LFI22:units}{12\parcm87}
\setsymbol{LFI:FWHM:LFI23:units}{13\parcm27}
\setsymbol{LFI:FWHM:LFI24:units}{22\parcm98}
\setsymbol{LFI:FWHM:LFI25:units}{30\parcm46}
\setsymbol{LFI:FWHM:LFI26:units}{30\parcm31}
\setsymbol{LFI:FWHM:LFI27:units}{32\parcm65}
\setsymbol{LFI:FWHM:LFI28:units}{32\parcm66}

\setsymbol{LFI:FWHM:LFI18}{13.39}
\setsymbol{LFI:FWHM:LFI19}{13.01}
\setsymbol{LFI:FWHM:LFI20}{12.75}
\setsymbol{LFI:FWHM:LFI21}{12.74}
\setsymbol{LFI:FWHM:LFI22}{12.87}
\setsymbol{LFI:FWHM:LFI23}{13.27}
\setsymbol{LFI:FWHM:LFI24}{22.98}
\setsymbol{LFI:FWHM:LFI25}{30.46}
\setsymbol{LFI:FWHM:LFI26}{30.31}
\setsymbol{LFI:FWHM:LFI27}{32.65}
\setsymbol{LFI:FWHM:LFI28}{32.66}

\setsymbol{LFI:FWHM:uncertainty:LFI18:units}{0.170\arcm}
\setsymbol{LFI:FWHM:uncertainty:LFI19:units}{0.174\arcm}
\setsymbol{LFI:FWHM:uncertainty:LFI20:units}{0.170\arcm}
\setsymbol{LFI:FWHM:uncertainty:LFI21:units}{0.156\arcm}
\setsymbol{LFI:FWHM:uncertainty:LFI22:units}{0.164\arcm}
\setsymbol{LFI:FWHM:uncertainty:LFI23:units}{0.171\arcm}
\setsymbol{LFI:FWHM:uncertainty:LFI24:units}{0.652\arcm}
\setsymbol{LFI:FWHM:uncertainty:LFI25:units}{1.075\arcm}
\setsymbol{LFI:FWHM:uncertainty:LFI26:units}{1.131\arcm}
\setsymbol{LFI:FWHM:uncertainty:LFI27:units}{1.266\arcm}
\setsymbol{LFI:FWHM:uncertainty:LFI28:units}{1.287\arcm}

\setsymbol{LFI:FWHM:uncertainty:LFI18}{0.170}
\setsymbol{LFI:FWHM:uncertainty:LFI19}{0.174}
\setsymbol{LFI:FWHM:uncertainty:LFI20}{0.170}
\setsymbol{LFI:FWHM:uncertainty:LFI21}{0.156}
\setsymbol{LFI:FWHM:uncertainty:LFI22}{0.164}
\setsymbol{LFI:FWHM:uncertainty:LFI23}{0.171}
\setsymbol{LFI:FWHM:uncertainty:LFI24}{0.652}
\setsymbol{LFI:FWHM:uncertainty:LFI25}{1.075}
\setsymbol{LFI:FWHM:uncertainty:LFI26}{1.131}
\setsymbol{LFI:FWHM:uncertainty:LFI27}{1.266}
\setsymbol{LFI:FWHM:uncertainty:LFI28}{1.287}

\setsymbol{HFI:center:frequency:100GHz:units}{100\,GHz}
\setsymbol{HFI:center:frequency:143GHz:units}{143\,GHz}
\setsymbol{HFI:center:frequency:217GHz:units}{217\,GHz}
\setsymbol{HFI:center:frequency:353GHz:units}{353\,GHz}
\setsymbol{HFI:center:frequency:545GHz:units}{545\,GHz}
\setsymbol{HFI:center:frequency:857GHz:units}{857\,GHz}

\setsymbol{HFI:center:frequency:100GHz}{100}
\setsymbol{HFI:center:frequency:143GHz}{143}
\setsymbol{HFI:center:frequency:217GHz}{217}
\setsymbol{HFI:center:frequency:353GHz}{353}
\setsymbol{HFI:center:frequency:545GHz}{545}
\setsymbol{HFI:center:frequency:857GHz}{857}

\setsymbol{HFI:Ndetectors:100GHz}{8}
\setsymbol{HFI:Ndetectors:143GHz}{11}
\setsymbol{HFI:Ndetectors:217GHz}{12}
\setsymbol{HFI:Ndetectors:353GHz}{12}
\setsymbol{HFI:Ndetectors:545GHz}{3}
\setsymbol{HFI:Ndetectors:857GHz}{4}

\setsymbol{HFI:FWHM:Maps:100GHz:units}{9\parcm88}
\setsymbol{HFI:FWHM:Maps:143GHz:units}{7\parcm18}
\setsymbol{HFI:FWHM:Maps:217GHz:units}{4\parcm87}
\setsymbol{HFI:FWHM:Maps:353GHz:units}{4\parcm65}
\setsymbol{HFI:FWHM:Maps:545GHz:units}{4\parcm72}
\setsymbol{HFI:FWHM:Maps:857GHz:units}{4\parcm39}
\setsymbol{HFI:FWHM:Maps:100GHz}{9.88}
\setsymbol{HFI:FWHM:Maps:143GHz}{7.18}
\setsymbol{HFI:FWHM:Maps:217GHz}{4.87}
\setsymbol{HFI:FWHM:Maps:353GHz}{4.65}
\setsymbol{HFI:FWHM:Maps:545GHz}{4.72}
\setsymbol{HFI:FWHM:Maps:857GHz}{4.39}

\setsymbol{HFI:beam:ellipticity:Maps:100GHz}{1.15}
\setsymbol{HFI:beam:ellipticity:Maps:143GHz}{1.01}
\setsymbol{HFI:beam:ellipticity:Maps:217GHz}{1.06}
\setsymbol{HFI:beam:ellipticity:Maps:353GHz}{1.05}
\setsymbol{HFI:beam:ellipticity:Maps:545GHz}{1.14}
\setsymbol{HFI:beam:ellipticity:Maps:857GHz}{1.19}

\setsymbol{HFI:FWHM:Mars:100GHz:units}{9\parcm37}
\setsymbol{HFI:FWHM:Mars:143GHz:units}{7\parcm04}
\setsymbol{HFI:FWHM:Mars:217GHz:units}{4\parcm68}
\setsymbol{HFI:FWHM:Mars:353GHz:units}{4\parcm43}
\setsymbol{HFI:FWHM:Mars:545GHz:units}{3\parcm80}
\setsymbol{HFI:FWHM:Mars:857GHz:units}{3\parcm67}

\setsymbol{HFI:FWHM:Mars:100GHz}{9.37}
\setsymbol{HFI:FWHM:Mars:143GHz}{7.04}
\setsymbol{HFI:FWHM:Mars:217GHz}{4.68}
\setsymbol{HFI:FWHM:Mars:353GHz}{4.43}
\setsymbol{HFI:FWHM:Mars:545GHz}{3.80}
\setsymbol{HFI:FWHM:Mars:857GHz}{3.67}

\setsymbol{HFI:beam:ellipticity:Mars:100GHz}{1.18}
\setsymbol{HFI:beam:ellipticity:Mars:143GHz}{1.03}
\setsymbol{HFI:beam:ellipticity:Mars:217GHz}{1.14}
\setsymbol{HFI:beam:ellipticity:Mars:353GHz}{1.09}
\setsymbol{HFI:beam:ellipticity:Mars:545GHz}{1.25}
\setsymbol{HFI:beam:ellipticity:Mars:857GHz}{1.03}

\setsymbol{HFI:CMB:relative:calibration:100GHz}{$\lsim 1\%$}
\setsymbol{HFI:CMB:relative:calibration:143GHz}{$\lsim 1\%$}
\setsymbol{HFI:CMB:relative:calibration:217GHz}{$\lsim 1\%$}
\setsymbol{HFI:CMB:relative:calibration:353GHz}{$\lsim 1\%$}
\setsymbol{HFI:CMB:relative:calibration:545GHz}{}
\setsymbol{HFI:CMB:relative:calibration:857GHz}{}

\setsymbol{HFI:CMB:absolute:calibration:100GHz}{$\lsim 2\%$}
\setsymbol{HFI:CMB:absolute:calibration:143GHz}{$\lsim 2\%$}
\setsymbol{HFI:CMB:absolute:calibration:217GHz}{$\lsim 2\%$}
\setsymbol{HFI:CMB:absolute:calibration:353GHz}{$\lsim 2\%$}
\setsymbol{HFI:CMB:absolute:calibration:545GHz}{}
\setsymbol{HFI:CMB:absolute:calibration:857GHz}{}

\setsymbol{HFI:FIRAS:gain:calibration:accuracy:statistical:100GHz}{}
\setsymbol{HFI:FIRAS:gain:calibration:accuracy:statistical:143GHz}{}
\setsymbol{HFI:FIRAS:gain:calibration:accuracy:statistical:217GHz}{}
\setsymbol{HFI:FIRAS:gain:calibration:accuracy:statistical:353GHz}{2.5\%}
\setsymbol{HFI:FIRAS:gain:calibration:accuracy:statistical:545GHz}{1\%}
\setsymbol{HFI:FIRAS:gain:calibration:accuracy:statistical:857GHz}{0.5\%}

\setsymbol{HFI:FIRAS:gain:calibration:accuracy:systematic:100GHz}{}
\setsymbol{HFI:FIRAS:gain:calibration:accuracy:systematic:143GHz}{}
\setsymbol{HFI:FIRAS:gain:calibration:accuracy:systematic:217GHz}{}
\setsymbol{HFI:FIRAS:gain:calibration:accuracy:systematic:353GHz}{}
\setsymbol{HFI:FIRAS:gain:calibration:accuracy:systematic:545GHz}{7\%}
\setsymbol{HFI:FIRAS:gain:calibration:accuracy:systematic:857GHz}{7\%}

\setsymbol{HFI:FIRAS:zero:point:accuracy:100GHz:units}{0.8\MJysr}
\setsymbol{HFI:FIRAS:zero:point:accuracy:143GHz:units}{}
\setsymbol{HFI:FIRAS:zero:point:accuracy:217GHz:units}{}
\setsymbol{HFI:FIRAS:zero:point:accuracy:353GHz:units}{1.4\MJysr}
\setsymbol{HFI:FIRAS:zero:point:accuracy:545GHz:units}{2.2\MJysr}
\setsymbol{HFI:FIRAS:zero:point:accuracy:857GHz:units}{1.7\MJysr}

\setsymbol{HFI:FIRAS:zero:point:accuracy:100GHz}{0.8}
\setsymbol{HFI:FIRAS:zero:point:accuracy:143GHz}{}
\setsymbol{HFI:FIRAS:zero:point:accuracy:217GHz}{}
\setsymbol{HFI:FIRAS:zero:point:accuracy:353GHz}{1.4}
\setsymbol{HFI:FIRAS:zero:point:accuracy:545GHz}{2.2}
\setsymbol{HFI:FIRAS:zero:point:accuracy:857GHz}{1.7}

\setsymbol{HFI:unit:conversion:100GHz:units}{0.2415\MJysrmK}
\setsymbol{HFI:unit:conversion:143GHz:units}{0.3694\MJysrmK}
\setsymbol{HFI:unit:conversion:217GHz:units}{0.4811\MJysrmK}
\setsymbol{HFI:unit:conversion:353GHz:units}{0.2883\MJysrmK}
\setsymbol{HFI:unit:conversion:545GHz:units}{0.05826\MJysrmK}
\setsymbol{HFI:unit:conversion:857GHz:units}{0.002238\MJysrmK}

\setsymbol{HFI:unit:conversion:100GHz}{0.2415}
\setsymbol{HFI:unit:conversion:143GHz}{0.3694}
\setsymbol{HFI:unit:conversion:217GHz}{0.4811}
\setsymbol{HFI:unit:conversion:353GHz}{0.2883}
\setsymbol{HFI:unit:conversion:545GHz}{0.05826}
\setsymbol{HFI:unit:conversion:857GHz}{0.002238}

\setsymbol{HFI:colour:correction:alpha=-2:V1.01:100GHz}{0.9893}
\setsymbol{HFI:colour:correction:alpha=-2:V1.01:143GHz}{0.9759}
\setsymbol{HFI:colour:correction:alpha=-2:V1.01:217GHz}{1.0007}
\setsymbol{HFI:colour:correction:alpha=-2:V1.01:353GHz}{1.0028}
\setsymbol{HFI:colour:correction:alpha=-2:V1.01:545GHz}{1.0019}
\setsymbol{HFI:colour:correction:alpha=-2:V1.01:857GHz}{0.9889}

\setsymbol{HFI:colour:correction:alpha=0:V1.01:100GHz}{1.0008}
\setsymbol{HFI:colour:correction:alpha=0:V1.01:143GHz}{1.0148}
\setsymbol{HFI:colour:correction:alpha=0:V1.01:217GHz}{0.9909}
\setsymbol{HFI:colour:correction:alpha=0:V1.01:353GHz}{0.9888}
\setsymbol{HFI:colour:correction:alpha=0:V1.01:545GHz}{0.9878}
\setsymbol{HFI:colour:correction:alpha=0:V1.01:857GHz}{1.0014}

\usepackage{graphicx}
\usepackage{txfonts}
\usepackage[breaklinks, colorlinks, citecolor=cyan]{hyperref}
\usepackage{url}
\usepackage{natbib}
\usepackage{color}
\usepackage{amssymb}
\usepackage{booktabs}
\usepackage{subfig}
\usepackage[table]{xcolor}
\definecolor{tableShade}{HTML}{F1F5FA}   
\definecolor{tableShade2}{HTML}{ECF3FE} 
\definecolor{light-grey}{gray}{0.90}

\newfont{\gwpfont}{cmssq8 scaled 1000}
\newcommand{\rexcess}{{\gwpfont REXCESS}}
\def\msol{{M$_{\odot}$}}

\def\xmm{{\it XMM-Newton}}
\def\planck{{\it Planck}}

\def\M500{M_{500}}
\def\R500{R_{500}}
\def\Mgv{M_{\rm g,500}}

\def\YX {Y_{\rm X, 500}}
\def\TX {T_{\rm X}}
\def\YSZ {Y_{\rm SZ}}

\def\Mv {M_{\rm 500}}
\def \Rv {R_{500}}
\def\keV {\rm keV}
\def\Yv {Y_{500}}

\def\DAY{D_{\rm A}^2\, Y_{500}}

\def\msol {{\rm M_{\odot}}}

\def\lesssim{\mathrel{\hbox{\rlap{\hbox{\lower4pt\hbox{$\sim$}}}\hbox{$<$}}}}
\def\gtrsim{\mathrel{\hbox{\rlap{\hbox{\lower4pt\hbox{$\sim$}}}\hbox{$>$}}}}

\newcommand{\propsim}{\lower 3pt \hbox{$\, \buildrel {\textstyle
       \propto}\over {\textstyle \sim}\,$}}

\begin{document}
%This author list corresponds to \title{Author list for PIP 09, Proj. Ref. 5.1/5.2/5.5: The relation between galaxy cluster mass and Sunyaev-Zeldovich signal}
%Prepared by R. Leonardi (rleonardi@sciops.esa.int), ESAC/ESA
%This version is from Wed Jul 11 15:42:59 2012 CET
%\subtitle{There are 185 co-authors in this list}
\author{\small
Planck Collaboration:
P.~A.~R.~Ade\inst{76}
\and
N.~Aghanim\inst{53}
\and
M.~Arnaud\inst{68}
\and
M.~Ashdown\inst{65, 6}
\and
F.~Atrio-Barandela\inst{19}
\and
J.~Aumont\inst{53}
\and
C.~Baccigalupi\inst{75}
\and
A.~Balbi\inst{34}
\and
A.~J.~Banday\inst{84, 9}
\and
R.~B.~Barreiro\inst{61}
% ...
}
\institute{\small
APC, AstroParticule et Cosmologie, Universit\'{e} Paris Diderot, CNRS/IN2P3, CEA/lrfu, Observatoire de Paris, Sorbonne Paris Cit\'{e}, 10, rue Alice Domon et L\'{e}onie Duquet, 75205 Paris Cedex 13, France\\
\and
Aalto University Mets\"{a}hovi Radio Observatory, Mets\"{a}hovintie 114, FIN-02540 Kylm\"{a}l\"{a}, Finland\\
\and
Academy of Sciences of Tatarstan, Bauman Str., 20, Kazan, 420111, Republic of Tatarstan, Russia\\
\and
Agenzia Spaziale Italiana Science Data Center, c/o ESRIN, via Galileo Galilei, Frascati, Italy\\
\and
Agenzia Spaziale Italiana, Viale Liegi 26, Roma, Italy\\
\and
Astrophysics Group, Cavendish Laboratory, University of Cambridge, J J Thomson Avenue, Cambridge CB3 0HE, U.K.\\
\and
Atacama Large Millimeter/submillimeter Array, ALMA Santiago Central Offices, Alonso de Cordova 3107, Vitacura, Casilla 763 0355, Santiago, Chile\\
\and
CITA, University of Toronto, 60 St. George St., Toronto, ON M5S 3H8, Canada\\
\and
CNRS, IRAP, 9 Av. colonel Roche, BP 44346, F-31028 Toulouse cedex 4, France\\
\and
California Institute of Technology, Pasadena, California, U.S.A.\\
\and
Centre of Mathematics for Applications, University of Oslo, Blindern, Oslo, Norway\\
\and
Centro de Astrof\'{\i}sica, Universidade do Porto, Rua das Estrelas, 4150-762 Porto, Portugal\\
\and
Centro de Estudios de F\'{i}sica del Cosmos de Arag\'{o}n (CEFCA), Plaza San Juan, 1, planta 2, E-44001, Teruel, Spain\\
\and
Computational Cosmology Center, Lawrence Berkeley National Laboratory, Berkeley, California, U.S.A.\\
\and
Consejo Superior de Investigaciones Cient\'{\i}ficas (CSIC), Madrid, Spain\\
\and
DSM/Irfu/SPP, CEA-Saclay, F-91191 Gif-sur-Yvette Cedex, France\\
\and
DTU Space, National Space Institute, Juliane Mariesvej 30, Copenhagen, Denmark\\
\and
D\'{e}partement de Physique Th\'{e}orique, Universit\'{e} de Gen\`{e}ve, 24, Quai E. Ansermet,1211 Gen\`{e}ve 4, Switzerland\\
\and
Departamento de F\'{\i}sica Fundamental, Facultad de Ciencias, Universidad de Salamanca, 37008 Salamanca, Spain\\
\and
Departamento de F\'{\i}sica, Universidad de Oviedo, Avda. Calvo Sotelo s/n, Oviedo, Spain\\
\and
Department of Astronomy and Geodesy, Kazan Federal University,  Kremlevskaya Str., 18, Kazan, 420008, Russia\\
\and
Department of Astrophysics, IMAPP, Radboud University, P.O. Box 9010, 6500 GL Nijmegen,  The Netherlands\\
\and
Department of Physics \& Astronomy, University of British Columbia, 6224 Agricultural Road, Vancouver, British Columbia, Canada\\
\and
Department of Physics and Astronomy, Dana and David Dornsife College of Letter, Arts and Sciences, University of Southern California, Los Angeles, CA 90089, U.S.A.\\
\and
Department of Physics and Astronomy, University of Sussex, Brighton BN1 9QH, U.K.\\
\and
Department of Physics, Gustaf H\"{a}llstr\"{o}min katu 2a, University of Helsinki, Helsinki, Finland\\
\and
Department of Physics, University of California, Santa Barbara, California, U.S.A.\\
\and
Department of Physics, University of Illinois at Urbana-Champaign, 1110 West Green Street, Urbana, Illinois, U.S.A.\\
\and
Dipartimento di Fisica e Astronomia G. Galilei, Universit\`{a} degli Studi di Padova, via Marzolo 8, 35131 Padova, Italy\\
\and
Dipartimento di Fisica, Universit\`{a} La Sapienza, P. le A. Moro 2, Roma, Italy\\
\and
Dipartimento di Fisica, Universit\`{a} degli Studi di Milano, Via Celoria, 16, Milano, Italy\\
\and
Dipartimento di Fisica, Universit\`{a} degli Studi di Trieste, via A. Valerio 2, Trieste, Italy\\
\and
Dipartimento di Fisica, Universit\`{a} di Ferrara, Via Saragat 1, 44122 Ferrara, Italy\\
\and
Dipartimento di Fisica, Universit\`{a} di Roma Tor Vergata, Via della Ricerca Scientifica, 1, Roma, Italy\\
\and
Dipartimento di Matematica, Universit\`{a} di Roma Tor Vergata, Via della Ricerca Scientifica, 1, Roma, Italy\\
\and
Dpto. Astrof\'{i}sica, Universidad de La Laguna (ULL), E-38206 La Laguna, Tenerife, Spain\\
\and
European Southern Observatory, ESO Vitacura, Alonso de Cordova 3107, Vitacura, Casilla 19001, Santiago, Chile\\
\and
European Space Agency, ESAC, Planck Science Office, Camino bajo del Castillo, s/n, Urbanizaci\'{o}n Villafranca del Castillo, Villanueva de la Ca\~{n}ada, Madrid, Spain\\
\and
European Space Agency, ESTEC, Keplerlaan 1, 2201 AZ Noordwijk, The Netherlands\\
\and
Haverford College Astronomy Department, 370 Lancaster Avenue, Haverford, Pennsylvania, U.S.A.\\
\and
Helsinki Institute of Physics, Gustaf H\"{a}llstr\"{o}min katu 2, University of Helsinki, Helsinki, Finland\\
\and
INAF - Osservatorio Astronomico di Roma, via di Frascati 33, Monte Porzio Catone, Italy\\
\and
INAF - Osservatorio Astronomico di Trieste, Via G.B. Tiepolo 11, Trieste, Italy\\
\and
INAF/IASF Bologna, Via Gobetti 101, Bologna, Italy\\
\and
INAF/IASF Milano, Via E. Bassini 15, Milano, Italy\\
\and
INFN, Sezione di Roma 1, Universit`{a} di Roma Sapienza, Piazzale Aldo Moro 2, 00185, Roma, Italy\\
\and
INRIA, Laboratoire de Recherche en Informatique, Universit\'{e} Paris-Sud 11, B\^{a}timent 490, 91405 Orsay Cedex, France\\
\and
IPAG: Institut de Plan\'{e}tologie et d'Astrophysique de Grenoble, Universit\'{e} Joseph Fourier, Grenoble 1 / CNRS-INSU, UMR 5274, Grenoble, F-38041, France\\
\and
IUCAA, Post Bag 4, Ganeshkhind, Pune University Campus, Pune 411 007, India\\
\and
Imperial College London, Astrophysics group, Blackett Laboratory, Prince Consort Road, London, SW7 2AZ, U.K.\\
\and
Infrared Processing and Analysis Center, California Institute of Technology, Pasadena, CA 91125, U.S.A.\\
\and
Institut Universitaire de France, 103, bd Saint-Michel, 75005, Paris, France\\
\and
Institut d'Astrophysique Spatiale, CNRS (UMR8617) Universit\'{e} Paris-Sud 11, B\^{a}timent 121, Orsay, France\\
\and
Institut d'Astrophysique de Paris, CNRS (UMR7095), 98 bis Boulevard Arago, F-75014, Paris, France\\
\and
Institute for Space Sciences, Bucharest-Magurale, Romania\\
\and
Institute of Astro and Particle Physics, Technikerstrasse 25/8, University of Innsbruck, A-6020, Innsbruck, Austria\\
\and
Institute of Astronomy and Astrophysics, Academia Sinica, Taipei, Taiwan\\
\and
Institute of Astronomy, University of Cambridge, Madingley Road, Cambridge CB3 0HA, U.K.\\
\and
Institute of Theoretical Astrophysics, University of Oslo, Blindern, Oslo, Norway\\
\and
Instituto de Astrof\'{\i}sica de Canarias, C/V\'{\i}a L\'{a}ctea s/n, La Laguna, Tenerife, Spain\\
\and
Instituto de F\'{\i}sica de Cantabria (CSIC-Universidad de Cantabria), Avda. de los Castros s/n, Santander, Spain\\
\and
Istituto di Fisica del Plasma, CNR-ENEA-EURATOM Association, Via R. Cozzi 53, Milano, Italy\\
\and
Jet Propulsion Laboratory, California Institute of Technology, 4800 Oak Grove Drive, Pasadena, California, U.S.A.\\
\and
Jodrell Bank Centre for Astrophysics, Alan Turing Building, School of Physics and Astronomy, The University of Manchester, Oxford Road, Manchester, M13 9PL, U.K.\\
\and
Kavli Institute for Cosmology Cambridge, Madingley Road, Cambridge, CB3 0HA, U.K.\\
\and
LAL, Universit\'{e} Paris-Sud, CNRS/IN2P3, Orsay, France\\
\and
LERMA, CNRS, Observatoire de Paris, 61 Avenue de l'Observatoire, Paris, France\\
\and
Laboratoire AIM, IRFU/Service d'Astrophysique - CEA/DSM - CNRS - Universit\'{e} Paris Diderot, B\^{a}t. 709, CEA-Saclay, F-91191 Gif-sur-Yvette Cedex, France\\
\and
Laboratoire Traitement et Communication de l'Information, CNRS (UMR 5141) and T\'{e}l\'{e}com ParisTech, 46 rue Barrault F-75634 Paris Cedex 13, France\\
\and
Laboratoire de Physique Subatomique et de Cosmologie, Universit\'{e} Joseph Fourier Grenoble I, CNRS/IN2P3, Institut National Polytechnique de Grenoble, 53 rue des Martyrs, 38026 Grenoble cedex, France\\
\and
Laboratoire de Physique Th\'{e}orique, Universit\'{e} Paris-Sud 11 \& CNRS, B\^{a}timent 210, 91405 Orsay, France\\
\and
Max-Planck-Institut f\"{u}r Astrophysik, Karl-Schwarzschild-Str. 1, 85741 Garching, Germany\\
\and
Max-Planck-Institut f\"{u}r Extraterrestrische Physik, Giessenbachstra{\ss}e, 85748 Garching, Germany\\
\and
Optical Science Laboratory, University College London, Gower Street, London, U.K.\\
\and
SISSA, Astrophysics Sector, via Bonomea 265, 34136, Trieste, Italy\\
\and
School of Physics and Astronomy, Cardiff University, Queens Buildings, The Parade, Cardiff, CF24 3AA, U.K.\\
\and
Space Research Institute (IKI), Profsoyuznaya 84/32, Moscow, Russia\\
\and
Space Research Institute (IKI), Russian Academy of Sciences, Profsoyuznaya Str, 84/32, Moscow, 117997, Russia\\
\and
Space Sciences Laboratory, University of California, Berkeley, California, U.S.A.\\
\and
Special Astrophysical Observatory, Russian Academy of Sciences, Nizhnij Arkhyz, Zelenchukskiy region, Karachai-Cherkessian Republic, 369167, Russia\\
\and
Stanford University, Dept of Physics, Varian Physics Bldg, 382 Via Pueblo Mall, Stanford, California, U.S.A.\\
\and
T\"{U}B\.{I}TAK National Observatory, Akdeniz University Campus, 07058, Antalya, Turkey\\
\and
UPMC Univ Paris 06, UMR7095, 98 bis Boulevard Arago, F-75014, Paris, France\\
\and
Universit\'{e} de Toulouse, UPS-OMP, IRAP, F-31028 Toulouse cedex 4, France\\
% ...
}

\title{\textit{Planck} intermediate results. III. The relation between galaxy cluster mass and Sunyaev-Zeldovich signal} 
\authorrunning{Planck Collaboration}
\titlerunning{The relation between galaxy cluster mass and Sunyaev-Zeldovich signal}

%\date{Received XXXX; accepted YYYY}

\abstract{We examine the relation between the galaxy cluster mass $M$ and Sunyaev-Zeldovich (SZ) effect signal $\DAY$ for a sample of 19 objects for which weak lensing (WL) mass measurements obtained from Subaru Telescope data are available in the literature. Hydrostatic X-ray masses are derived from \xmm\ archive data, and the SZ effect signal is measured from \planck\ all-sky survey data. We find an $M_{\rm WL}$--$\DAY$ relation that is consistent in slope and normalisation with previous determinations using weak lensing masses; however, there is a normalisation offset with respect to previous measures based on hydrostatic X-ray mass-proxy relations. We verify that our SZ effect measurements are in excellent agreement with previous determinations from \planck\ data. For the present sample, the hydrostatic X-ray masses at $\Rv$ are on average $\sim 20$ per cent larger than the corresponding weak lensing masses, which is contrary to expectations. We show that the mass discrepancy is driven by a difference in mass concentration as measured by the two methods and, for the present sample, that the mass discrepancy and difference in mass concentration are especially large for disturbed systems. The mass discrepancy is also linked to the offset in centres used by the X-ray and weak lensing analyses, which again is most important in disturbed systems. We outline several approaches that are needed to help achieve convergence in cluster mass measurement with X-ray and weak lensing observations. }


   \maketitle
%
%________________________________________________________________

\section{Introduction}
Although the Sunyaev-Zeldovich (SZ) effect was discovered in 1972, it has taken almost until the present day for its potential to be fully realised. Our observational and theoretical understanding of galaxy clusters has improved immeasurably in the last 40 years, of course. But recent advances in detection sensitivity, together with the advent of large-area survey capability, have revolutionised the SZ field, allowing vast improvements in sensitivity and dynamic range to be obtained \citep[e.g.,][]{poi99,kom99,kor11} and catalogues of tens to hundreds of SZ-detected clusters to be compiled \citep[e.g.,][]{van10,mar11,planck2011-5.1a,rei12}. 

The SZ signal is of singular interest because it is not affected by cosmological dimming and because the total SZ flux or integrated Compton parameter, $Y_{\rm SZ}$, is expected to correlate particularly tightly with mass \citep[e.g.,][]{bar96,das04,mot05,nag06,wik08,agh09}. SZ-detected cluster samples are thus expected to range to high redshift and be as near as possible to mass-selected, making them potentially very powerful cosmological probes. Notwithstanding, a well-calibrated relationship between the total mass and the observed SZ signal is needed to leverage the statistical potential of these new cluster samples.

In fact the relationship between mass and $Y_{\rm SZ}$ is still poorly determined, owing in large part to the difficulty of making sufficiently precise measurements of either quantity. Moreover, the majority of mass measurements used to date \citep[e.g.,][]{ben04,bon08,and10,planck2011-5.2b} have relied on X-ray observations that assume hydrostatic equilibrium, which many theoretical studies tell us is likely to result in a mass that is systematically underestimated by about $10$--$15$ percent due to neglect of bulk motions in the intracluster medium   \citep[ICM; e.g.][]{nag07,pif08,mene10}. This effect is now commonly referred to in the literature as the ``hydrostatic mass bias''. 

In this context, weak lensing observations offer an alternative way of measuring the total mass. As the weak lensing effect is due directly to the gravitational potential, it is generally thought to be unbiased. However, it is a technique that is sensitive to all the mass along the line of sight, so that projection effects may play an important role in adding scatter to any observed relation. In addition, as it only measures the projected (2D) mass, analytical models are needed to transform into the more physically motivated spherical (3D) mass, and this is likely to add further noise because of cluster triaxiality \citep[e.g.,][]{cor07,mene10}. Furthermore, recent theoretical work suggests that some bias may in fact be present in weak lensing observations. The systematic 5--10 percent underestimate of the true mass in the simulations of \citet{bec11} is apparently due to the use of a Navarro-Frenk-White (NFW) model that does not describe the data correctly at large radii. Notwithstanding, the most recent observational results from small samples of clusters for which both X-ray and weak lensing data are available indicate either that there is good agreement between X-ray and weak lensing masses \citep{zha10,vik09}, or that the X-ray mass is systematically lower than the weak lensing mass by up to 20 percent, with the underestimate being more important at larger radii \citep[e.g.,][]{mah08}. 

The only investigations of the mass-$Y_{\rm SZ}$ relation using weak lensing masses published to date have been those of \citet{mar09,marr11}, using data from the Local Cluster Substructure Survey (LoCuSS)\footnote{\url{http://www.sr.bham.ac.uk/locuss/index.php}}. The first directly compared 2D quantites (i.e., cylindrical SZ effect vs projected mass) within a fixed physical radius of 350 kpc, while the second compared the spherically integrated Compton parameter against deprojected mass. In the latter case, a much larger scatter than expected was found, which the authors attributed to line of sight projection effects in weak lensing mass estimates.

In the present paper we make use of the same weak lensing data set from LoCuSS, high quality \xmm\ archival X-ray data, and SZ observations from the \planck\footnote{\planck\ (\url{http://www.esa.int/Planck}) is a project of the European Space Agency (ESA) with instruments provided by two scientific consortia funded by ESA member 
states (in particular the lead countries France and Italy), with contributions from NASA (USA) and telescope reflectors provided by a collaboration between ESA and a scientific consortium led and funded by Denmark.} All-Sky Survey to investigate the interplay between the different mass measures and the spherically integrated Compton parameter $Y_{\rm SZ}$ in 19 clusters of galaxies. We find that for this particular sample, the weak lensing mass-$Y_{\rm SZ}$ relation at large radii has a slightly higher normalisation than that expected from studies based on hydrostatic X-ray mass estimates. We show that this is due to the hydrostatic X-ray masses being, on average, {\it larger} than the corresponding weak lensing masses, in contradiction with the expectations from numerical simulations. We show that the problem is particularly acute for merging systems and appears to be due, at least in part, to a systematic difference in the concentration as measured by the two methods. In addition, an offset between the centres used for the X-ray and weak lensing mass determinations appears to introduce a secondary systematic effect.

We adopt a $\Lambda$CDM cosmology with $H_0=70$~km~s$^{-1}$~Mpc$^{-1}$, $\Omega_{\rm M}=0.3$ and $\Omega_\Lambda=0.7$.  The factor $E(z) = \sqrt{\Omega_{\rm M} (1+z)^3+\Omega_\Lambda}$ is the ratio of the Hubble constant at redshift $z$ to its present day value.  The variables $M_{\Delta}$ and $R_{\Delta}$ are the total mass and radius corresponding to a total density contrast $\Delta \, \rho_{\rm c}(z)$, where $\rho_{\rm c}(z)$ is the critical 
density of the Universe at the cluster redshift; thus, e.g., $M_{500} = (4\pi/3)\,500\,\rho_{\rm c}(z)\,R_{500}^3$.  The quantity $Y_{\rm X}$ is defined as the product of  $\Mgv$, the gas mass within $\Rv$, and $\TX$, the spectroscopic temperature measured in the $[0.15$--$0.75]~\Rv$ aperture. The SZ flux is characterised by $Y_{\Delta}$, where $Y_{\Delta}\,D_{\rm A}^2$ is the spherically integrated Compton parameter within $R_{\Delta}$, and $D_{\rm A}$ is the angular-diameter distance to the cluster. %Note that $Y_{\Delta}$ is the directly observed `apparent' quantity, while $D_{\rm A}^2\, Y_{\Delta}$ is the corresponding `absolute' quantity, intrinsic to the cluster. 
All uncertainties are given at the 68 percent confidence level.  

%%%%%%%%%%%%%%%%%%%%%%%%%%%%%%%%%%%%%

\section{Sample selection and data sets}

%______________________________________________________________
% Table of basic properties
%
\begin{table*}[ht]
\begin{centering}
\caption{{\footnotesize Basic properties of the sample. }\label{tab:basic}} 
\newcolumntype{L}{>{\columncolor{white}[0pt][\tabcolsep]}l}
\newcolumntype{R}{>{\columncolor{white}[\tabcolsep][0pt]}l}
%\rowcolors{3}{light-grey}{white}
%\resizebox{\textwidth}{!} {
\begin{tabular}{@{}LrrrrrrrrR@{}}
\toprule
\toprule

\multicolumn{1}{l}{} & \multicolumn{1}{l}{} & \multicolumn{2}{c}{X-ray} & \multicolumn{2}{c}{Weak lensing} & \multicolumn{1}{c}{Offset} & \multicolumn{1}{c}{Relaxed} & \multicolumn{1}{c}{Disturbed} \\
\cmidrule[0.5pt](lr){3-4}
\cmidrule[0.5pt](lr){5-6}
\cmidrule[0.5pt](lr){7-7}
\noalign{\smallskip}

\multicolumn{1}{l}{Cluster} & \multicolumn{1}{l}{$z$} & \multicolumn{1}{c}{R.A.} & 
\multicolumn{1}{c }{Dec.} & \multicolumn{1}{c}{R.A.} &
\multicolumn{1}{c }{Dec.} & \multicolumn{1}{c}{(arcmin)} &
\multicolumn{1}{c}{} \\

\midrule
            A68  & 0.255  & $00:37:06.7$ & $+09:09:24.6$ & $00:37:06.9$ & $+09:09:24.5 $ &  0.05  &  \ldots & \ldots\\
           A209  & 0.206  & $01:31:52.6$ & $-13:36:40.4$ & $01:31:52.5$ & $-13:36:40.5 $ &  0.01  &  \ldots & \checkmark \\
           A267  & 0.230  & $01:52:42.3$ & $+01:00:33.7$ & $01:52:41.9$ & $+01:00:25.7 $ &  0.15  &   \ldots & \checkmark \\
           A291  & 0.196  & $02:01:43.1$ & $-02:11:48.4$ & $02:01:43.1$ & $-02:11:50.4 $ &  0.03  &  \checkmark & \ldots \\
           A383  & 0.188  & $02:48:03.4$ & $-03:31:44.0$ & $02:48:03.4$ & $-03:31:44.7 $ &  0.01  &  \checkmark & \ldots \\
           A521  & 0.248  & $04:54:08.4$ & $-10:14:15.9$ & $04:54:06.8$ & $-10:13:25.8 $ &  0.88  &   \ldots & \checkmark \\
           A520  & 0.203  & $04:54:09.6$ & $+02:55:16.4$ & $04:54:14.0$ & $+02:57:11.6 $ &  2.19  &   \ldots & \checkmark \\
           A963  & 0.206  & $10:17:03.7$ & $+39:02:53.4$ & $10:17:03.6$ & $+39:02:50.0 $ &  0.06  &  \ldots & \ldots \\
          A1835  & 0.253  & $14:01:02.2$ & $+02:52:41.7$ & $14:01:02.1$ & $+02:52:42.8 $ &  0.03  & \checkmark & \ldots  \\
          A1914  & 0.171  & $14:26:02.5$ & $+37:49:27.6$ & $14:25:56.7$ & $+37:48:58.9 $ &  1.52  &  \ldots & \ldots \\
ZwCl1454.8+2233  & 0.258  & $14:57:15.1$ & $+22:20:32.7$ & $14:57:15.2$ & $+22:20:33.6 $ &  0.02  & \checkmark & \ldots  \\
ZwCl1459.4+4240  & 0.290  & $15:01:22.7$ & $+42:20:47.4$ & $15:01:23.1$ & $+42:20:38.0 $ &  0.16  &   \ldots & \checkmark \\
          A2034  & 0.113  & $15:10:12.7$ & $+33:30:37.6$ & $15:10:11.8$ & $+33:29:12.3 $ &  1.43  &   \ldots & \checkmark \\
          A2219  & 0.228  & $16:40:20.1$ & $+46:42:38.4$ & $16:40:19.7$ & $+46:42:42.0 $ &  0.11  &  \ldots & \ldots \\
 RXJ1720.1+2638  & 0.164  & $17:20:10.0$ & $+26:37:30.1$ & $17:20:10.1$ & $+26:37:30.5 $ &  0.01  &  \checkmark & \ldots  \\
          A2261  & 0.224  & $17:22:27.0$ & $+32:07:56.5$ & $17:22:27.2$ & $+32:07:57.1 $ &  0.03  &  \ldots & \ldots \\
 RXJ2129.6+0005  & 0.235  & $21:29:40.0$ & $+00:05:19.0$ & $21:29:40.0$ & $+00:05:21.8 $ &  0.05  &  \checkmark & \ldots  \\
          A2390  & 0.231  & $21:53:36.7$ & $+17:41:41.8$ & $21:53:36.8$ & $+17:41:43.3 $ &  0.03  &  \checkmark & \ldots  \\
          A2631  & 0.278  & $23:37:37.5$ & $+00:16:00.4$ & $23:37:39.7$ & $+00:16:17.0 $ &  0.59  &  \ldots & \ldots \\
\bottomrule
\end{tabular}
%}

\tablefoot{X-ray coordinates correspond to the peak of the X-ray emission. The weak lensing coordinates correspond to the position of the BCG. Cluster morphological classification is described in Sect.~\ref{sec:morpho}.}
\end{centering}
\end{table*}
%______________________________________________________________

The present investigation requires three fundamental data sets. The first is a homogeneously analysed weak lensing data set with published NFW mass model parameters to enable calculation of the mass at the radius corresponding to  any desired density contrast. The second is a good quality X-ray observation data set that allows detection of the X-ray emission to large radius (i.e., at least up to $R_{500}$). The third is a good quality SZ data set including high signal-to-noise SZ flux measurements for all systems. 

While there are many weak lensing investigations of individual objects in the literature, lensing observations of moderately large cluster {\it samples} are comparatively rare. For the present comparison, we chose to use published results from LoCuSS, which is an all-sky X-ray-selected sample of 100 massive galaxy clusters at $0.1< z < 0.3$ drawn from the REFLEX \citep{boe04} and eBCS \citep{ebe00} catalogues, for which gravitational lensing data from the \textit{Hubble Space Telescope} and Subaru Telescope are being accumulated. At the time of writing, relevant data from only part of the full sample have been published, as detailed below. 

Results from a Subaru weak lensing analysis of 30 LoCuSS clusters have been published by \citet{oka10}, who provide NFW mass model parameters for 26 systems. A similar lensing analysis has been undertaken on a further seven merging systems by \citet{oka08}. Excluding the two of these merging clusters that are bimodal and  thus not resolved in the \planck\ beam, we have a total of 31 objects for which the best-fitting NFW profile mass model is available in the literature. These are ideal for our study, given the object selection process (massive X-ray clusters) and the fact that the lensing analysis procedure is the same for all systems.

High-quality \xmm\ X-ray data with at least 10\,ks EMOS exposure time are available for 21 of these clusters. Since we wished to undertake a fully homogeneous analysis of the X-ray data, we excluded two systems, A754 and A2142, whose observations consist of a mosaic of several pointings each. For the remaining 19 systems, a full hydrostatic X-ray mass analysis is possible using the approach described below in Sect.~\ref{sec:xmass}.

SZ observations of the full sample are available from \planck, \citep{tauber2010a, Planck2011-1.1} the third-generation space mission to measure the anisotropy of the cosmic microwave background (CMB). \planck\ observes the sky in nine frequency bands covering 30--857\,GHz with high sensitivity and angular resolution from 31\arcm\ to 5\arcm. The Low Frequency Instrument \citep[LFI;][]{Mandolesi2010, Bersanelli2010,Planck2011-1.4} covers the 30, 44, and 70\,GHz bands with amplifiers cooled to 20\,\hbox{K}. The High Frequency Instrument (HFI; \citealt{Lamarre2010, Planck2011-1.5}) covers the 100, 143, 217, 353, 545, and 857\,GHz bands with bolometers cooled to 0.1\,\hbox{K}.  Polarisation is measured in all but the two highest frequency bands \citep{Leahy2010, Rosset2010}. A combination of radiative cooling and three mechanical coolers produces the temperatures needed for the detectors and optics
\citep{Planck2011-1.3}. Two data processing centres (DPCs) check and calibrate the data and make maps of the sky \citep{Planck2011-1.7, Planck2011-1.6}.  \Planck's sensitivity, angular resolution, and frequency coverage make it a powerful instrument for Galactic and extragalactic astrophysics as well as cosmology. Early astrophysics results, based on data taken between 13~August 2009 and 7~June 2010, are given in Planck Collaboration VIII--XXVI 2011.  Intermediate astrophysics results are now being presented in a series of papers based on data taken between 13~August 2009 and 27~November 2010. All of the 19 systems considered in this paper have been observed by \planck\ as part of this survey, and indeed their characteristics are such that they are almost all strongly detected, having a median signal-to-noise ratio of $\sim 7$. 

%%%%%%%%%%%%%%%%%%%%%%%%%%%%%%%%%%%%%

\section{Data preparation and analysis}

Table~\ref{tab:basic} lists basic details of the cluster sample, including name, redshift, and the coordinates of the X-ray and weak lensing centres.

\subsection{Weak lensing}

As mentioned above, spherical weak lensing masses for the sample are given by \cite{oka08} and \citet{oka10}. These were derived from fitting a projected NFW model to a tangential distortion profile centred on the position of the brightest cluster galaxy (BCG). In all cases we converted the best-fitting NFW profile model to our chosen cosmology and obtained $M_\Delta$ by interpolating to the density contrast of interest\footnote{For A963, only one-band imaging data are available, which may lead to an underestimate of the weak lensing mass (Okabe et al. 2012, priv. communication).}. For the 16 clusters in \citet{oka10}, we used the published fractional uncertainties at $\Delta = 500$ and 2500. Uncertainties at $\Delta = 1000$ were obtained from Okabe et al. (2012, priv. communication).
For the three clusters published by \citet[][A520, A1914, and A2034]{oka08}, only $M_{\rm vir}$ and $c_{\rm vir}$ are available. Here we estimated the uncertainties at each density contrast by multiplying the fractional uncertainty on $M_{\rm vir}$ given in \citet{oka08} by the median fractional uncertainty, relative to $M_{\rm vir}$, of all other clusters in \citet{oka10} at this density contrast. 

%%%%%%%%%%%%%%%%%%%%%%%%%%%%%%%%%%%%%

\subsection{X-ray}

%%%%%%%%%%%%%%%%%%%%%%%%%%%%%%%%%%%%%

\subsubsection{Data analysis}

The preliminary X-ray data analysis follows that described in \citet{pra07}, \citet{pra10}, and \citet{planck2011-5.2b}. In brief, surface brightness profiles centred on the X-ray emission peak were extracted in the 0.3--2\,keV band and used to derive the regularised gas density profiles, $n_{\rm e}(r)$, using the non-parametric deprojection and PSF-correction method of \citet{cro06}. The projected temperature was measured in annuli as described in \citet{pra10}. The 3D temperature profiles, $T(r)$, were calculated by convolving a suitable parametric model with a response matrix that takes into account projection and PSF effects, projecting this model accounting for the bias introduced by fitting isothermal models to multi-temperature plasma emission \citep{maz04,vik06w}, and fitting to the projected annular profile. Note that in addition to point sources, obvious X-ray sub-structures (corresponding to, e.g., prominent secondary maxima in the X-ray surface brightness) were excised before calculating the density and temperature profiles discussed above.

%%%%%%%%%%%%%%%%%%%%%%%%%%%%%%%%%%%%%

\subsubsection{X-ray mass profile}
\label{sec:xmass}

The X-ray mass was calculated for each cluster as described in \citet{dem10}. Using the gas density $n_{\rm e}(r)$ and temperature $T(r)$ profiles, and assuming  hydrostatic equilibrium, the total mass is given by:
\begin{equation}
M\,(\leq R) = -\frac{kT(r)\, r}{G\, \mu m_{\rm p}}\left[\frac{d\ln{n_{\rm e}(r)}}{d\ln{r}}+ \frac{d\ln{T(r)}}{d\ln{r}}\right].\label{eqn:xhe}
\end{equation}
To suppress noise due to structure in the regularised gas density profiles, we fitted them with the parametric model described by \citet{vik06} and used the radial derivative $d \ln{n_{\rm e}}/ d \ln{r}$  given by this parametric function fit. The corresponding uncertainties were given by differentiation of the regularised density profile at each point corresponding to the effective radius of the deconvolved temperature profile.

Uncertainties on each X-ray mass point were calculated using a Monte Carlo approach based on that of \citet{pra03}, where a random temperature was generated at each radius at which the temperature profile is measured, and a cubic spline used to compute the derivative. We only kept random profiles that were physical, meaning that the mass profile must increase monotonically with radius and the randomised temperature profiles must be convectively stable, assuming the standard Schwarzschild criterion in the abscence of strong heat conductivity, i.e., $d \ln{T} / d\ln{n_e} < 2/3$. The number of rejected profiles varied on a cluster-by-cluster basis, with morphologically disturbed clusters generally requiring more discards. The final mass profiles were built from a minimum of 100 and a maximum of 1000 Monte Carlo realisations. The mass at each density contrast relative to the critical density of the Universe, $M_{\Delta}$, was calculated via interpolation in the $\log{M}  - \log{\Delta}$ plane. The uncertainty on the resulting mass value was then calculated from the region containing 68 percent of the realisations on each side. 

For two clusters, the hydrostatic X-ray mass determinations should be treated with some caution. The first is A521, which is a well-known merging system. Here the gas density profile at large radius declines precipitously, yielding a $d\ln{n_e}/ d\ln{r}$ value that results in an integrated mass profile that is practically a pure power law at large cluster-centric distances. Although we excised the obvious substructure to the north-west before the X-ray mass analysis, the complex nature of this system precludes a precise X-ray mass analysis. The second cluster for which the X-ray mass determination is suspect is A2261, for the more prosaic reason that the X-ray temperature profile is only detected up to $R_{\rm det,max} \sim 0.6\,\Rv \sim 0.8\,R_{1000}$. In the following, we exclude these clusters in cases when the hydrostatic X-ray mass is under discussion.

%______________________________________________________________
% Table of all parameters
%
\begin{table*}[]
\begin{centering}
\caption{{\footnotesize Masses, X-ray and SZ properties. X-ray masses are calculated as described in Sect.~\ref{sec:xmass}; weak lensing masses were published in the LoCuSS weak lensing analysis papers \citep{oka08,oka10}. }\label{tab:xray}} 
\newcolumntype{L}{>{\columncolor{white}[0pt][\tabcolsep]}l}
\newcolumntype{R}{>{\columncolor{white}[\tabcolsep][0pt]}l}
%\rowcolors{3}{light-grey}{white}
\resizebox{\textwidth}{!} {
\begin{tabular}{@{}LrrrrrrrrrR@{}}
\toprule
\toprule

\multicolumn{1}{c}{} & \multicolumn{2}{c}{2500} & \multicolumn{2}{c}{1000} & \multicolumn{5}{c}{500} \\
\cmidrule[0.5pt](rl){2-3}
\cmidrule[0.5pt](rl){4-5}
\cmidrule[0.5pt](rl){6-10}

\noalign{\smallskip}

\multicolumn{1}{l}{Name} & \multicolumn{1}{c}{$M_{2500}$} & 
\multicolumn{1}{c}{$D_{\rm A}^2\, Y_{2500}$} & \multicolumn{1}{c}{$M_{1000}$} &
\multicolumn{1}{c}{$D_{\rm A}^2\, Y_{1000}$} & \multicolumn{1}{c}{$M_{500}$} &
\multicolumn{1}{c}{$D_{\rm A}^2\, Y_{500}$} &
\multicolumn{1}{c}{$M_{\rm gas,500}$} &
\multicolumn{1}{c}{$T_{\rm X}$} &
\multicolumn{1}{c}{$Y_{\rm X, 500}$} \\

\multicolumn{1}{l}{} & \multicolumn{1}{c}{($10^{14}$ M$_{\odot}$)} & 
\multicolumn{1}{c}{($10^{-5}$ Mpc$^2$)} & \multicolumn{1}{c}{($10^{14}$ M$_{\odot}$)} &
\multicolumn{1}{c}{($10^{-5}$ Mpc$^2$)} & \multicolumn{1}{c}{($10^{14}$ M$_{\odot}$)} &
\multicolumn{1}{c}{($10^{-5}$ Mpc$^2$)} &
\multicolumn{1}{c}{($10^{13}$ M$_{\odot}$)} &
\multicolumn{1}{c}{(keV)} &
\multicolumn{1}{c}{(M$_{\odot}$ keV)} \\

\noalign{\smallskip}
\midrule

& &  & & & & & & & &\\
\multicolumn{11}{c}{X-ray}\\
& &  & & & & & & & &\\

            A68 & $ 4.3_{- 0.6}^{+ 0.6}$ & $ 3.8_{- 0.8}^{+ 0.8}$ & $ 5.9_{- 0.8}^{+ 0.9}$ & $ 5.5_{- 1.2}^{+ 1.2}$ & $ 6.9_{- 1.1}^{+ 1.2}$ & $ 6.5_{- 1.4}^{+ 1.4}$ & $ 7.9_{- 0.1}^{+ 0.1}$ & $ 8.1_{- 0.3}^{+ 0.3}$ & $ 6.4_{- 0.3}^{+ 0.3}$ &  \\
           A209 & $ 1.6_{- 0.5}^{+ 0.6}$ & $ 2.3_{- 0.2}^{+ 0.2}$ & $ 4.3_{- 1.0}^{+ 0.9}$ & $ 6.6_{- 0.5}^{+ 0.5}$ & $ 6.3_{- 1.0}^{+ 1.0}$ & $10.7_{- 0.9}^{+ 0.9}$ & $10.6_{- 0.1}^{+ 0.1}$ & $ 6.7_{- 0.1}^{+ 0.1}$ & $ 7.1_{- 0.2}^{+ 0.2}$ &  \\
           A267 & $ 1.9_{- 0.4}^{+ 0.4}$ & $ 2.1_{- 0.5}^{+ 0.5}$ & $ 3.0_{- 0.5}^{+ 0.5}$ & $ 3.6_{- 0.8}^{+ 0.8}$ & $ 3.6_{- 0.5}^{+ 0.5}$ & $ 4.5_{- 1.1}^{+ 1.1}$ & $ 4.0_{- 0.0}^{+ 0.0}$ & $ 5.5_{- 0.1}^{+ 0.1}$ & $ 2.2_{- 0.1}^{+ 0.1}$ &  \\
$\dagger$  A291 & $ 1.5_{- 0.3}^{+ 0.3}$ & $ 0.9_{- 0.4}^{+ 0.4}$ & $ 2.3_{- 0.3}^{+ 0.4}$ & $ 1.5_{- 0.7}^{+ 0.7}$ & $ 2.7_{- 0.4}^{+ 0.4}$ & $ 1.9_{- 0.9}^{+ 0.9}$ & $ 3.8_{- 0.0}^{+ 0.0}$ & $ 4.0_{- 0.1}^{+ 0.1}$ & $ 1.5_{- 0.0}^{+ 0.0}$ &  \\
$\dagger$  A383 & $ 1.6_{- 0.2}^{+ 0.2}$ & $ 0.5_{- 0.4}^{+ 0.4}$ & $ 2.2_{- 0.3}^{+ 0.3}$ & $ 0.8_{- 0.7}^{+ 0.7}$ & $ 2.6_{- 0.4}^{+ 0.4}$ & $ 0.9_{- 0.8}^{+ 0.8}$ & $ 3.7_{- 0.0}^{+ 0.0}$ & $ 4.2_{- 0.1}^{+ 0.1}$ & $ 1.6_{- 0.0}^{+ 0.0}$ &  \\
$\ddagger$ A521 & $ 0.7_{- 0.6}^{+ 0.3}$ & $ 1.0_{- 0.1}^{+ 0.1}$ & $ 5.6_{- 1.5}^{+ 2.0}$ & $ 8.1_{- 1.2}^{+ 1.2}$ & $39.1_{-35.1}^{+15.0}$ & $14.1_{- 2.1}^{+ 2.1}$ & $ 6.5_{- 0.0}^{+ 0.0}$ & $ 5.6_{- 0.1}^{+ 0.1}$ & $ 3.7_{- 0.1}^{+ 0.1}$ &  \\
           A520 & $ 2.9_{- 2.3}^{+ 0.8}$ & $ 3.9_{- 0.5}^{+ 0.5}$ & $ 6.0_{- 1.1}^{+ 1.1}$ & $ 8.1_{- 1.0}^{+ 1.0}$ & $ 8.3_{- 1.8}^{+ 2.3}$ & $10.6_{- 1.3}^{+ 1.3}$ & $11.4_{- 0.1}^{+ 0.1}$ & $ 8.0_{- 0.2}^{+ 0.2}$ & $ 9.1_{- 0.3}^{+ 0.4}$ &  \\
           A963 & $ 2.0_{- 0.4}^{+ 0.4}$ & $ 2.1_{- 0.3}^{+ 0.3}$ & $ 3.9_{- 0.7}^{+ 0.7}$ & $ 4.0_{- 0.6}^{+ 0.6}$ & $ 4.9_{- 0.9}^{+ 1.0}$ & $ 5.3_{- 0.8}^{+ 0.8}$ & $ 6.7_{- 0.0}^{+ 0.0}$ & $ 5.6_{- 0.1}^{+ 0.1}$ & $ 3.7_{- 0.1}^{+ 0.1}$ &  \\
          A1835 & $ 5.9_{- 0.6}^{+ 0.7}$ & $ 9.0_{- 0.9}^{+ 0.9}$ & $ 7.7_{- 0.7}^{+ 0.7}$ & $12.8_{- 1.3}^{+ 1.3}$ & $ 8.2_{- 0.7}^{+ 0.7}$ & $14.7_{- 1.5}^{+ 1.5}$ & $11.6_{- 0.0}^{+ 0.0}$ & $ 8.3_{- 0.1}^{+ 0.1}$ & $ 9.7_{- 0.2}^{+ 0.2}$ &  \\
          A1914 & $ 4.1_{- 0.4}^{+ 0.5}$ & $ 5.0_{- 0.4}^{+ 0.4}$ & $ 5.5_{- 0.8}^{+ 0.9}$ & $ 7.1_{- 0.6}^{+ 0.6}$ & $ 6.9_{- 1.2}^{+ 1.4}$ & $ 8.5_{- 0.7}^{+ 0.7}$ & $10.8_{- 0.1}^{+ 0.0}$ & $ 8.3_{- 0.2}^{+ 0.2}$ & $ 8.9_{- 0.2}^{+ 0.2}$ &  \\
ZwCl1454.8+2233 & $ 1.8_{- 0.3}^{+ 0.3}$ & $ 1.9_{- 0.5}^{+ 0.5}$ & $ 2.7_{- 0.4}^{+ 0.5}$ & $ 2.9_{- 0.7}^{+ 0.7}$ & $ 3.4_{- 0.6}^{+ 0.6}$ & $ 3.7_{- 0.9}^{+ 0.9}$ & $ 4.9_{- 0.0}^{+ 0.0}$ & $ 4.6_{- 0.1}^{+ 0.1}$ & $ 2.3_{- 0.0}^{+ 0.1}$ &  \\
ZwCl1459.4+4240 & $ 2.3_{- 0.5}^{+ 0.6}$ & $ 1.9_{- 0.5}^{+ 0.5}$ & $ 4.1_{- 0.8}^{+ 0.8}$ & $ 3.4_{- 0.8}^{+ 0.8}$ & $ 5.4_{- 1.1}^{+ 1.1}$ & $ 4.5_{- 1.1}^{+ 1.1}$ & $ 7.0_{- 0.1}^{+ 0.1}$ & $ 6.3_{- 0.2}^{+ 0.2}$ & $ 4.4_{- 0.2}^{+ 0.2}$ &  \\
          A2034 & $ 3.3_{- 0.6}^{+ 0.4}$ & $ 2.9_{- 0.3}^{+ 0.3}$ & $ 4.1_{- 0.6}^{+ 0.6}$ & $ 4.4_{- 0.4}^{+ 0.4}$ & $ 5.7_{- 0.6}^{+ 0.6}$ & $ 5.5_{- 0.5}^{+ 0.5}$ & $ 7.0_{- 0.1}^{+ 0.0}$ & $ 6.3_{- 0.2}^{+ 0.2}$ & $ 4.4_{- 0.2}^{+ 0.1}$ &  \\
          A2219 & $ 4.5_{- 0.6}^{+ 0.7}$ & $ 7.5_{- 0.4}^{+ 0.4}$ & $ 7.2_{- 1.2}^{+ 1.5}$ & $14.5_{- 0.7}^{+ 0.7}$ & $10.3_{- 2.5}^{+ 3.0}$ & $21.2_{- 1.0}^{+ 1.0}$ & $17.6_{- 0.1}^{+ 0.2}$ & $ 9.5_{- 0.2}^{+ 0.3}$ & $16.8_{- 0.5}^{+ 0.7}$ &  \\
 RXJ1720.1+2638 & $ 2.4_{- 0.5}^{+ 0.6}$ & $ 2.6_{- 0.3}^{+ 0.3}$ & $ 4.4_{- 0.7}^{+ 0.8}$ & $ 4.7_{- 0.5}^{+ 0.5}$ & $ 6.0_{- 1.2}^{+ 1.3}$ & $ 6.5_{- 0.7}^{+ 0.7}$ & $ 6.9_{- 0.0}^{+ 0.0}$ & $ 5.8_{- 0.1}^{+ 0.1}$ & $ 4.0_{- 0.1}^{+ 0.1}$ &  \\
$\ddagger$A2261 & $ 2.5_{- 0.5}^{+ 0.5}$ & $ 4.9_{- 0.5}^{+ 0.5}$ & $ 3.1_{- 0.4}^{+-0.1}$ & $ 7.1_{- 0.7}^{+ 0.7}$ & $ 3.9_{- 0.6}^{+ 0.6}$ & $ 9.0_{- 0.8}^{+ 0.8}$ & $ 9.4_{- 0.2}^{+ 0.2}$ & $ 6.7_{- 0.5}^{+ 0.5}$ & $ 6.3_{- 0.7}^{+ 0.7}$ &  \\
 RXJ2129.6+0005 & $ 2.5_{- 0.5}^{+ 0.4}$ & $ 1.9_{- 0.5}^{+ 0.5}$ & $ 3.7_{- 0.4}^{+ 0.4}$ & $ 3.1_{- 0.8}^{+ 0.8}$ & $ 4.3_{- 0.5}^{+ 0.5}$ & $ 3.9_{- 1.0}^{+ 1.0}$ & $ 7.3_{- 0.0}^{+ 0.0}$ & $ 5.6_{- 0.1}^{+ 0.1}$ & $ 4.1_{- 0.1}^{+ 0.1}$ &  \\
          A2390 & $ 4.9_{- 1.1}^{+ 1.3}$ & $ 6.2_{- 0.4}^{+ 0.4}$ & $ 7.8_{- 1.5}^{+ 1.4}$ & $11.5_{- 0.8}^{+ 0.8}$ & $ 9.7_{- 1.6}^{+ 1.7}$ & $15.7_{- 1.1}^{+ 1.1}$ & $15.8_{- 0.1}^{+ 0.1}$ & $ 9.0_{- 0.2}^{+ 0.2}$ & $14.3_{- 0.5}^{+ 0.5}$ &  \\
          A2631 & $ 2.6_{- 1.1}^{+ 1.2}$ & $ 3.0_{- 0.5}^{+ 0.5}$ & $ 6.6_{- 1.5}^{+ 1.9}$ & $ 7.2_{- 1.1}^{+ 1.1}$ & $ 9.8_{- 2.9}^{+ 3.8}$ & $ 9.9_{- 1.5}^{+ 1.5}$ & $ 9.8_{- 0.1}^{+ 0.1}$ & $ 7.4_{- 0.3}^{+ 0.3}$ & $ 7.2_{- 0.3}^{+ 0.4}$ &  \\

& &  & & & & & & & &\\
\multicolumn{11}{c}{Weak lensing} \\
& &  & & & & & & & &\\

            A68 & $ 1.4_{- 0.6}^{+ 0.6}$ & $ 2.3_{- 0.5}^{+ 0.5}$ & $ 2.7_{- 0.8}^{+ 0.8}$ & $ 4.4_{- 1.0}^{+ 1.0}$ & $ 4.1_{- 1.0}^{+ 1.2}$ & $ 6.0_{- 1.3}^{+ 1.3}$ & $ 8.0_{- 0.1}^{+ 0.1}$ & $ 8.3_{- 0.3}^{+ 0.3}$ & $ 6.6_{- 0.3}^{+ 0.3}$ &  \\
           A209 & $ 2.1_{- 0.5}^{+ 0.5}$ & $ 2.7_{- 0.2}^{+ 0.2}$ & $ 5.1_{- 0.7}^{+ 0.7}$ & $ 7.3_{- 0.6}^{+ 0.6}$ & $ 8.6_{- 1.2}^{+ 1.3}$ & $12.3_{- 1.0}^{+ 1.0}$ & $10.6_{- 0.1}^{+ 0.1}$ & $ 6.6_{- 0.2}^{+ 0.2}$ & $ 6.9_{- 0.2}^{+ 0.2}$ &  \\
           A267 & $ 1.4_{- 0.3}^{+ 0.3}$ & $ 1.8_{- 0.4}^{+ 0.4}$ & $ 2.4_{- 0.4}^{+ 0.4}$ & $ 3.2_{- 0.8}^{+ 0.8}$ & $ 3.2_{- 0.6}^{+ 0.7}$ & $ 4.3_{- 1.0}^{+ 1.0}$ & $ 4.1_{- 0.0}^{+ 0.0}$ & $ 5.6_{- 0.1}^{+ 0.1}$ & $ 2.3_{- 0.1}^{+ 0.1}$ &  \\
$\dagger$  A291 & $ 0.9_{- 0.4}^{+ 0.4}$ & $ 0.7_{- 0.3}^{+ 0.3}$ & $ 2.3_{- 0.6}^{+ 0.6}$ & $ 1.5_{- 0.7}^{+ 0.7}$ & $ 4.0_{- 0.9}^{+ 1.0}$ & $ 2.1_{- 1.0}^{+ 1.0}$ & $ 3.7_{- 0.0}^{+ 0.0}$ & $ 3.9_{- 0.1}^{+ 0.1}$ & $ 1.4_{- 0.1}^{+ 0.1}$ &  \\
$\dagger$  A383 & $ 1.7_{- 0.2}^{+ 0.2}$ & $ 0.5_{- 0.4}^{+ 0.4}$ & $ 2.6_{- 0.4}^{+ 0.4}$ & $ 0.8_{- 0.7}^{+ 0.7}$ & $ 3.3_{- 0.6}^{+ 0.7}$ & $ 1.0_{- 0.9}^{+ 0.9}$ & $ 3.7_{- 0.0}^{+ 0.0}$ & $ 4.1_{- 0.1}^{+ 0.1}$ & $ 1.5_{- 0.0}^{+ 0.0}$ &  \\
$\ddagger$ A521 & $ 1.1_{- 0.3}^{+ 0.3}$ & $ 1.2_{- 0.2}^{+ 0.2}$ & $ 2.4_{- 0.5}^{+ 0.5}$ & $ 4.4_{- 0.7}^{+ 0.7}$ & $ 3.9_{- 0.7}^{+ 0.7}$ & $ 8.0_{- 1.3}^{+ 1.3}$ & $ 6.6_{- 0.0}^{+ 0.0}$ & $ 6.1_{- 0.1}^{+ 0.1}$ & $ 4.0_{- 0.1}^{+ 0.1}$ &  \\
           A520 & $ 1.1_{- 0.4}^{+ 0.3}$ & $ 2.1_{- 0.3}^{+ 0.3}$ & $ 2.5_{- 0.7}^{+ 0.8}$ & $ 5.8_{- 0.7}^{+ 0.7}$ & $ 4.1_{- 1.2}^{+ 1.1}$ & $ 9.1_{- 1.1}^{+ 1.1}$ & $11.3_{- 0.1}^{+ 0.1}$ & $ 7.9_{- 0.2}^{+ 0.2}$ & $ 9.0_{- 0.3}^{+ 0.3}$ &  \\
           A963 & $ 1.0_{- 0.2}^{+ 0.2}$ & $ 1.4_{- 0.2}^{+ 0.2}$ & $ 2.4_{- 0.4}^{+ 0.6}$ & $ 3.4_{- 0.5}^{+ 0.5}$ & $ 4.2_{- 0.7}^{+ 0.9}$ & $ 5.2_{- 0.8}^{+ 0.8}$ & $ 6.7_{- 0.0}^{+ 0.0}$ & $ 5.6_{- 0.1}^{+ 0.1}$ & $ 3.8_{- 0.1}^{+ 0.1}$ &  \\
          A1835 & $ 2.8_{- 0.6}^{+ 0.6}$ & $ 6.2_{- 0.7}^{+ 0.7}$ & $ 6.0_{- 0.9}^{+ 0.9}$ & $11.2_{- 1.2}^{+ 1.2}$ & $ 9.5_{- 1.5}^{+ 1.7}$ & $14.1_{- 1.5}^{+ 1.5}$ & $11.6_{- 0.0}^{+ 0.0}$ & $ 8.4_{- 0.1}^{+ 0.1}$ & $ 9.7_{- 0.2}^{+ 0.2}$ &  \\
          A1914 & $ 1.6_{- 0.7}^{+ 0.6}$ & $ 3.1_{- 0.3}^{+ 0.3}$ & $ 3.1_{- 1.2}^{+ 1.3}$ & $ 5.4_{- 0.5}^{+ 0.5}$ & $ 4.7_{- 1.9}^{+ 1.6}$ & $ 7.0_{- 0.7}^{+ 0.7}$ & $10.9_{- 0.0}^{+ 0.0}$ & $ 8.5_{- 0.2}^{+ 0.2}$ & $ 9.2_{- 0.2}^{+ 0.2}$ &  \\
ZwCl1454.8+2233 & $ 0.9_{- 0.4}^{+ 0.4}$ & $ 1.5_{- 0.4}^{+ 0.4}$ & $ 1.7_{- 0.5}^{+ 0.6}$ & $ 2.6_{- 0.7}^{+ 0.7}$ & $ 2.6_{- 0.8}^{+ 1.0}$ & $ 3.5_{- 0.9}^{+ 0.9}$ & $ 4.9_{- 0.0}^{+ 0.0}$ & $ 4.6_{- 0.1}^{+ 0.1}$ & $ 2.3_{- 0.0}^{+ 0.0}$ &  \\
ZwCl1459.4+4240 & $ 1.8_{- 0.4}^{+ 0.4}$ & $ 2.0_{- 0.4}^{+ 0.4}$ & $ 2.9_{- 0.6}^{+ 0.7}$ & $ 3.6_{- 0.7}^{+ 0.7}$ & $ 3.9_{- 0.9}^{+ 1.0}$ & $ 5.0_{- 1.0}^{+ 1.0}$ & $ 7.0_{- 0.1}^{+ 0.1}$ & $ 6.4_{- 0.2}^{+ 0.2}$ & $ 4.5_{- 0.2}^{+ 0.2}$ &  \\
          A2034 & $ 1.6_{- 0.9}^{+ 0.7}$ & $ 2.1_{- 0.2}^{+ 0.2}$ & $ 3.4_{- 1.5}^{+ 1.6}$ & $ 4.1_{- 0.4}^{+ 0.4}$ & $ 5.1_{- 2.4}^{+ 2.1}$ & $ 5.4_{- 0.5}^{+ 0.5}$ & $ 7.0_{- 0.1}^{+ 0.1}$ & $ 6.4_{- 0.2}^{+ 0.2}$ & $ 4.5_{- 0.1}^{+ 0.2}$ &  \\
          A2219 & $ 3.7_{- 0.6}^{+ 0.6}$ & $ 6.7_{- 0.3}^{+ 0.3}$ & $ 6.0_{- 0.9}^{+ 0.9}$ & $13.4_{- 0.6}^{+ 0.6}$ & $ 8.0_{- 1.3}^{+ 1.5}$ & $19.6_{- 0.9}^{+ 0.9}$ & $17.6_{- 0.1}^{+ 0.2}$ & $ 9.6_{- 0.2}^{+ 0.3}$ & $16.9_{- 0.5}^{+ 0.7}$ &  \\
 RXJ1720.1+2638 & $ 1.9_{- 0.4}^{+ 0.4}$ & $ 2.3_{- 0.2}^{+ 0.2}$ & $ 2.9_{- 0.6}^{+ 0.7}$ & $ 4.0_{- 0.4}^{+ 0.4}$ & $ 3.7_{- 0.9}^{+ 1.1}$ & $ 5.5_{- 0.6}^{+ 0.6}$ & $ 7.0_{- 0.0}^{+ 0.0}$ & $ 5.9_{- 0.1}^{+ 0.1}$ & $ 4.1_{- 0.1}^{+ 0.1}$ &  \\
$\ddagger$A2261 & $ 3.5_{- 0.4}^{+ 0.4}$ & $ 5.5_{- 0.5}^{+ 0.5}$ & $ 5.9_{- 0.7}^{+ 0.7}$ & $ 8.4_{- 0.8}^{+ 0.8}$ & $ 8.0_{- 1.1}^{+ 1.2}$ & $10.4_{- 1.0}^{+ 1.0}$ & $ 9.2_{- 0.2}^{+ 0.3}$ & $ 6.1_{- 0.5}^{+ 0.6}$ & $ 5.6_{- 0.6}^{+ 0.8}$ &  \\
 RXJ2129.6+0005 & $ 1.3_{- 0.5}^{+ 0.5}$ & $ 1.4_{- 0.3}^{+ 0.3}$ & $ 2.9_{- 0.7}^{+ 0.7}$ & $ 2.9_{- 0.7}^{+ 0.7}$ & $ 4.6_{- 1.0}^{+ 1.1}$ & $ 4.0_{- 1.0}^{+ 1.0}$ & $ 7.3_{- 0.0}^{+ 0.0}$ & $ 5.6_{- 0.1}^{+ 0.1}$ & $ 4.1_{- 0.1}^{+ 0.1}$ &  \\
          A2390 & $ 3.1_{- 0.4}^{+ 0.4}$ & $ 4.9_{- 0.3}^{+ 0.3}$ & $ 5.1_{- 0.8}^{+ 0.8}$ & $ 9.8_{- 0.7}^{+ 0.7}$ & $ 7.0_{- 1.2}^{+ 1.3}$ & $14.3_{- 1.0}^{+ 1.0}$ & $15.8_{- 0.1}^{+ 0.1}$ & $ 9.1_{- 0.2}^{+ 0.2}$ & $14.4_{- 0.5}^{+ 0.5}$ &  \\
          A2631 & $ 2.4_{- 0.4}^{+ 0.3}$ & $ 3.1_{- 0.4}^{+ 0.4}$ & $ 3.7_{- 0.5}^{+ 0.5}$ & $ 6.3_{- 0.9}^{+ 0.9}$ & $ 4.8_{- 0.7}^{+ 0.7}$ & $ 9.0_{- 1.3}^{+ 1.3}$ & $ 9.8_{- 0.1}^{+ 0.1}$ & $ 7.5_{- 0.2}^{+ 0.4}$ & $ 7.4_{- 0.3}^{+ 0.4}$ & \\

\bottomrule
\end{tabular}
}

\tablefoot{$T_{\rm X}$ is the spectroscopic temperature within $\Rv$. \\
$\dagger$A291 and A383 were excluded from scaling relation fits involving SZ quantities (see Sect.~\ref{sec:szquan} for details).  \\
$\ddagger$A521 and A2261 were excluded from scaling relation fits involving hydrostatic X-ray mass estimates (see Sect.~\ref{sec:xmass} for details).}
\end{centering}
\end{table*}
%______________________________________________________________

%%%%%%%%%%%%%%%%%%%%%%%%%%%%%%%%%%%%%

\subsubsection{X-ray pressure profile}
\label{sec:xpress}

Using the radial density and temperature information, we also calculated the X-ray pressure profile $P(r) = n_{\rm e}(r)\, kT(r)$. We then fitted the pressure profile of each cluster with the generalised Navarro, Frank \& White (GNFW) model introduced by \citet{nag07}, viz.,
\begin{equation}
P(x) = \frac{P_{0} } { (c_{\rm 500,p}\,x)^{\gamma}\left[1+(c_{\rm 500,p}\,x)^\alpha\right]^{(\beta-\gamma)/\alpha} }. 
\label{eq:pgnfw}
\end{equation}
Here the parameters $(\alpha,\beta,\gamma)$ are the intermediate, outer, and central slopes, respectively, $c_{\rm 500,p}$ is a concentration parameter,  $r_{\rm s} = R_{500}/c_{\rm 500,p}$, and $x = r/\Rv$. In the fitting, the outer slope was fixed at $\beta = 5.49$, a choice that is motivated by simulations  since it is essentially unconstrained by the X-ray data \citep[see][for discussion]{arn10}. The best-fitting X-ray pressure profile parameters are listed in Table~\ref{tab:pnfw}, and the observed profiles and best-fitting models are plotted in Fig.~\ref{fig:pprof}.

%%%%%%%%%%%%%%%%%%%%%%%%%%%%%%%%%%%%%

\subsubsection{Morphological classification}
\label{sec:morpho}

We divided the 19 clusters into three morphological sub-classes based on the scaled central density $E(z)^{-2}\, n_{\rm e,0}$, which is a good proxy for the overall dynamical state \citep[see, e.g.,][]{pra10,arn10}. The scaled central density was obtained from a $\beta$-model fit to the inner $R < 0.05\,\Rv$ region. The seven clusters with the highest scaled central density values were classed as relaxed\footnote{As these systems have the highest scaled central density, they are fully equivalent to a cool core sub-sample.}; the six with the lowest values were classed as disturbed; the six with intermediate values were classed as intermediate (i.e., neither relaxed nor disturbed). Strict application of the \rexcess\ morphological classification criteria based on scaled central density and centroid shift parameter $\langle w \rangle$ \citep{pra09} results in a similar classification scheme.

Images of the cluster sample ordered by $E(z)^{-2}\, n_{\rm e,0}$ are shown Appendix~\ref{appx:imgal}. Henceforth, in all figures dealing with morphological classification, relaxed systems are plotted in blue, unrelaxed in red and intermediate in black. Scaled X-ray profiles resulting from the analysis described below, colour-coded by morphological sub-class, are shown in Fig.~\ref{fig:profiles}.

%%%%%%%%%%%%%%%%%%%%%%%%%%%%%%%%%%%%%

\subsection{SZ}\label{sec:szquan}

The SZ signal was extracted from the six High Frequency Instrument (HFI) temperature channel maps corresponding to the nominal \planck\ survey (i.e., slightly more than 2.5 full sky surveys). We used full resolution maps of HEALPix \citep{gor05}\footnote{\url{http://healpix.jpl.nasa.gov}} $N_{\rm side}=2048$ and assumed that the  beams were described by circular Gaussians. We adopted beam FWHM values of 9.88, 7.18, 4.87, 4.65, 4.72, and 4.39 arcmin for channel frequencies 100, 143, 217, 353, 545, and 857 GHz, respectively. Flux extraction was undertaken using the full relativistic treatment of the SZ spectrum \citep{ito98}, assuming the global temperature $T_{\rm X}$ given in Table~\ref{tab:xray}. Bandpass uncertainties were taken into account in the flux measurement. Uncertainties  due to beam corrections and map calibrations are expected to be small, as discussed extensively in \citet{planck2011-5.1a}, \citet{planck2011-5.1b}, \citet{planck2011-5.2a}, \citet{planck2011-5.2b}, and \citet{planck2011-5.2c}.

We extracted the SZ signal using multi-frequency matched filters \citep[MMF,][]{her02,mel06}, which optimally filter and combine maps to estimate the SZ signal. As input, the MMF requires information on the instrumental beam, the SZ frequency spectrum, and a cluster profile; noise auto- and cross-spectra are estimated directly from the data. The algorithm can be run in a blind mode, where the position, normalisation and extent are all determined by the MMF \citep[e.g.,][]{planck2011-5.1a}, or in a targeted mode, where the position and size are estimated using external data, and only the normalisation (or SZ flux) is determined by the MMF \citep[e.g.,][]{planck2011-5.2b}. Here we adopted the latter mode, using the position, size, and SZ profile of each cluster determined from external X-ray and/or weak lensing data. In this case the MMF thus returns only the integrated SZ flux and its associated statistical uncertainty.

%_______________
%% Figure: SZ vs M
%%
\begin{figure*}[]
\begin{centering}
\includegraphics[scale=1.,angle=0,keepaspectratio,width=0.325\textwidth]{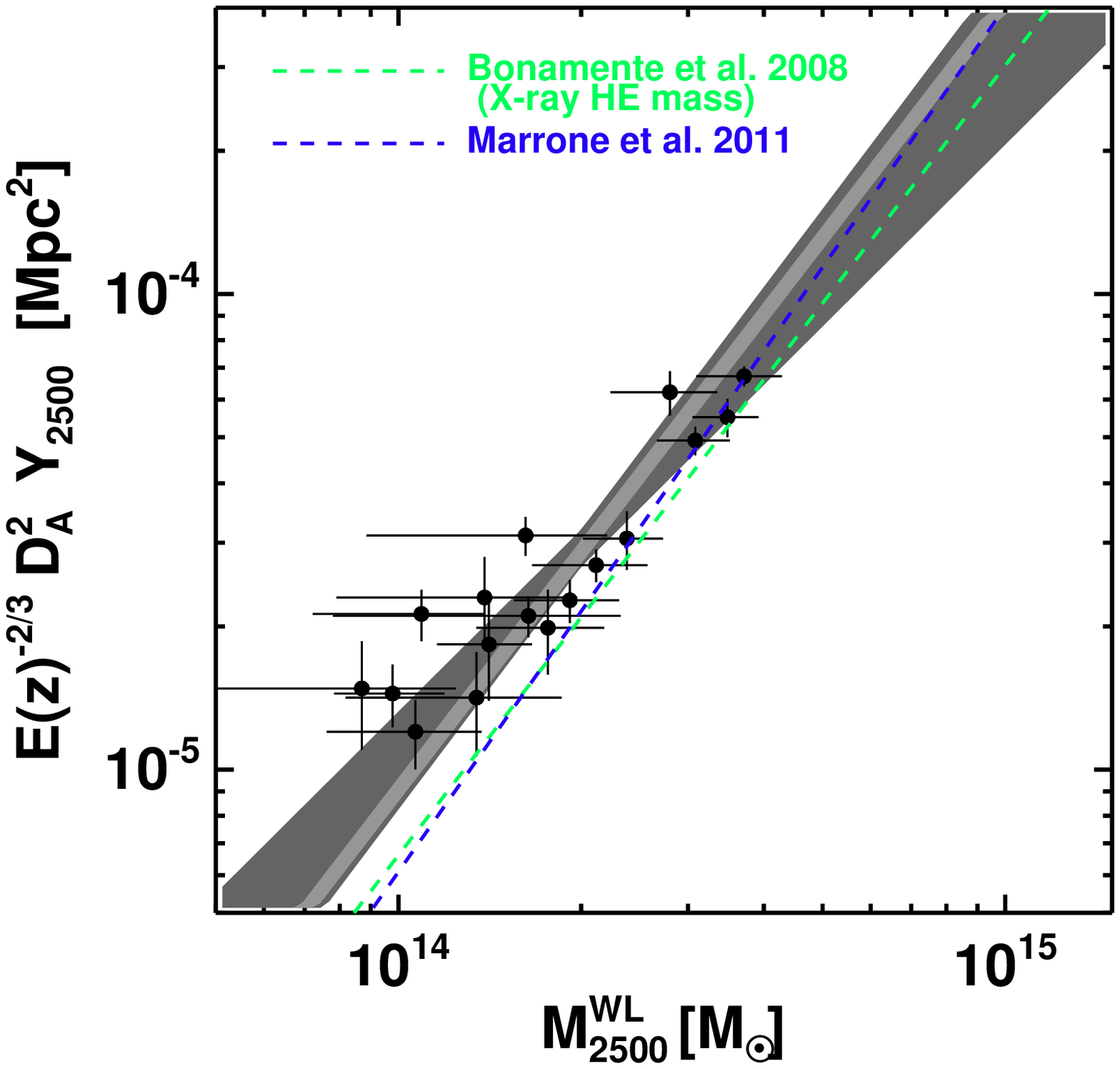}
\hfill
\includegraphics[scale=1.,angle=0,keepaspectratio,width=0.325\textwidth]{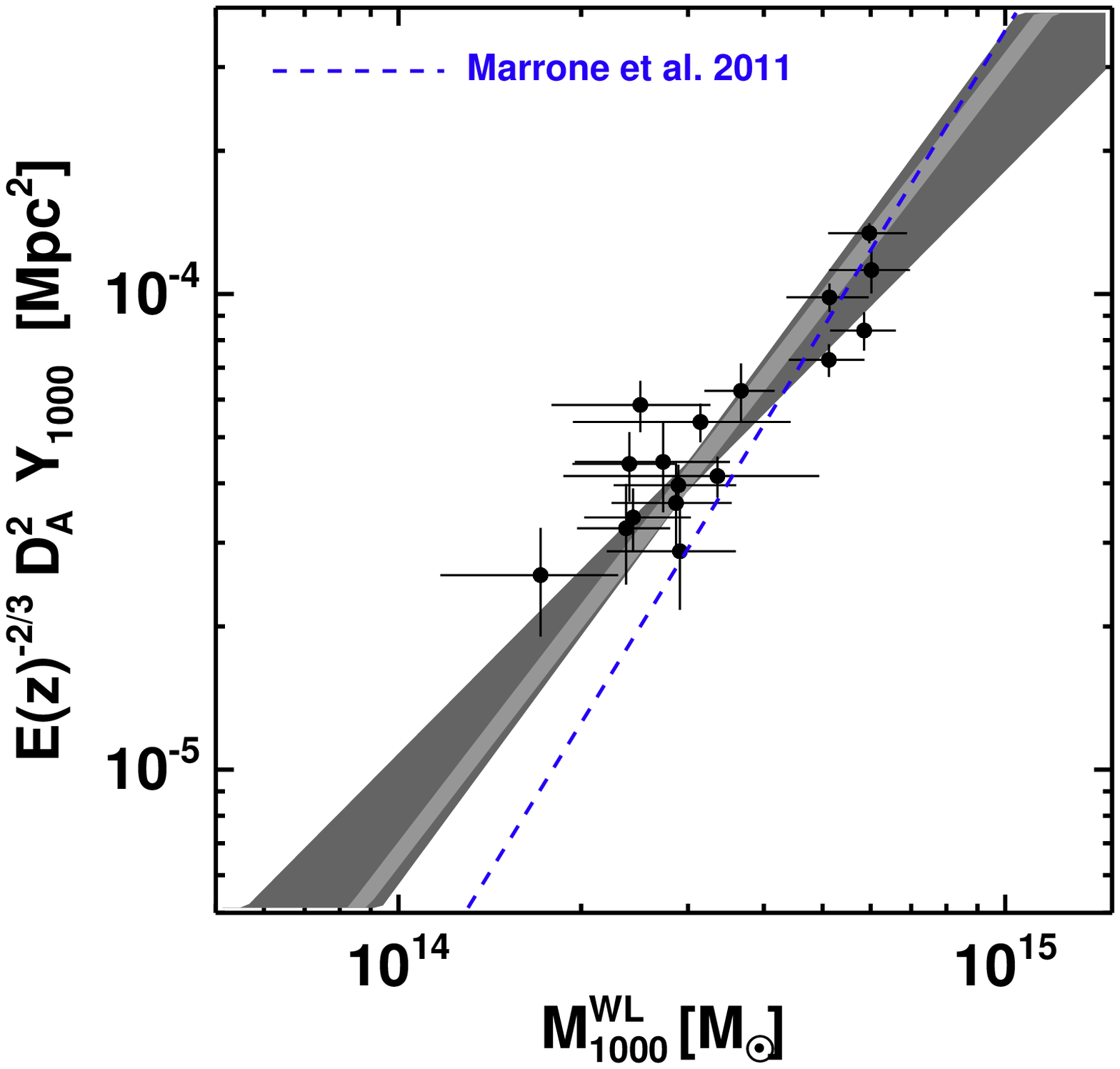}
\hfill
\includegraphics[scale=1.,angle=0,keepaspectratio,width=0.325\textwidth]{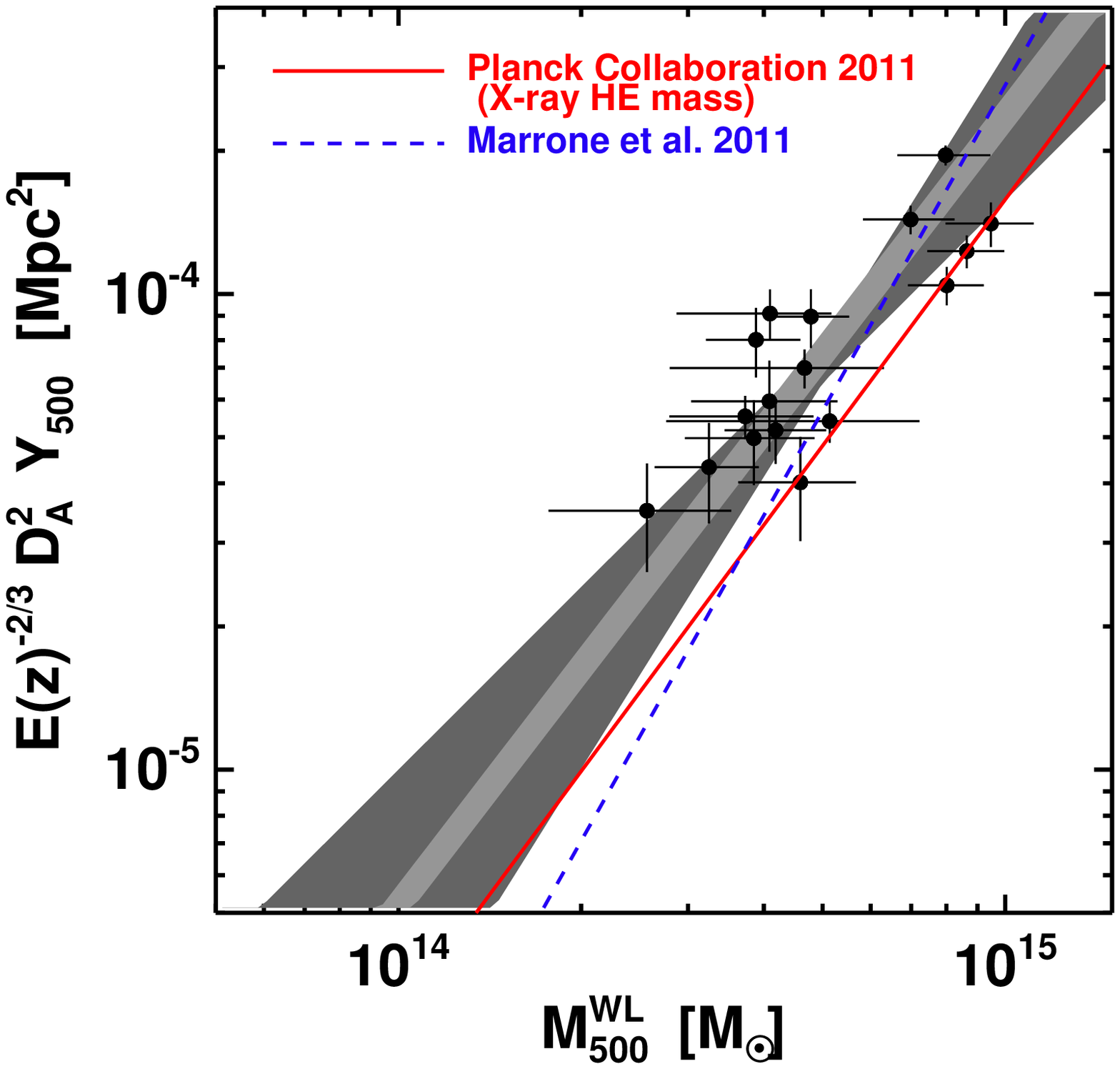}
\end{centering}
\caption{{\footnotesize  Relations between $\YSZ$ and total mass for apertures determined from weak lensing mass profiles corresponding to density contrasts of $\Delta = 2500$ (\textit{left}), 1000 (\textit{middle}), and 500 (\textit{right}). In all panels the dark grey region represents the best-fitting relation obtained with slope and normalisation as free parameters, and the light grey region denotes the best-fitting relation obtained with the slope fixed to the self-similar value of 5/3. Previous results from \citet{marr11}, \citet{bon08}, and the analysis of 62 nearby systems by \citet{planck2011-5.2b} are shown for comparison. The masses in the latter two studies were derived from X-ray analyses. The original cylindrically integrated SZ signal measurement in \citeauthor{bon08} has been converted to a spherically integrated measurement assuming an \citep{arn10} profile.}}\label{fig:szrel}
\end{figure*}
%%_______________

%______________________________________________________________
% Table of results for M-Y relation
\begin{table}[]
\newcolumntype{L}{>{\columncolor{white}[0pt][\tabcolsep]}l}
\newcolumntype{R}{>{\columncolor{white}[\tabcolsep][0pt]}l}
%\rowcolors{0}{light-grey}{white}
\caption{{\footnotesize best-fitting parameters for the weak lensing mass--$\DAY$ scaling relations.} 
\label{tab:szmass}} 
\centering  
\begin{tabular}{@{}Lcccccc}
\toprule
\toprule

\multicolumn{1}{c}{$\Delta$} & \multicolumn{1}{c}{$M_0$} & 
\multicolumn{1}{c}{$A$} &
\multicolumn{1}{c}{$B$} & \multicolumn{1}{c}{$\sigma_{\perp}$} &\multicolumn{1}{c}{$\sigma_{Y|M}$} \\

\multicolumn{1}{c}{} & \multicolumn{1}{c}{(M$_{\odot}$)} & 
\multicolumn{1}{c}{} &
\multicolumn{1}{c}{} & \multicolumn{1}{c}{($\%$)} &\multicolumn{1}{c}{($\%$)} \\

\midrule

2500 & $2 \times 10^{14}$ & $-4.53\pm0.04$ & $1.48\pm0.21$ & $36\pm7$ & $20\pm4$ \\
1000 & $3 \times 10^{14}$ & $-4.38\pm0.03$ & $1.51\pm0.22$ & $33\pm6$ & $18\pm3$ \\
500 & $5 \times 10^{14}$ & $-4.15\pm0.04$ & $1.65\pm0.38$ & $33\pm8$ & $17\pm4$ \\

\midrule

2500 & $2 \times 10^{14}$ & $-4.53\pm0.03$ & 5/3 & $38\pm4$ & $19\pm2$ \\
1000 & $3 \times 10^{14}$ & $-4.38\pm0.02$ & 5/3 & $35\pm2$ & $18\pm2$ \\
500 & $5 \times 10^{14}$ & $-4.13\pm0.04$  & 5/3 & $38\pm3$ & $20\pm2$ \\

\bottomrule
\end{tabular}
\tablefoot{Relations are expressed as $E(z)^\gamma\, [D_{\rm A}^2\, \Yv] =10^A\, [E(z)^\kappa\, M/M_{0}]^B$, with $\gamma = -2/3$, $\kappa = 1$.  \\
$\sigma_{\perp}$ is the orthogonal dispersion about the best-fitting relation.\\
$\sigma_{Y|M}$ is the dispersion in $Y$ at given $M$ for the best-fitting relation.}
\end{table}
%______________________________________________________________

Our baseline SZ measurement involved extraction of the SZ flux from a position centred on the X-ray emission peak using the observed X-ray pressure profile of each cluster described above in Sect.~\ref{sec:xpress} as a spatial template. Apertures were determined independently either from the weak lensing or the X-ray mass analysis. The extraction was achieved by excising a $10^\circ \times 10^\circ$ patch with  pixel size $1\farcm72$, centred on the X-ray (or weak lensing) position, from the six HFI maps, and estimating the SZ flux using the MMF. The profiles were truncated at $5\,R_{\rm 500}$ to ensure integration of the total SZ signal. The flux and corresponding error were then scaled to smaller apertures ($R_{\rm 500}$, $R_{\rm 1000}$, $R_{\rm 2500}$) using the profile assumed for extraction.

We undertook two further tests of the SZ flux extraction process. First, we  measured the SZ flux using the ``universal'' pressure profile as a spatial template. Here we find that the error-weighted mean ratio between these measurements and those using the X-ray pressure profile as a spatial template is $Y_{\rm GNFW}/Y_{\rm univ} = 1.02\pm0.05$, with no trend with morphological sub-class. Second, we measured the SZ flux with the position left free. In this case the mean error-weighted ratio between the flux measurements is  $Y_{\rm free}/Y_{\rm fix} = 1.04\pm0.05$, again with no trend with morphological sub-class. 

The SZ flux measurements for two systems are suspect. One object, A291, appears to be strongly contaminated by a radio source. The other, A383, while not obviously contaminated by a radio source and appearing to be a very relaxed system in X-rays, exhibits an offset exceeding 4 arcmin ($\sim 0.8\,\Rv$) between the SZ and X-ray positions. This cluster is detected at a rather low signal-to-noise ratio, and we also note that \citet{marr11}, in their Sunyaev-Zeldovich Array (SZA)  observations, found that this system has an unusually low SZ flux for its apparent mass. In addition, \citet{zit12} find that A383 is a cluster-cluster lens system, where the nearby $z =0.19$ cluster is lensing a more distant $z=0.9$ object that lies about 4\arcm\ to the north-east from the centre of the main system. \citeauthor{zit12} also mention the presence of at least two other well-defined optical structures within 15\arcm\ of A383. In view of the complexity of these systems, we exclude them from any analysis that follows involving discussion of the SZ signal.

%%%%%%%%%%%%%%%%%%%%%%%%%%%%%%%%%%%%%

\section{Results}

\subsection{Fitting procedure}

We obtained the parameters governing scaling relations between various quantities by fitting each set of observables $(X,Y)$ with a power law of the form 
\begin{equation}
E(z)^\gamma\, Y = 10^A\, [E(z)^\kappa\, (X/X_0)]^B,  
\end{equation}
where $E(z)$ is the Hubble constant normalised to its present-day value and $\gamma$ and $\kappa$ were fixed to their expected self-similar scalings with redshift. The fits were undertaken using linear regression in the log-log plane, taking the uncertainties in both variables into account, and the scatter was computed as described in \citet{pra09} and \citet{planck2011-5.2b}. The fitting procedure used the BCES orthogonal regression method \citep{akr96}. In addition to fitting with the slope and normalisation free, we also investigated the scaling relations obtained with the slope fixed to the self-similar values. All uncertainties on fitted parameters were estimated using bootstrap resampling.

%%%%%%%%%%%%%%%%%%%%%%%%%%%%%%%%%%%%%

%_______________
%% Figure: SZ vs Yx, SZ vs MYx
%%
\begin{figure*}[]
\begin{centering}
\includegraphics[scale=1.,angle=0,keepaspectratio,width=0.95\columnwidth]{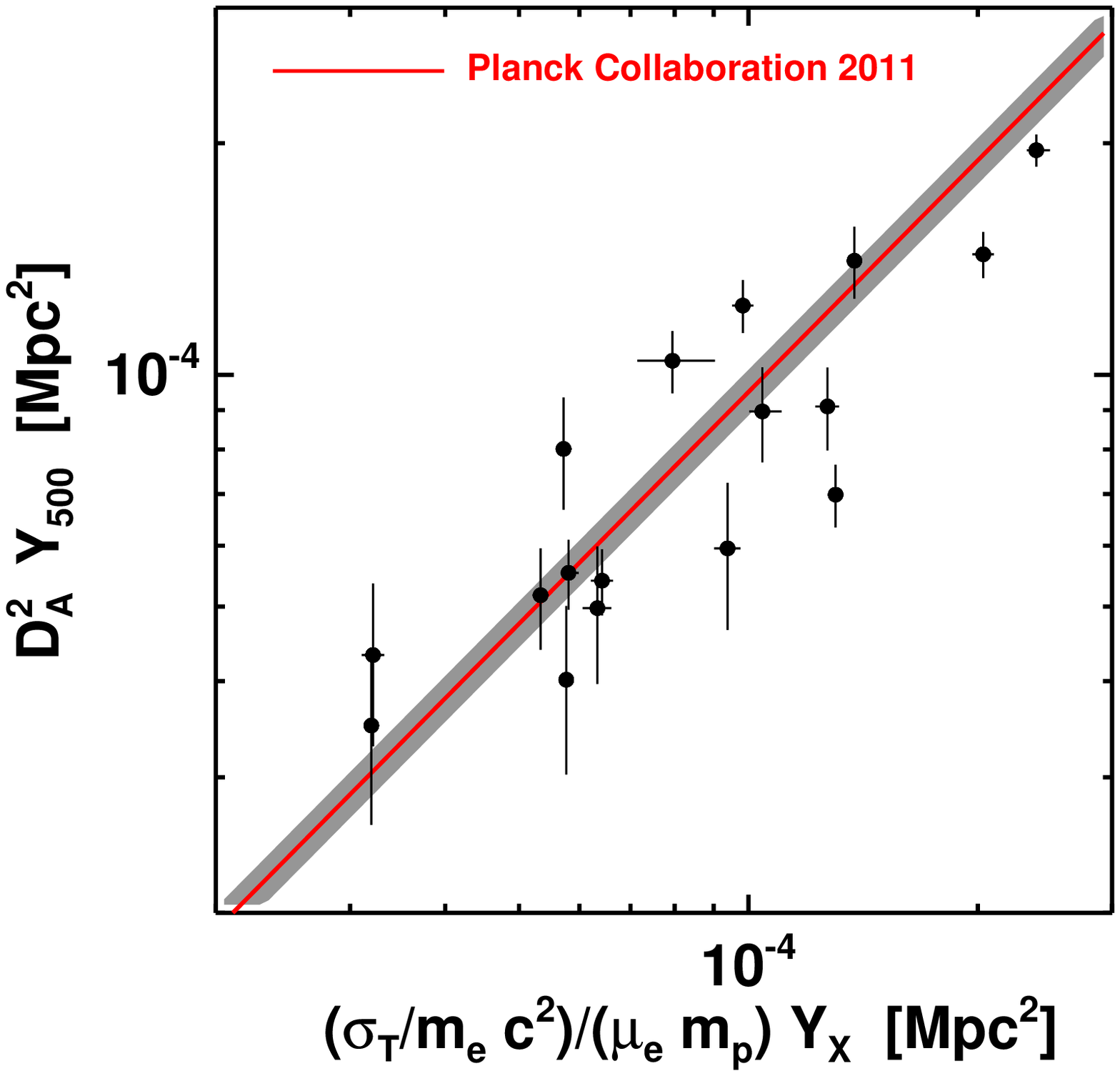}
\hfill
\includegraphics[scale=1.,angle=0,keepaspectratio,width=0.95\columnwidth]{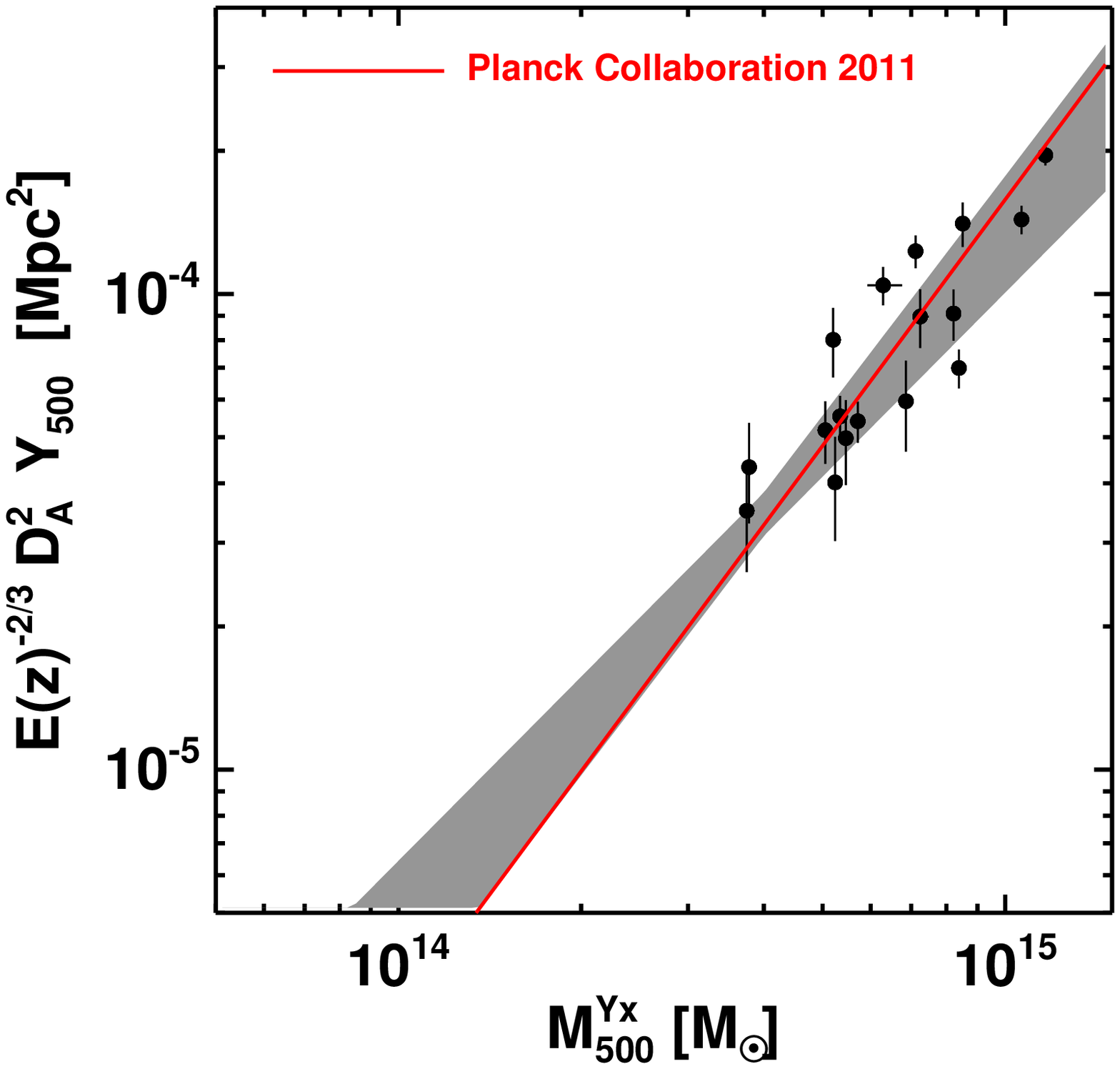}
\end{centering}
\caption{{\footnotesize Comparison of present SZ flux measurements to our previous results. Quantities are measured within the $\Rv$ derived from weak lensing. {\it Left panel:} Relation between $\DAY$ and $C_{\rm XSZ}\, \YX = M_{\rm g,500}  T_{\rm X}$, where $T_{\rm X}$ is the spectroscopic temperature in the $[0.15-0.75]\,\Rv$ region. The grey shaded region is the best-fitting power-law relation obtained with slope fixed to 1; the red line shows the results from our previous analysis of 62 local systems \citep{planck2011-5.2b}. {\it Right panel:} Correlation between $\DAY$ and $M_{500}^{\rm Yx}$ derived from the relation of \citet{arn10}, compared to the results from \citet{planck2011-5.2b}. The shaded region illustrates the best-fitting BCES orthogonal regression and associated $\pm 1\sigma$ uncertainties. }}\label{fig:YxYsz}
\end{figure*}
%%_______________

\subsection{SZ -- weak lensing mass scaling relation}

Figure~\ref{fig:szrel} shows the relation between the weak lensing mass $M^{\rm WL}_{\Delta}$ and the SZ flux $\DAY$ measured using our baseline method. All quantities have been integrated in spheres corresponding to $\Delta = 2500, 1000,$ and 500, as determined from the weak lensing mass profiles. The best-fitting power-law relations are overplotted, both for regression with the slope fixed to the self-similar value of 5/3 (light grey region), and for regression with the slope and normalisation free (dark grey region). Numerical values for the best-fitting relations, including the dispersion about them, are given in Table~\ref{tab:szmass}. 

For fits where the slope and normalisation were left as free parameters, the  slope of the $M^{\rm WL}_{\Delta}$--$\DAY$ relation is compatible with the self-similar value of 5/3 at all values of the density contrast $\Delta$. The orthogonal scatter about the best-fitting relation is $\sigma_{\perp} \sim 30$--$35$ percent, and the scatter in $\DAY$ for a given $M^{\rm WL}_{\Delta}$ is $\sigma_{Y|M} \sim 20$ percent, with no significant trend with density contrast. 

Similar fits to the $\DAY$--$M^{\rm WL}_{\Delta}$ relation for different SZ extractions, e.g., with the ``universal'' pressure profile, or with the SZ position left as a free parameter, did not yield results significantly different from those described above.

%%%%%%%%%%%%%%%%%%%%%%%%%%%%%%%%%%%%%

\section{Discussion}

\subsection{Comparison to previous results}

For the $\DAY$--$M^{\rm WL}_{\Delta}$ relation, our results are in good agreement with earlier determinations at all values of $\Delta$, albeit within the relatively large uncertainties of both our analysis and those of previous investigations. A comparison to the most recent results of \citet{marr11}, who also use weak lensing masses shows that, while the normalisations are in agreement, the slopes are slightly (although not significantly) shallower. This is easily explained by our exclusion of A383 from the regression analysis (see Sect.~\ref{sec:szquan} for details); this object was not excluded in the  \citeauthor{marr11} regression fits (see discussion in their Sect.~4.3). A fit of their data excluding A383 yields a slope of $1.77 \pm 0.16$, in good agreement with our value (Marrone 2012, priv. communication).

The scatter we observe ($\sigma_{Y|M} \sim 20$ percent) is also in excellent agreement with that seen by \citet{marr11}. Although numerical simulations predict that there is intrinsically only of order ten percent scatter between the mass and the integrated Compton parameter \citep[e.g.,][]{das04}, observational measurement uncertainties and complications due to, e.g., mass along the line of sight or cluster triaxiality introduce a further source of scatter. Simulations that take an observational approach to measurement uncertainties \citep{bec11} predict a dispersion of order 20 percent, as observed.

Perhaps the most interesting outcome from the present analysis concerns the normalisation of the $\DAY$--$M_{500}^{\rm WL}$ relation. As shown in the right-hand panel of Fig.~\ref{fig:szrel}, there is a normalisation offset when the slope is fixed to the self-similar value of 5/3, which is significant at $1.4\sigma$ with respect to our previous investigation of 62 local ($z < 0.5$) systems using masses estimated from the $\Mv$--$\YX$ relation \citep{planck2011-5.2b}. Interestingly, a similar offset was found by \citet{marr11} when comparing their scaling relations to those of \citet{and10}. This normalisation offset could be due either to a larger than expected SZ flux, or to a difference in mass measurements between studies (or indeed, both effects may contribute). Notably, both \citeauthor{planck2011-5.2b} and \citeauthor{and10} used cluster masses estimated from the $\Mv$--$\YX$ relation calibrated using X-ray observations. In the following, we first verify the consistency between the present results and our previous work, finding excellent agreement. We then examine in more detail the reasons behind the observed normalisation offset.

%%%%%%%%%%%%%%%%%%%%%%%%%%%%%%%%%%%%%

%_______________
%% Figure: SZ vs M
%%
\begin{figure*}[]
\begin{centering}
\includegraphics[scale=1.,angle=0,keepaspectratio,width=0.95\columnwidth]{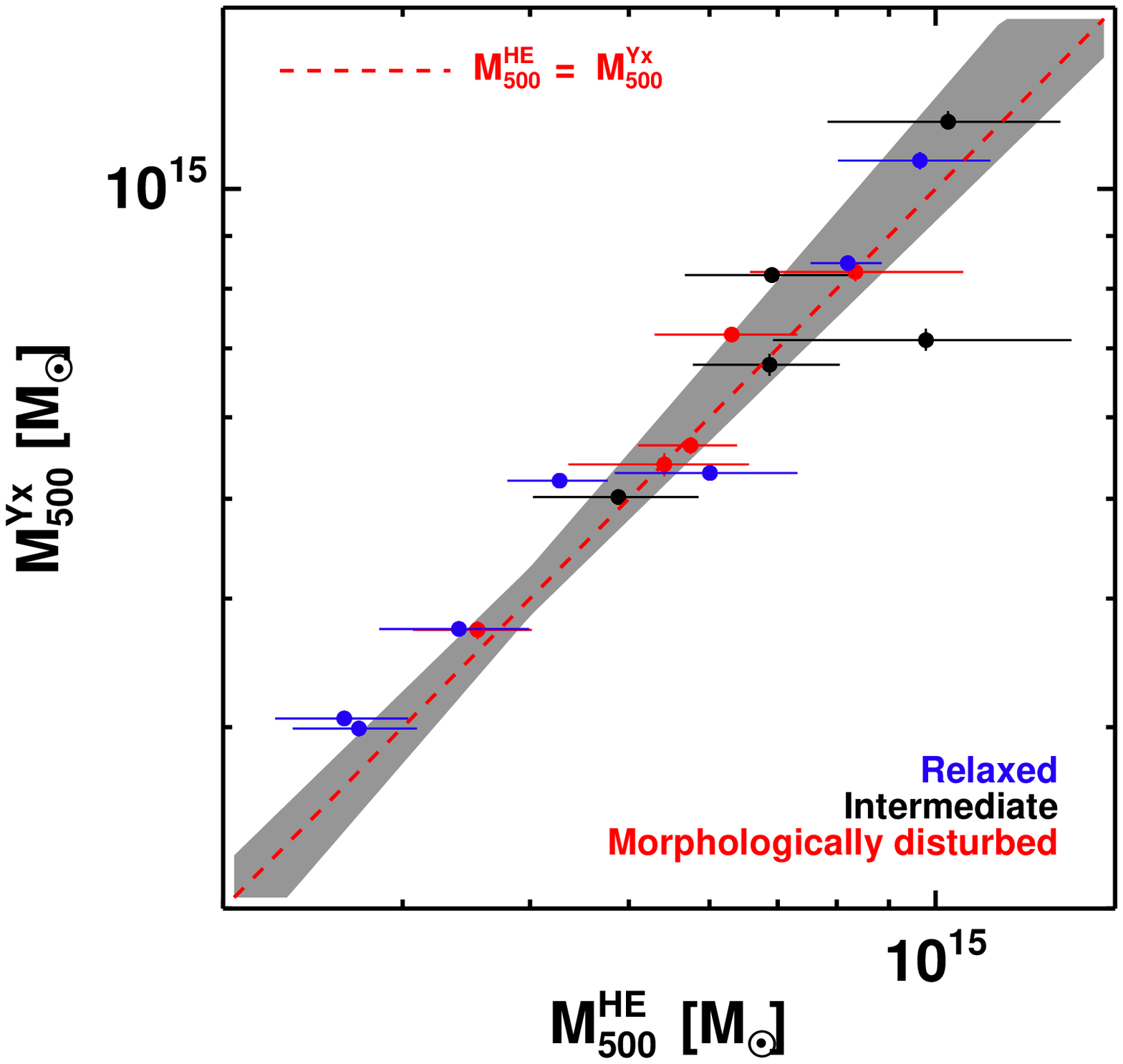}
\hfill
\includegraphics[scale=1.,angle=0,keepaspectratio,width=0.95\columnwidth]{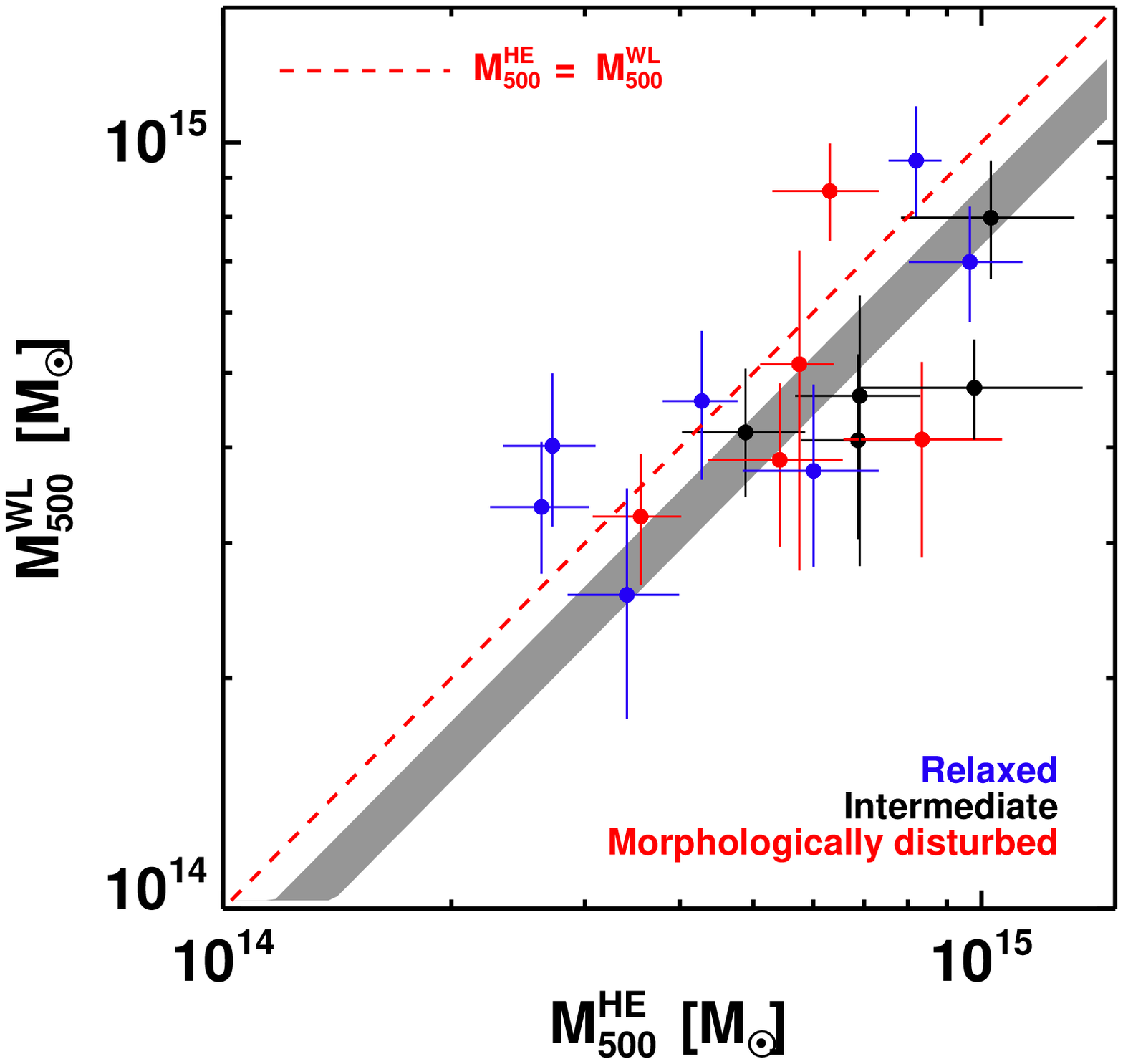}
\end{centering}
\caption{{\footnotesize {\it Left panel:} Relation between the mass derived from the hydrostatic X-ray analysis (Sect.~\ref{sec:xmass}) compared to the mass derived from the $\Mv$--$\YX$ proxy relation of \citet{arn10}. The shaded region shows the best-fitting BCES orthogonal fit to the data and associated $\pm 1\sigma$ uncertainties, and the dashed line denotes equality. {\it Right panel:} Relation between the  mass derived from the hydrostatic X-ray analysis and the weak lensing mass of \citet{oka08} and \citet{oka10}. The shaded region is the best-fitting regression between the two quantities with the slope fixed at 1. The dashed line denotes equality.}}\label{fig:MxvsM}
\end{figure*}
%%_______________

\subsubsection{SZ measurements}

Here we wish to verify the consistency between the present SZ flux measurements and those we published in \citet{planck2011-5.2b}. We compare the measurements in a statistical sense since not all of the present sample appear in the 62 ESZ clusters published in that paper. We first compare the SZ flux measurement to its X-ray analogue $\YX$. This quantity, which was first introduced in \citet{kra06}, is defined as the product of the gas mass  and the temperature. For consistency with our previous work, we define $\YX = M_{\rm g,500} T_{\rm X}$, where $T_{\rm X}$ is the spectroscopic temperature in the $[0.15-0.75]\,\Rv$ region. The left-hand panel of Fig.~\ref{fig:YxYsz} shows the relation between the SZ flux and $\YX$, The latter has been normalised by
\begin{equation}
C_{\rm XSZ} =\frac{\sigma_{\rm T}}{m_{\rm e} c^2} \frac{1}{\mu_{\rm e} m_{\rm p}} =   1.416 \times 10^{-19}~~\frac{{\rm Mpc}^2}{\rm \msol\, \keV} 
\end{equation}
for $\mu_{\rm e}=1.148$, the mean molecular weight of electrons for a plasma of 0.3 times solar abundance. The grey shaded area shows the best-fitting power-law relation between the two quantities obtained with the slope fixed to 1\footnote{A fit with the slope left free is shallower, but compatible with unity at the 1-sigma level.}. For comparison, we plot the $C_{\rm XSZ}\,\YX$--$\DAY$ relation obtained by \citet{planck2011-5.2b}: $\DAY /C_{\rm XSZ}\, \YX = 0.95\pm0.04$. As can be seen, the present SZ flux measurements are in excellent agreement with our previous determination.\footnote{The ratio is also consistent with that predicted solely from \rexcess\ X-ray observations: $\DAY / C_{\rm XSZ}\,\YX = 0.924\pm0.004$  \citep{arn10}.} 

We recall that in \citet{planck2011-5.2b} the mass was estimated from the $\Mv$--$\YX$ relation of \citet{arn10}. As a second test, we thus calculated $M_{500}^{Y_{\rm X}}$ for all objects using the \citeauthor{arn10} relation and compared the resulting correlation between $\DAY$-$M_{500}^{Y_{\rm X}}$ to our previous measurements. The right-hand panel of Fig.~\ref{fig:YxYsz} shows this comparison, where the grey shaded area is the best-fitting power-law relation between the two quantities obtained using orthogonal BCES regression with the slope and normalisation as free parameters. The $\DAY$--$M_{500}^{Y_{\rm X}}$ relation from our previous investigation \citep{planck2011-5.2b} is overplotted. Once again, the SZ flux measurements are in very good agreement with our previous determination. 

Given the excellent agreement with previous results, we thus conclude that the normalisation offset in the $\DAY$--$M^{\rm WL}_{500}$ relation is not due to a systematic overestimation of the SZ flux with respect to our previous measurements.

%%%%%%%%%%%%%%%%%%%%%%%%%%%%%%%%%%%%%
%_______________
%% Figure: mass ratio vs concentration
%%
\begin{figure*}[]
\begin{centering}
\includegraphics[scale=1.,angle=0,keepaspectratio,width=0.95\columnwidth]{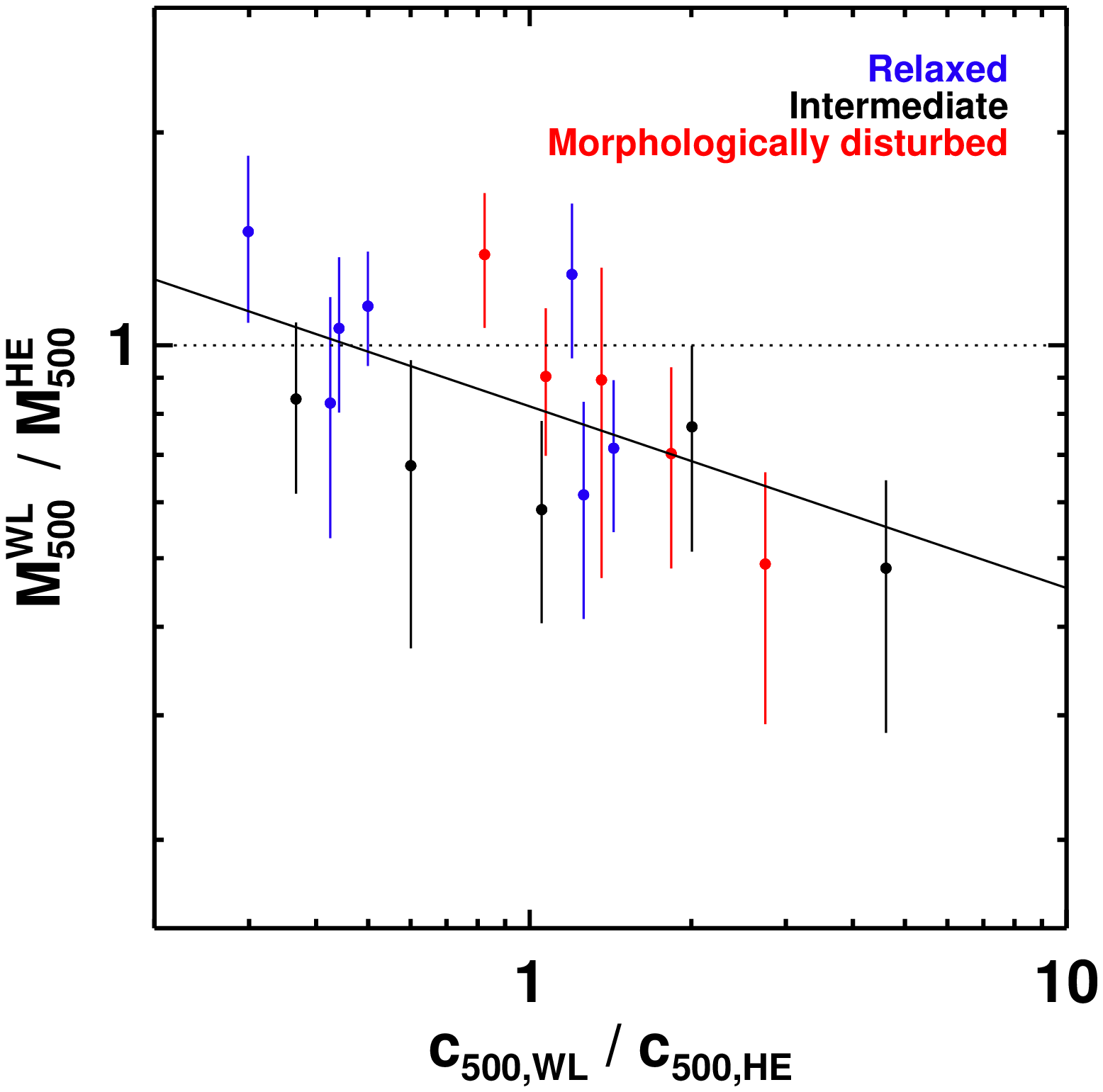}
\hfill
\includegraphics[scale=1.,angle=0,keepaspectratio,width=0.95\columnwidth]{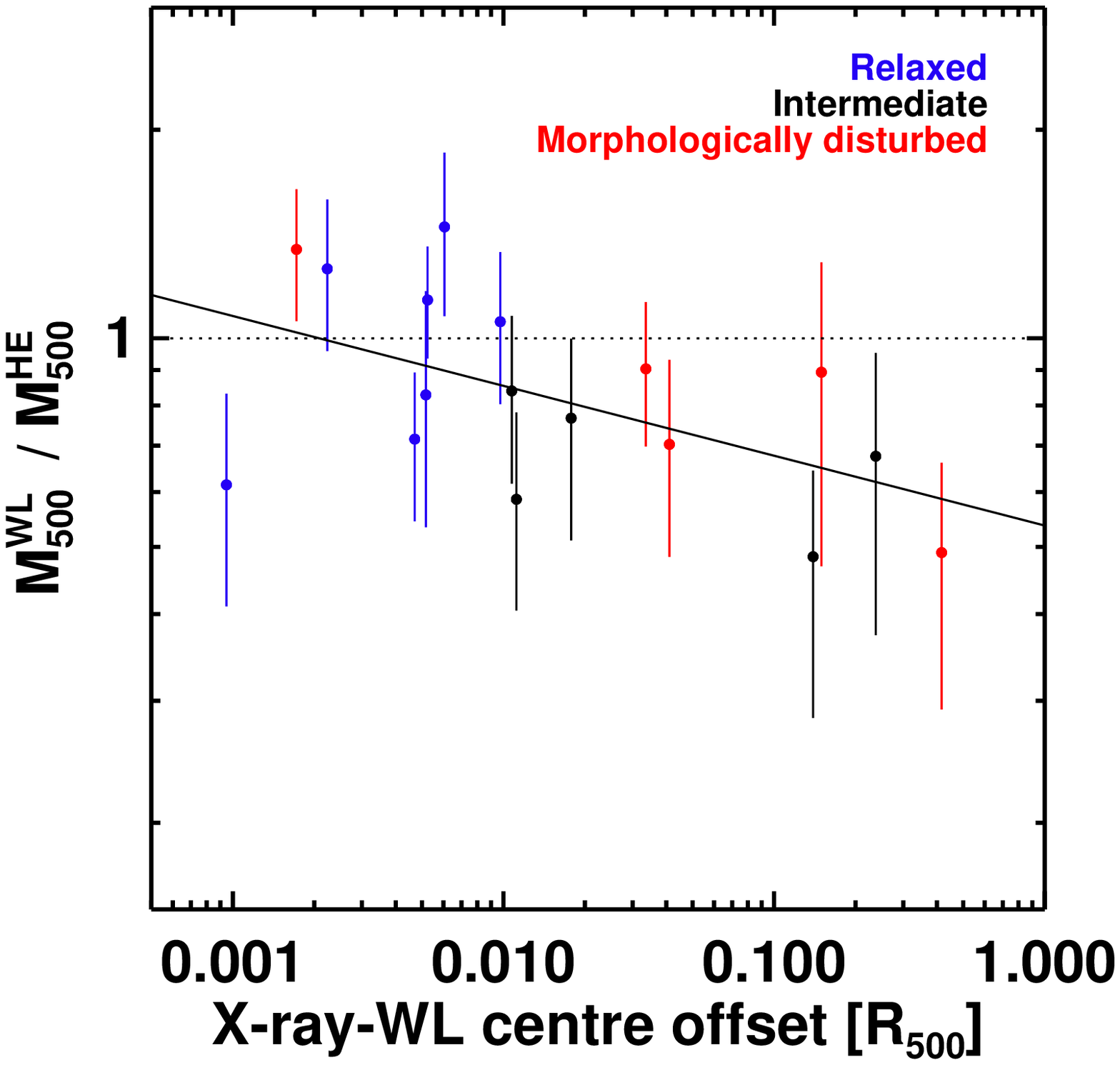}
\end{centering}
\caption{{\footnotesize {\it Left panel:} This plot shows the ratio of weak lensing mass to hydrostatic X-ray mass as a function of the ratio of NFW mass profile concentration parameter from weak lensing and X-ray analyses. {\it Right panel:} The ratio of weak lensing mass to hydrostatic X-ray mass is a function of offset between X-ray and weak lensing centres. In both panels, the solid line is the best-fitting orthogonal BCES power-law relation between the quantities, and the different sub-samples are colour coded.}}\label{fig:Mratio}
\end{figure*}
%%_______________

\subsubsection{Mass measurements}

The normalisation offset of the $\DAY$--$M^{\rm WL}_{500}$ relation may also be due to a systematic difference in mass measurements. We first need to verify that the hydrostatic X-ray mass estimates detailed in Sect.~\ref{sec:xmass} ($M_{500}^{\rm HE}$) are in agreement with the expectations from the mass proxy relation ($M_{500}^{Y{\rm x}}$). The comparison between these two quantities is shown in the left-hand panel of Fig.~\ref{fig:MxvsM}. The shaded region enclosed by the BCES orthogonal regression fit and its uncertainties is entirely consistent with equality between the two quantities. 

This leaves us with only one remaining possibility to explain the normalisation discrepancy in the $\DAY$--$M_{500}$ relation: a systematic difference in X-ray and weak lensing masses. The right-hand panel of Fig.~\ref{fig:MxvsM} shows the comparison between the hydrostatic X-ray mass $M_{500}^{\rm HE}$ and the weak lensing mass $M_{500}^{\rm WL}$. A clear offset can indeed be seen. 

However, contrary to expectations, the offset indicates that on average the hydrostatic X-ray masses are {\it larger} than the weak lensing masses. A power-law fit with the slope fixed to 1, denoted by the grey region in Fig.~\ref{fig:MxvsM}, indicates that $M_{500}^{\rm WL} = (0.78\pm0.08)\, M_{500}^{\rm HE}$. In other words, for this sample, the weak lensing masses are $\sim 20$ percent {\it smaller} than the hydrostatic X-ray masses at the $2.6\sigma$ significance level.

The mass discrepancy is clearly dependent on morphological sub-class. For relaxed systems, a power-law fit with the slope fixed to 1 yields a mean ratio of $M_{500}^{\rm WL} = (0.94\pm0.10)\, M_{500}^{\rm HE}$, indicating relatively good agreement between weak lensing and X-ray mass estimates. In contrast, the mean ratio for the intermediate and disturbed systems is $M_{500}^{\rm WL} = (0.72\pm0.12)\, M_{500}^{\rm HE}$. So the mass discrepancy is essentially driven by the difference between the hydrostatic X-ray and weak lensing masses of the intermediate and disturbed systems (although there is still a slight offset even for relaxed systems).

\citet{zha10} compared hydrostatic X-ray and LoCuSS Subaru weak lensing data for 12 clusters, finding excellent agreement between the different mass measures [$M_{500}^{\rm WL} = (1.01\pm0.07)\, M_{500}^{\rm HE}$]. As part of their X-ray--weak lensing study, \citet{zha10} analysed the same \xmm\ data for ten of the clusters presented here. For the clusters we have in common, the ratio of X-ray masses measured at $\Rv$ is $M_{\rm Zhang, 500}^{\rm HE}/M_{500}^{\rm HE} = 0.83\pm0.13$. This offset is similar to the offset we find between the X-ray and weak lensing masses discussed above, as expected since \citet{zha10} found $M_{500}^{\rm WL} \sim M_{500}^{\rm HE}$. However for relaxed systems (four in total), we find good agreement between hydrostatic mass estimates, with a ratio of $M_{\rm Zhang, 500}^{\rm HE}/M_{500}^{\rm HE} = 1.00\pm0.09$. It is not clear where the difference in masses comes from, although we note that for some clusters \citeauthor{zha10} centred their profiles on the weak lensing centre. This point is discussed in more detail below.

%_______________
%% Figure: A2631
%%
\begin{figure*}[]
\begin{centering}
\includegraphics[scale=1.,angle=0,keepaspectratio,width=0.95\columnwidth]{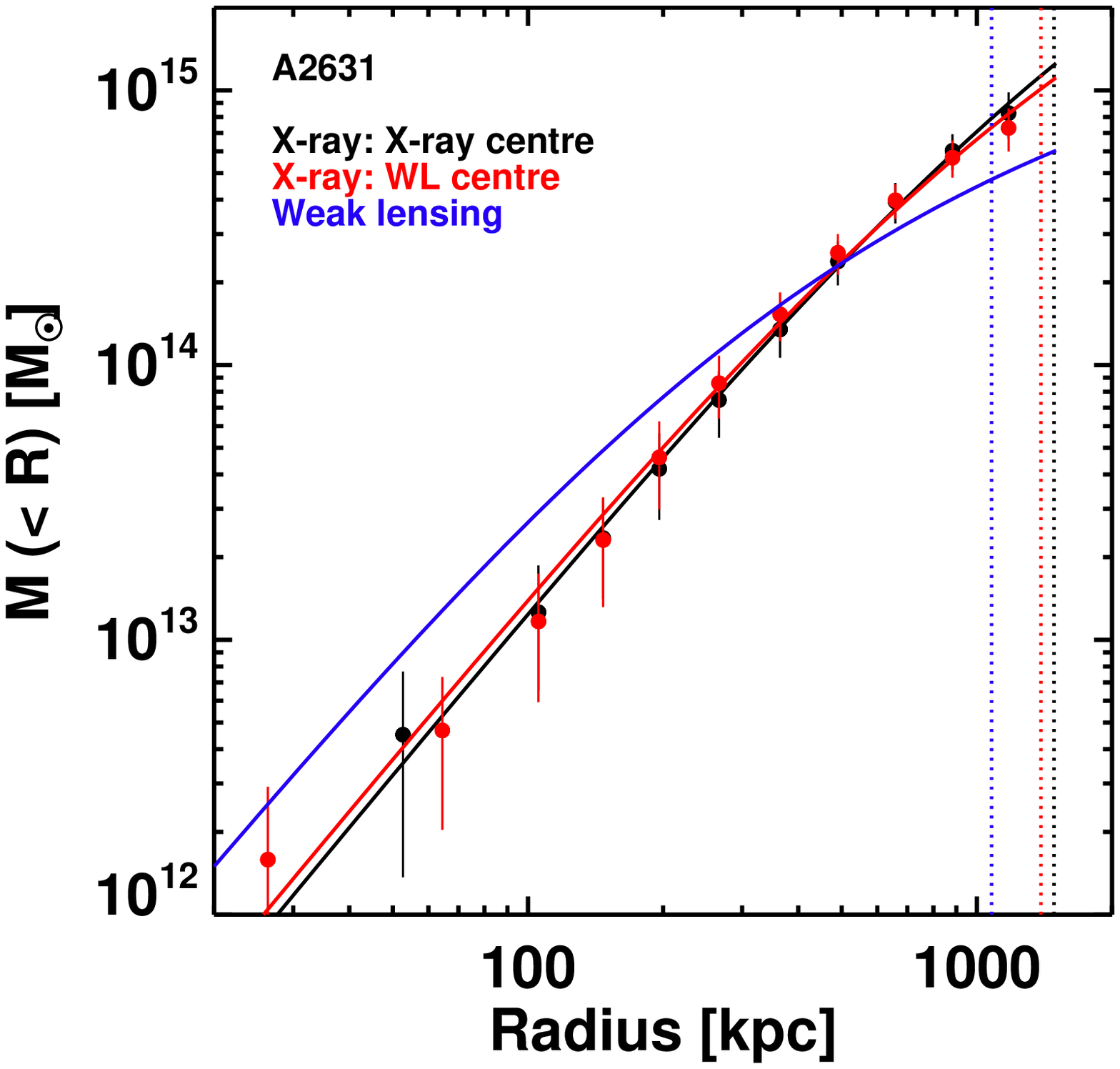}
\hfill
\includegraphics[scale=1.,angle=0,keepaspectratio,width=0.95\columnwidth]{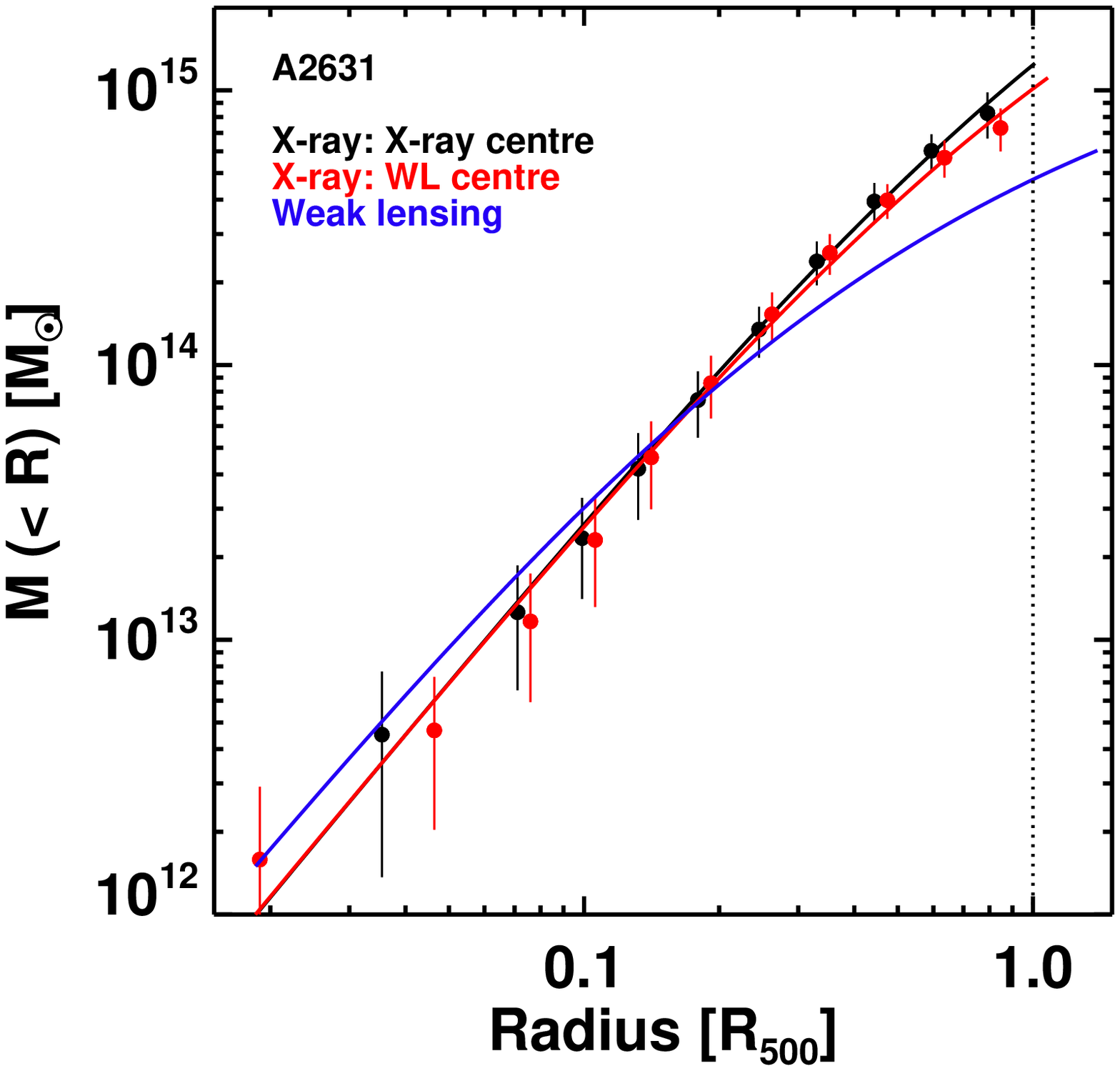}
\end{centering}
\caption{{\footnotesize Lensing and X-ray mass profiles for A2631 in $h_{70}^{-1}$ kpc ({\it left panel}) and in terms of $\Rv$ ({\it right panel}).  }}\label{fig:a2631}
\end{figure*}
%%_______________

%%%%%%%%%%%%%%%%%%%%%%%%%%%%%%%%%%%%%

\subsection{The mass discrepancy}
\label{sec:massdiscrep}

Our finding that the hydrostatic X-ray masses are larger than the weak lensing masses contradicts the results from many recent numerical simulations, all of which conclude that the hydrostatic assumption underestimates the true mass owing to its neglect of pressure support from gas bulk motions \citep[e.g.,][]{nag07,pif08,mene10}. If the weak lensing mass is indeed unbiased (or less biased) and thus, on average, more representative of the true mass, then one would expect the weak lensing masses to be {\it larger} than the hydrostatic X-ray masses. What could be the cause of this unexpected result? 

%%%%%%%%%%%%%%%%%%%%%%%%%%%%%%%%%%%%%

\subsubsection{Concentration}

To investigate further, we fitted the integrated X-ray mass profiles with an NFW model of the form
\begin{equation}
M(<r) = 4 \pi\, \rho_c(z)\, \delta_c\, r_{\rm s}^3\, \left[ \ln{(1+r/r_{\rm s})} - \frac{r/r_{\rm s}}{1+r/r_{\rm s}} \right], \label{eqn:nfw}
\end{equation}
\noindent where $\rho_c(z)$ is the critical density of the universe at redshift $z$, the quantity $r_{\rm s}$ is the scale radius where the logarithmic slope of the density profile reaches $-2$, and $\delta_{\rm c}$ is a characteristic dimensionless density. This model has been shown to be an adequate fit to the mass profiles of many morphologically relaxed systems \citep[e.g.,][]{pra02,poi05,vik06,gas07}. We emphasise that for the present investigation we use it only as a convenient fitting formula that allows direct comparison with the equivalent weak lensing parameterisations. The best-fitting NFW mass model parameters are listed in Table~\ref{tab:mnfw} and plots of the integrated mass profiles and the best-fitting models can be found in Fig.~\ref{fig:mprof}.

Figure~\ref{fig:Mratio} shows the ratio of weak lensing to hydrostatic X-ray mass at $\Rv$ in terms of the ratio of the concentration parameter of each NFW mass profile fit. There is a clear trend for the mass ratio to depend on the ratio of the concentration parameters. Indeed, a BCES orthogonal power-law fit to the relation yields
\begin{equation}
\frac{M_{500}^{\rm WL}}{M_{500}^{\rm HE}} = 10^{-0.09\pm0.03}\, \left(\frac{c_{\rm 500, WL}}{c_{\rm 500,HE}}\right)^{-0.25\pm0.07}.
\end{equation}
This result is extremely robust to the presence of outliers in the relation and to the radial range used to determine the X-ray NFW fit. The median ratio of scale radius to $R_{\rm 500, HE}$ is $r_{\rm s} / R_{\rm 500,HE} = 0.35$; excluding the three clusters for which $r_{\rm s} / R_{\rm 500,HE} > 1$ yields a slope of $-0.25 \pm 0.11$. The result indicates that the weak lensing analysis finds NFW mass profiles that are, on average, more concentrated than the corresponding hydrostatic X-ray NFW mass profiles in disturbed systems. As illustrated in Fig.~\ref{fig:a2631}, this in turn typically explains the trend for the weak lensing masses to be lower than the X-ray masses at $\Rv$.

Recent simulations (and some observations) have found that the X-ray ``hydrostatic mass bias'' is radially dependent \citep[e.g.,][]{mah08,mene10,zha10,ras12}, presumably due to the ICM becoming progressively less virialised the further one pushes into the cluster outskirts. The difference in concentration that we find here cannot be solved by appealing to such a radially dependent X-ray ``hydrostatic mass bias''. If this effect is real, then the hydrostatic X-ray mass estimates effectively ignore it, meaning that at each radius at which the X-ray mass profile is measured, the true mass would be underestimated and the underestimation would become worse with radius. The resulting hydrostatic X-ray mass profile would be over-concentrated relative to the true underlying mass distribution. Correcting for this effect would reduce even further the measured X-ray concentration, exacerbating the effect we see here.

%%%%%%%%%%%%%%%%%%%%%%%%%%%%%%%%%%%%%

\subsubsection{Centre offsets}

In the present work, the X-ray and weak lensing analyses are completely independent, extending even to the choice of centre for the various profiles under consideration. We recall that \citet{oka08} and \citet{oka10} centred their weak lensing shear profiles on the position of the BCG. In contrast, our hydrostatic X-ray analysis centres each profile on the X-ray peak after removal of obvious sub-structures.\footnote{This is in fact required, since otherwise the X-ray analysis would give unphysical results.}  The fact that at $\Rv$ the mass ratio vs. concentration ratio seems to be driven by the intermediate and disturbed systems  (see Fig.~\ref{fig:Mratio}) suggests that the different choice of centre could have a bearing on the results. 

We test this in the right-hand panel of Fig.~\ref{fig:Mratio}, which shows the weak lensing to hydrostatic X-ray mass ratio as a function of the offset between the BCG position and the X-ray peak. A clear trend is visible, in the sense that the larger the offset $R_{\rm X-WL}$ between centres in units of $\Rv$, the larger the mass discrepancy. Indeed, an orthogonal BCES power-law fit yields
\begin{equation}
\frac{M_{500}^{\rm WL}}{M_{500}^{\rm HE}} = 10^{-0.27\pm0.07}\, \left[\frac{R_{\rm X-WL}}{\Rv}\right]^{-0.10\pm0.04}.
\end{equation}
Thus at least part of the difference between X-ray and weak lensing mass estimates appears to be due to differences in centring between the two approaches. Although the trend is visible in each morphological sub-sample, the most extreme deviations occur in the intermediate and disturbed systems, which all have the largest offsets between X-ray and BCG positions. This is a well-known characteristic of the observed cluster population \citep[e.g.,][]{bil08,san09,haa10}.

We tested the effect of using a different centre on the X-ray mass for two systems. A2631 displays the largest difference in mass ratio as a function of concentration parameter (i.e., it is the right-most point in the left-hand panel of Fig.~\ref{fig:Mratio}) and a moderate X-ray--weak lensing centre offset $\sim 0.14\,\Rv$. A520 exhibits the largest difference in mass ratio as a function of X-ray--weak lensing centre offset (i.e., it is the right-most point in the right-hand panel of Fig.~\ref{fig:Mratio}), with an X-ray--weak lensing centre offset of $\sim 0.40\,\Rv$. For A2631, the choice of centre does not significantly change  either the X-ray mass profile or the parameters of the NFW model fitted to it, as can be seen in Fig.~\ref{fig:a2631}. However, when the X-ray profiles were centred on the weak lensing centre we were unable to find any physical solution to the hydrostatic X-ray mass equation (Eq.~\ref{eqn:xhe}) for A520.

We note that the dependence of the mass ratio on centre shift is qualitatively in agreement with the results of the simulations by \citet{ras12}. These authors found that the strongest weak lensing mass biases (with respect to the true mass) occurred in clusters with the largest X-ray centroid shift, $w$. This conclusion is supported by the clear correlation between $\langle w \rangle$ and $R_{\rm X-WL}$, for which we obtained a Spearman rank coefficient of $-0.70$ and a null hypothesis probability of $<0.001$.

%%%%%%%%%%%%%%%%%%%%%%%%%%%%%%%%%%%%%

\subsubsection{Other effects}

Several other effects could systematically influence the weak lensing mass measurements of \citet{oka08} and \citet{oka10}, and thus contribute to the offset between X-ray and weak lensing masses that we find here. 

Firstly, a potential bias could arise from dilution of the measured weak lensing signal produced by any cluster galaxies contained in the galaxy samples used to measure the tangential distortion profiles. As discussed in \citet{oka08} and \citet{oka10}, for all but one cluster considered here (A963), data in two passbands were available, enabling a separation of cluster and background galaxies based on their location in a colour-magnitude diagram. The sample of assumed background galaxies consisted of two components: a ``red sample'' with (depending on the available Subaru data) $V-i'$, $V-I_{\rm C}$, $V-R_{\rm C}$, or $g'-R_{\rm C}$ colours significantly greater than the colour index of the red sequence formed by early-type cluster galaxies; and a ``blue sample'' with significantly lower colour index than the red sequence. The red sample should have very little contamination, as all normal galaxies redder than the observed red sequence would be predicted to lie at higher redshifts than the cluster.  By contrast, the blue sample will be contaminated at some level by cluster dwarf galaxies undergoing significant star formation.  Interactions with other galaxies and the intracluster medium in the central regions of the cluster would tend to destroy dwarf galaxies or quench their star formation,  producing an observed radial distribution of dwarf galaxies which is much shallower than that of the bright early-type galaxies \citep{pra04}. Hence, while likely present, a small residual contamination of cluster galaxies in the ``blue'' background galaxy sample would be difficult to identify and remove without adding more photometric filters to the data set. A  dilution bias systematically lowering the measured weak lensing masses by a few percent cannot be excluded, and there would also be significant cluster-to-cluster variations in the strength of this effect.

Secondly, the derived cluster masses are sensitive to the estimated redshift distribution of the gravitationally lensed background galaxies. For the lensing mass measurements of \citet{oka08} and \citet{oka10}, the background galaxy redshifts were estimated from the photometric redshift catalogue of galaxies in the COSMOS survey field \citep{ilb09}. Given the depth of the Subaru data, mass measurements of clusters in this redshift range ($0.1 < z < 0.3$) are not very sensitive to percent level uncertainties in the photometric redshifts of the background galaxies. However, a potentially significant bias may arise from effectively excluding gravitational lensing measurements of galaxies that are smaller than the point spread functions (PSFs) of the Subaru images without imposing a similar size cut on the COSMOS galaxy catalogue. This would also tend to lower the lensing mass estimate by removing the apparently smallest (and thus on average more distant) galaxies, resulting in an overestimate of the mean effective background galaxy redshift compared to the true value.

Thirdly, the weak lensing distortion measurements of \citet{oka08} and \citet{oka10} are based on an implementation of the KSB+ method \citep{kai95, lup97, hoe98}. Tests against simulated lensing data \citep{hey06,mas07} indicate that this method is generally affected by a multiplicative calibration bias, underestimating the true lensing signal by up to $15$ percent, depending on details in the implementation of the method. This would result in an underestimate of the cluster masses by the same amount; however, tests of the particular KSB+ implementations of \citet{oka08} and \citet{oka10} against realistic simulated weak lensing data would be required to measure the calibration factor needed to correct for this effect. 

Finally, recent simulations predict that at large radii the true mass distribution departs from the NFW model that was used to fit the tangential shear profile  in \citet{oka08} and \citet{oka10}. For example, \citet{ogu11} recently investigated the use of smoothly truncated NFW profiles, finding them to be a more accurate description of their simulated clusters. They predicted that the use of a standard NFW model would produce an overestimate of the concentration, and corresponding underestimation of mass, of order five percent for clusters similar to those studied here. This bias can be minimised by including the effect of large-scale structure in the shear measurement uncertainties, thus giving lower statistical weight to the weak lensing measurements at large radii \citep[e.g.,][]{hoe03,dod04,hoe11b}.

Interestingly, while all these effects are individually well within the statistical errors of the mass measurement of a single cluster, they would 
all have a tendency to bias the measured weak lensing masses downwards with respect to their true value. Hence, their cumulative effect (combined with the centre offsets) could go a long way towards explaining the difference between X-ray and weak lensing masses. 

\subsection{Summary}

For the present sample, the mass discrepancy between hydrostatic X-ray and weak lensing mass measurements is such that, at $\Rv$, the X-ray masses are larger by $\sim 20$ percent. This appears to be due to a systematic difference in the mass concentration found by the different approaches, which in turn appears to be driven by the clusters that were classified as intermediate and/or disturbed. Such systems also show a clear tendency to have smaller weak lensing to X-ray mass ratios as a function of offset between the X-ray peak and the BCG position. On the other hand there is a mass normalisation offset for relaxed systems, but it is not significant, as found by previous studies \citep[e.g.,][]{vik09,zha10}.

The results discussed above bring to light three fundamental, interconnected  obstacles to a proper comparison of X-ray and weak lensing mass measurements. The first is how to define the ``centre'' of a cluster and determine what is the ``correct'' centre to use. For X-ray astronomers, the obvious choice, once clear sub-structures are excluded, is the X-ray peak or centroid (these are not necessarily the same). In contrast, since the weak lensing signal is most sensitive on large scales, many weak lensing analyses use the position of the BCG.

For relaxed clusters, the BCG position and the X-ray peak generally coincide, so that the question of what centre to choose does not arise. However, as numerical simulations will testify, for disturbed clusters neither of these choices is necessarily the ``correct'' centre, in the sense that often neither coincides with the true cluster centre of mass. Indeed, \citet{hoe11} show that the recovered weak lensing mass at a given density contrast depends on the offset of the BCG from the cluster centroid and that the mass can be underestimated by up to ten percent for reasonable values of centroid offset. This means that for unrelaxed systems with offsets between the position of the X-ray peak and the BCG position, both approaches will likely give incorrect results. 

The second obstacle is connected to the fact that X-rays measure close to 3D quantities, while lensing measures 2D quantities, and an analytical model is required to transform between the two. The choice of an NFW model is motivated by its simple functional form and from the fact that it provides a relatively good fit to both X-ray and weak lensing data. But we should not forget that the original NFW model profile was defined for {\it equilibrium} haloes \citep{nav96}, and many subsequent works have shown that the functional form is not a particularly good description of non-equilibrium haloes \citep[e.g.,][]{jin00}. Indeed, very recent work by \citet{bec11} and \citet{bah11} shows that the use of an NFW model profile in a weak lensing context can introduce non-negligible biases into the mass estimation procedure, primarily because of a departure of the mass distribution from the NFW form at large cluster-centric radii. 

This point is exacerbated by the third obstacle: the fact that X-rays and weak lensing observations have fundamentally different sensitivities to the mass distribution in a cluster. X-rays probe the central regions, and with present instruments at least, it is difficult to make very precise measurements of the logarithmic temperature gradient at and beyond $\Rv$. In contrast, this is just the radial range at which weak lensing starts to become most sensitive. Certainly, the combination of strong and weak lensing offers much tighter constraints on the inner mass distribution \citep[e.g.,][]{kne03}, but precise strong lensing measurements are difficult to achieve with ground-based instruments.

There are several requirements for future progress on this issue. The good agreement found here between mass estimates for relaxed systems is encouraging and needs to be confirmed for a much larger sample of objects, allowing a more precise observational constraint to be put on the ``hydrostatic mass bias''. Additionally, such a sample of relaxed objects will allow constraints to be put on the irreducible scatter between the different mass estimates due to the conversion from 2D to 3D quantities. Projects such as the ``Cluster Lensing and Supernova Survey with Hubble'' \citep[CLASH;][]{pos11} will surely make great progress on these questions. However, a new, or at least co-ordinated, approach is needed in the case of dynamically disturbed systems. Here, numerical simulations can also be used to inform the different analyses and to optimise the mass estimation procedure in each case. One possible approach would be to use the centroid of the projected X-ray pressure profile as the point around which both X-ray and weak lensing profiles could be centred. 

Eventually, X-ray and weak lensing mass estimates will be needed for a representative (or complete) sample of systems. We stress that data quality across such a sample must be as close to homogeneous as possible. In the case of the X-ray data set, the data must be sufficiently deep to measure the temperature profile to $\Rv$. Similarly stringent data quality will also be required for the weak lensing data set.

%%%%%%%%%%%%%%%%%%%%%%%%%%%%%%%%%%%%%

\section{Conclusions}

A well-calibrated relation between the direct observable and the underlying total mass is essential to leverage the statistical power of any cluster survey.  In this paper we presented an investigation of the relations between the SZ flux and the mass for a small sample of 19 clusters for which weak lensing mass measurements are available in the literature and high-quality X-ray observations are available in the \xmm\ archive. This ``holistic'' approach allowed us to investigate the interdependence of the different quantities and to attempt to square the circle regarding the different mass estimation methods.

Using weak lensing masses from the LoCuSS sample \citep{oka08,oka10}, we found  that the SZ flux is well correlated with the total mass, with a slope that is compatible with self-similar, and a dispersion about the best-fitting relation that is in agreement with both previous observational determinations \citep{marr11} and simulations that take into account observational measurement uncertainty \citep{bec11}. However, at $\Rv$, there was a normalisation offset with respect to that expected from previous measurements based on hydrostatic X-ray mass estimates. 

We verified that the SZ flux measurements and hydrostatic mass estimates of the present sample are in excellent agreement with our previous work \citep{planck2011-5.2b}. The normalisation offset is due to a systematic difference between hydrostatic X-ray and weak lensing masses, such that for this particular sample the weak lensing masses are $22\pm8$ percent {\it smaller} than the hydrostatic X-ray mass estimates. The difference is essentially driven by the intermediate and morphologically disturbed systems, for relaxed objects, the weak lensing mass measurements are in good agreement with the hydrostatic X-ray estimates.

We examined the possible causes of the mass discrepancy. At $\Rv$, the X-ray--weak lensing mass ratio is strongly correlated with the offset in X-ray and weak lensing centres. It is also strongly correlated with the ratio of NFW concentration parameters, indicating that the mass profiles determined from weak lensing are systematically more concentrated than the corresponding X-ray mass profiles in disturbed systems. We argued that a radially dependent ``hydrostatic mass bias'' in the X-ray observations would exacerbate this effect, and discussed several other alternative explanations, including dilution and uncertainties due to the use of NFW mass profiles to model the weak lensing data set. 

Significant progress on the mass calibration of clusters can only be achieved with a dedicated X-ray-lensing survey of a {\it representative} sample of clusters. Data of sufficient quality are an essential prerequisite for such a survey. X-ray observations with sufficiently deep exposures to measure the temperature at $\Rv$ are needed, as are optical observations of uniformly high quality. We expect such data to become available in the coming years. 

%%%%%%%%%%%%%%%%%%%%%%%%%%%%%%%%%%%%%

\begin{acknowledgements}
We thank N. Okabe and D. Marrone for useful discussions. The present work is partly based on observations obtained with \xmm, an ESA science mission with instruments and contributions directly funded by ESA Member States and the USA (NASA). The development of \Planck\ has been supported by: ESA; CNES and CNRS/INSU-IN2P3-INP (France); ASI, CNR, and INAF (Italy); NASA and DoE (USA); STFC and UKSA (UK); CSIC, MICINN and JA (Spain); Tekes, AoF and CSC (Finland); DLR and MPG (Germany); CSA (Canada); DTU Space (Denmark); SER/SSO (Switzerland); RCN (Norway); SFI (Ireland); FCT/MCTES (Portugal); and DEISA (EU). A description of the Planck Collaboration and a list of its members, including the technical or scientific activities in which they have been involved, can be found at \url{http://www.rssd.esa.int/Planck}.
\end{acknowledgements}

%%%%%%%%%%%%%%%%%%%%%%%%%%%%%%%%%%%%%

\bibliographystyle{aa}
\bibliography{19398.bib,Planck_bib.bib}

%%%%%%%%%%%%%%%%%%%%%%%%%%%%%%%%%%%%%%%%%%%%%%%%%%%%%%%%%%%%%%%%%%%%%%%%%%%%
\appendix

\section{Pressure profiles and best-fitting model }
\label{ap:pnfw}

\begin{figure*}[]
\begin{centering}
\includegraphics[scale=1.,angle=0,keepaspectratio,width=0.95\textwidth]{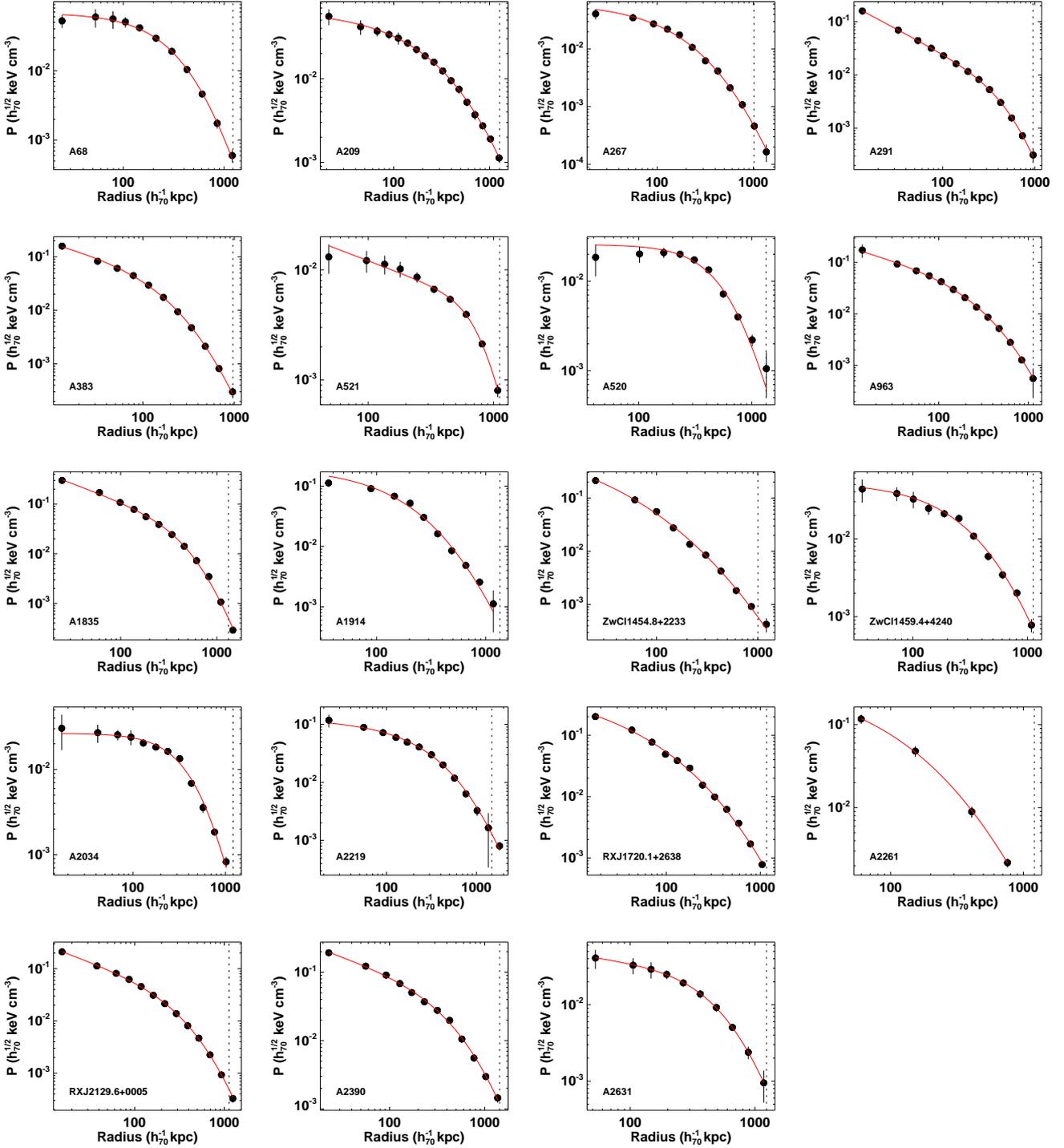}
\end{centering}
\caption{{\footnotesize  Pressure profiles of the sample with the best-fitting GNFW model overplotted (red line). The dotted vertical line indicates $\Rv$ for each cluster.  }}\label{fig:pprof}
\end{figure*}

The X-ray pressure profile of each cluster was fitted with a generalised-NFW (GNFW) model as described in Sect.~\ref{sec:xpress}. Table~\ref{tab:pnfw} gives the best-fitting model and Figure~\ref{fig:pprof} shows each X-ray profile with the best-fitting GNFW model overplotted.

%______________________________________________________________
% X-ray pressure profile fit parameters
%
\begin{table*}[t]
\newcolumntype{L}{>{\columncolor{white}[0pt][\tabcolsep]}l}
\newcolumntype{R}{>{\columncolor{white}[\tabcolsep][0pt]}l}
%\rowcolors{3}{light-grey}{white}
\caption[]{\footnotesize best-fitting X-ray pressure profile parameters. }
\label{tab:pnfw}
\centering
\begin{tabular}{@{}LccrcccrR@{}}

\toprule
\toprule

Cluster &  $\Rv $  & $P_{500}$ & $P_0$ &$c_{500}$ & $\alpha$ & $\gamma$ &$\chi^2$ & dof \\
%\cmidrule[0.5pt](lr){2-2}
%\cmidrule[0.5pt](lr){3-3}
     & (Mpc) & $(10^{-3}\,\keV$~cm$^{-3})$ & \\

\midrule

A68 & $ 1.030$ & $ 2.828$ & $ 24.01$ & $1.54$ & $1.31$ & $0.000 $ & $1.8$ & $7$ \\
A209 & $ 1.344$ & $ 4.328$ & $ 14.51$ & $0.78$ & $0.79$ & $0.008 $ & $0.9$ & $13$ \\
A267 & $ 0.964$ & $ 2.345$ & $ 26.66$ & $1.18$ & $0.89$ & $0.000 $ & $6.3$ & $8$ \\
A291 & $ 1.044$ & $ 2.558$ & $ 1.58$ & $1.41$ & $1.82$ & $0.919 $ & $0.5$ & $9$ \\
A383 & $ 0.989$ & $ 2.256$ & $ 14.61$ & $1.49$ & $0.97$ & $0.422 $ & $2.4$ & $7$ \\
A521 & $ 1.015$ & $ 2.699$ & $ 1.75$ & $1.17$ & $3.53$ & $0.434 $ & $1.7$ & $6$ \\
A520 & $ 1.056$ & $ 2.654$ & $ 9.67$ & $1.29$ & $1.88$ & $0.000 $ & $7.7$ & $6$ \\
A963 & $ 1.055$ & $ 2.667$ & $ 8.81$ & $0.98$ & $0.97$ & $0.463 $ & $1.0$ & $9$ \\
A1835 & $ 1.363$ & $ 4.934$ & $ 5.36$ & $1.81$ & $1.71$ & $0.721 $ & $8.5$ & $8$ \\
A1914 & $ 1.115$ & $ 2.769$ & $ 72.97$ & $1.56$ & $0.95$ & $0.000 $ & $30.1$ & $6$ \\
ZwCl1454.8+2233 & $ 0.914$ & $ 2.242$ & $ 706.26$ & $0.71$ & $0.44$ & $0.000 $ & $14.3$ & $6$ \\
ZwCl1459.4+4240 & $ 0.998$ & $ 2.870$ & $ 18.52$ & $1.20$ & $1.10$ & $0.000 $ & $5.3$ & $7$ \\
A2034 & $ 1.174$ & $ 2.726$ & $ 9.73$ & $1.78$ & $1.72$ & $0.000 $ & $10.0$ & $8$ \\
A2219 & $ 1.302$ & $ 4.258$ & $ 21.92$ & $1.03$ & $0.99$ & $0.056 $ & $4.5$ & $9$ \\
RXJ1720.1+2638 & $ 1.034$ & $ 2.344$ & $ 79.14$ & $0.71$ & $0.58$ & $0.178 $ & $10.9$ & $8$ \\
A2261 & $ 1.305$ & $ 4.244$ & $ 87.01$ & $1.54$ & $0.70$ & $0.000 $ & $0.1$ & $0$ \\
RXJ2129.6+0005 & $ 1.079$ & $ 2.967$ & $ 6.24$ & $1.23$ & $1.20$ & $0.612 $ & $1.4$ & $9$ \\
A2390 & $ 1.244$ & $ 3.913$ & $ 7.81$ & $0.91$ & $1.13$ & $0.467 $ & $3.1$ & $8$ \\
A2631 & $ 1.078$ & $ 3.260$ & $ 7.44$ & $1.24$ & $1.61$ & $0.200 $ & $0.0$ & $6$ \\

\bottomrule
\end{tabular}
\tablefoot{Column (2): $\Rv$, the radius corresponding to a density contrast of 500, estimated from the weak lensing mass analysis of \citet{oka08,oka10}. Column (3): $P_{500}$ as defined by Eq.~5 of \citet{arn10}. Columns (4) to (7)  give the best-fitting GNFW parameters for the pressure profiles (Eq.~\ref{eq:pgnfw}). The external slope parameter $\beta$ has been fixed to $5.49$ (see text). No uncertainties are given as the GNFW parameters are highly degenerate. The profiles and best-fitting GNFW models are illustrated in Fig.~\ref{fig:pprof}.}
\end{table*}

%_______________
%% Figure: Pressure profile fits
%%
%%_______________

%%%%%%%%%%%%%%%%%%%%%%%%%%%%%%%%%%%%%

\section{Image gallery}\label{appx:imgal}

\begin{figure*}[]
\begin{centering}
\includegraphics[scale=1.,angle=0,keepaspectratio,width=0.95\textwidth]{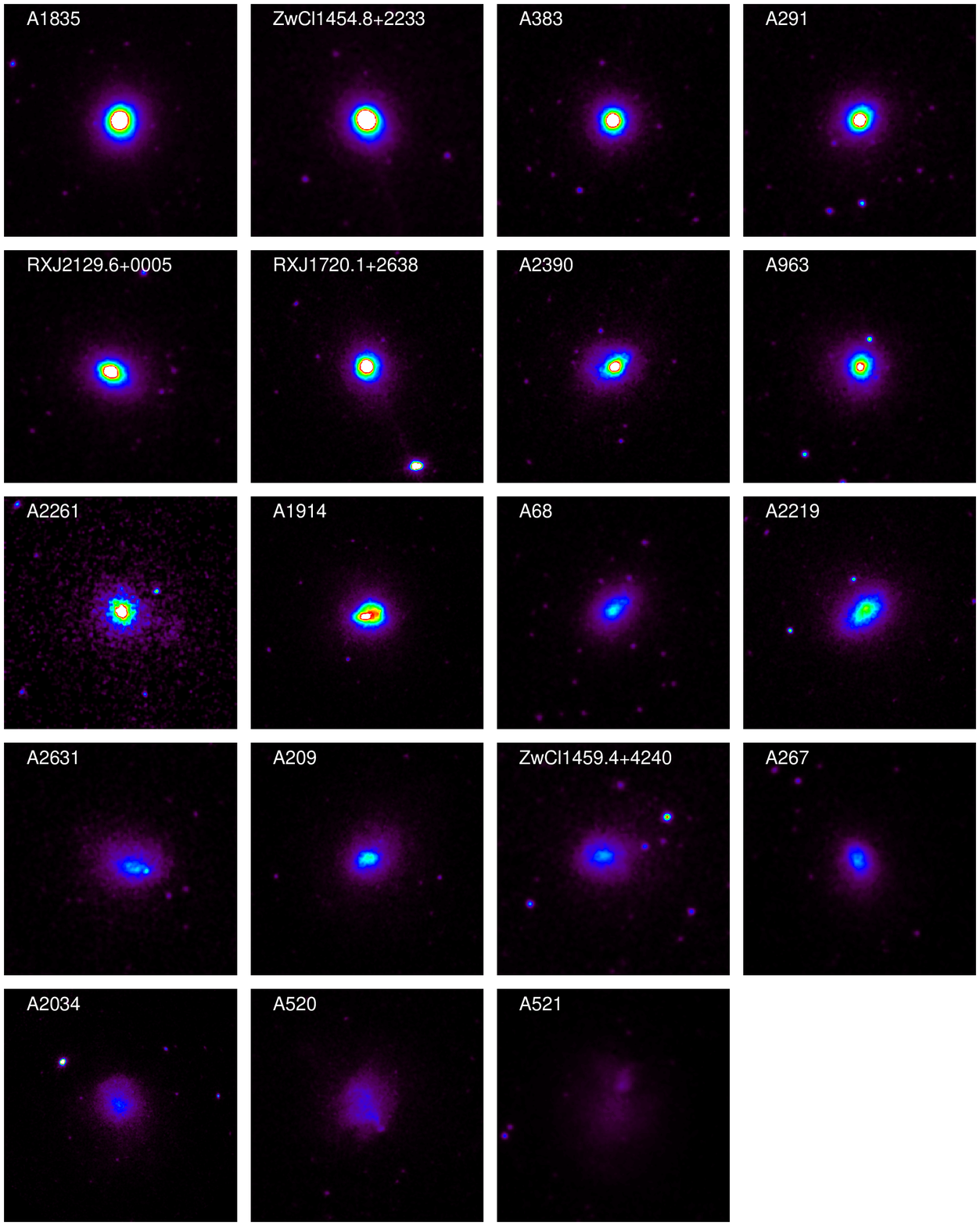}
\end{centering}
\caption{{\footnotesize  Image gallery. }}\label{fig:gallery}
\end{figure*}

Figure ~\ref{fig:gallery} shows the 0.3--2\,keV band X-ray image gallery of the cluster sample, arranged from top left to bottom right in order of the morphological characterisation parameter $n_{\rm e,0}$, the central density.  Images are corrected for surface brightness dimming with $z$, divided by the emissivity in the energy band, taking into account galactic absorption and instrument response, and scaled according to the self-similar model. The colour table is the same for all clusters, so that the images would be identical if clusters obeyed strict self-similarity, and each panel is $1.25\, \Rv$ on a side. 

%_______________
%% Figure: Images
%%
%%_______________

%%%%%%%%%%%%%%%%%%%%%%%%%%%%%%%%%%%%%

\section{Scaled X-ray profiles}

\begin{figure*}[!ht]
\begin{centering}
\includegraphics[width=0.45\textwidth]{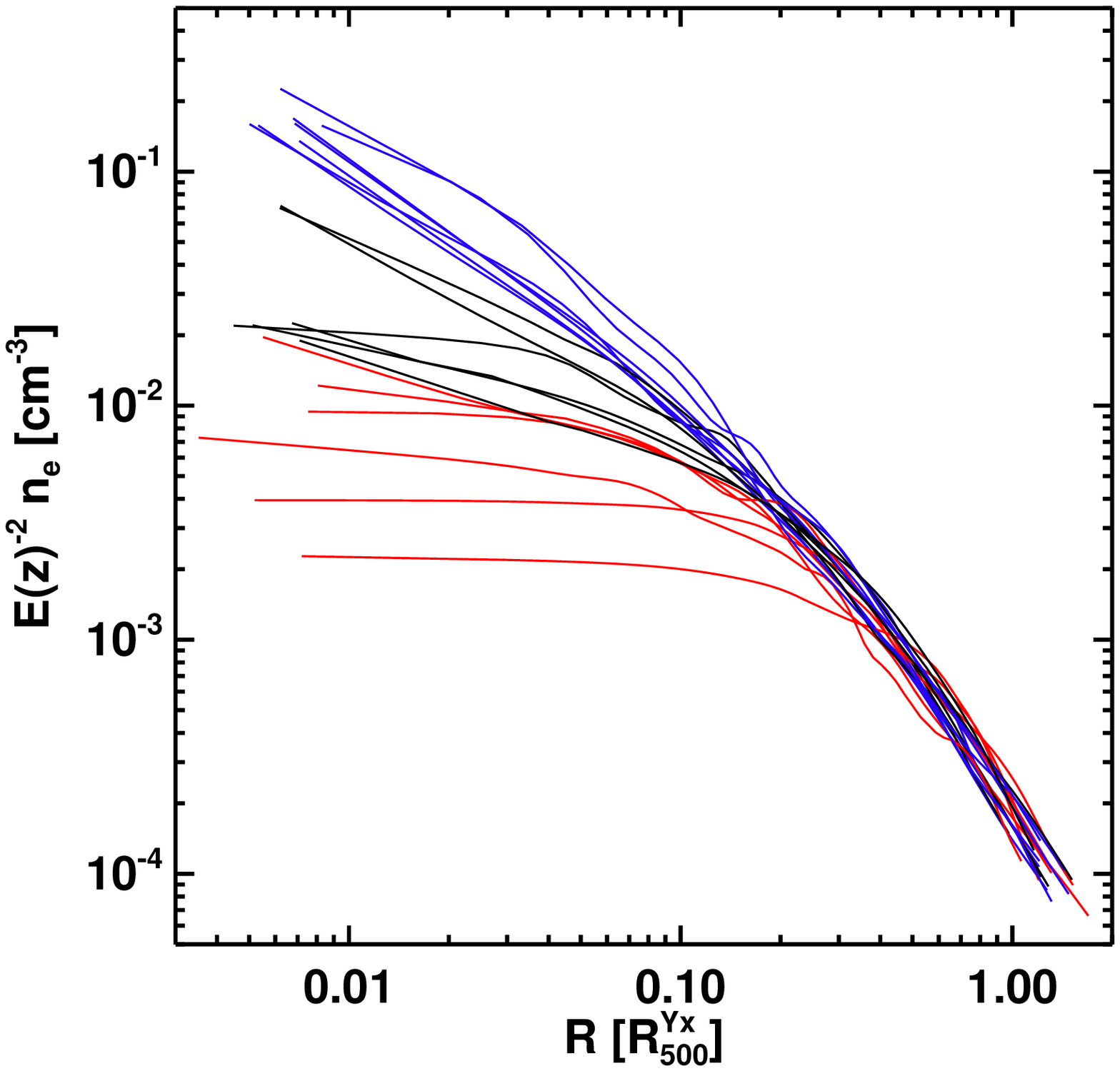}\qquad             
\includegraphics[width=0.45\textwidth]{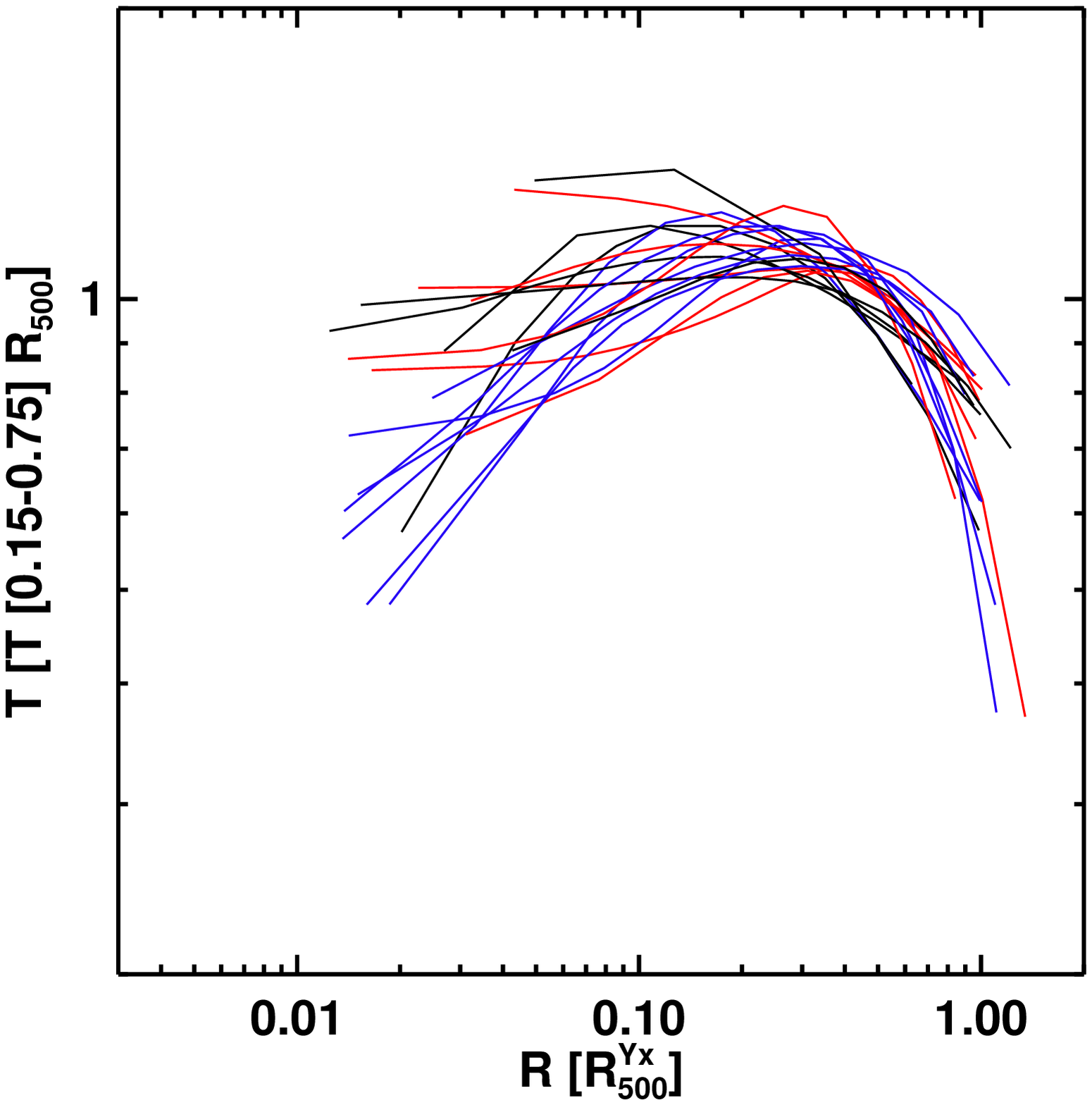}\\
\includegraphics[width=0.45\textwidth]{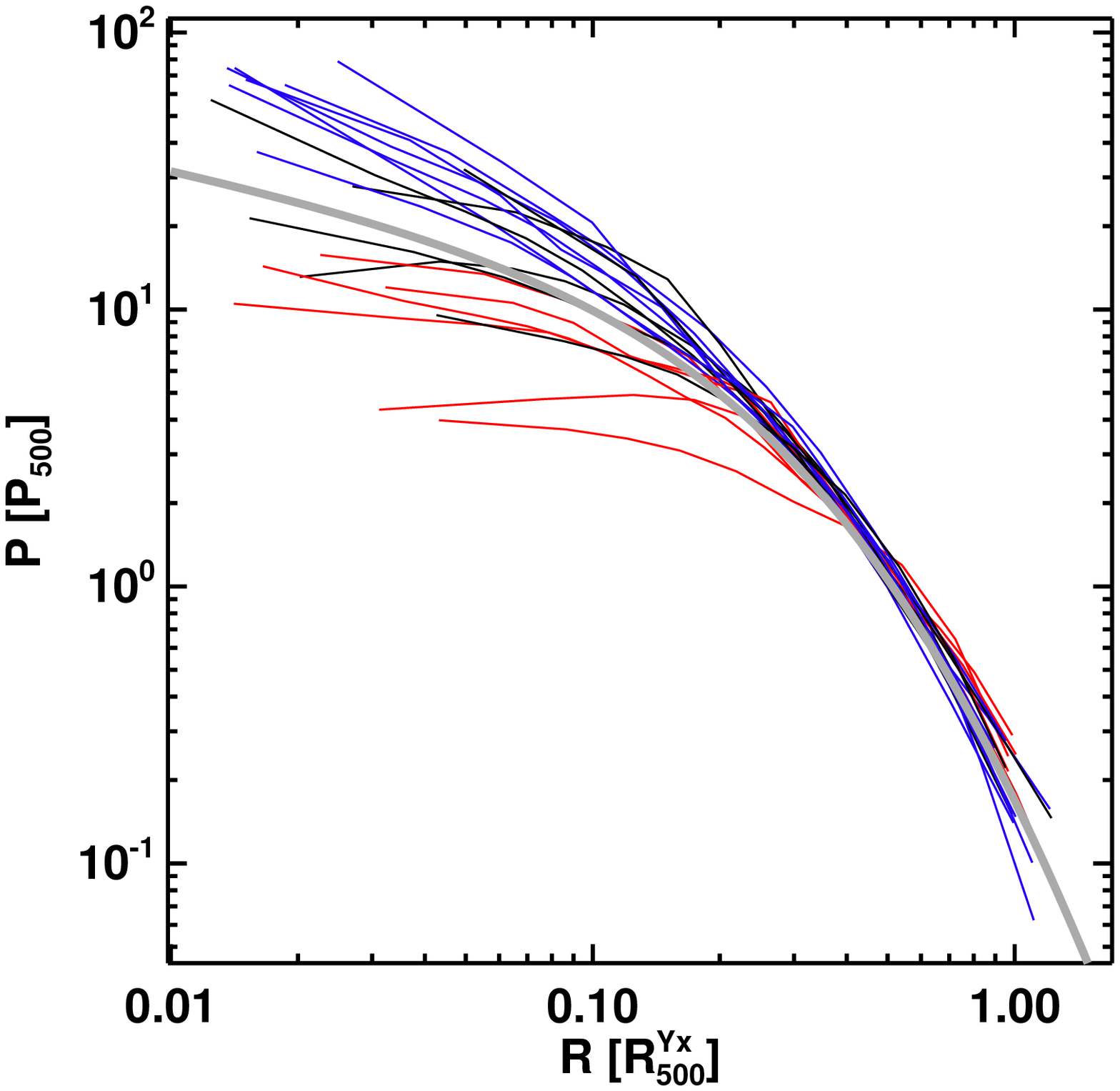}\qquad             
\includegraphics[width=0.45\textwidth]{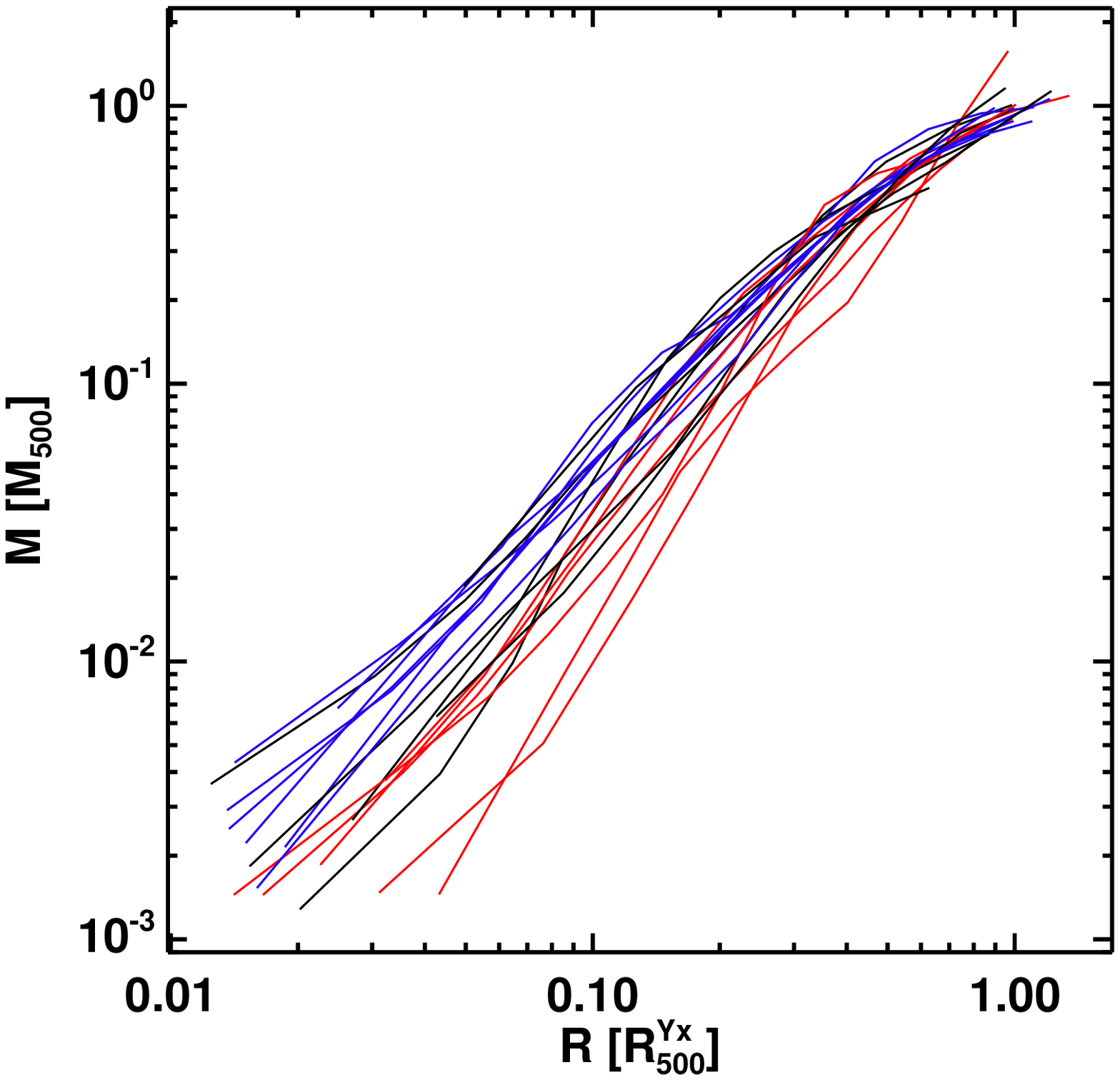}\\
\end{centering}
\caption{\footnotesize X-ray profiles scaled by $\Rv$ from the $\Mv$--$\YX$ relation of \citet{arn10}. From left to right, top to bottom: gas density, gas temperature, gas pressure, integrated mass. Cool core systems are plotted in blue, morphologically disturbed objects in red, and intermediate systems in black. The grey line in the pressure plot is the universal pressure profile of \citet{arn10}.}\label{fig:profiles}
\end{figure*}

Figure~\ref{fig:profiles} shows the X-ray profiles of the sample. They have been radially scaled by the $\Rv$ determined from the $\Mv$--$\YX$ relation of \citet[][see also \citealt{pra10}]{arn10}. Relaxed (or equivalently, cool core) systems are plotted in blue, disturbed systems in red, and intermediate objects in black. The gas density profiles are scaled by the expected self-similar evolution with redshift; the temperature profiles are scaled by the average spectroscopic temperature in the $[0.15-0.75]\,\Rv$ region; the pressure profiles are scaled by $P_{500}$; the mass profiles are scaled by $\Mv$ (also estimated from the $\Mv - \YX$ relation). 

%_______________
%% Figure: X-ray profiles
%%
%%_______________

%%%%%%%%%%%%%%%%%%%%%%%%%%%%%%%%%%%%%

\section{Mass profile fits}

\begin{figure*}[]
\begin{centering}
\includegraphics[scale=1.,angle=0,keepaspectratio,width=0.95\textwidth]{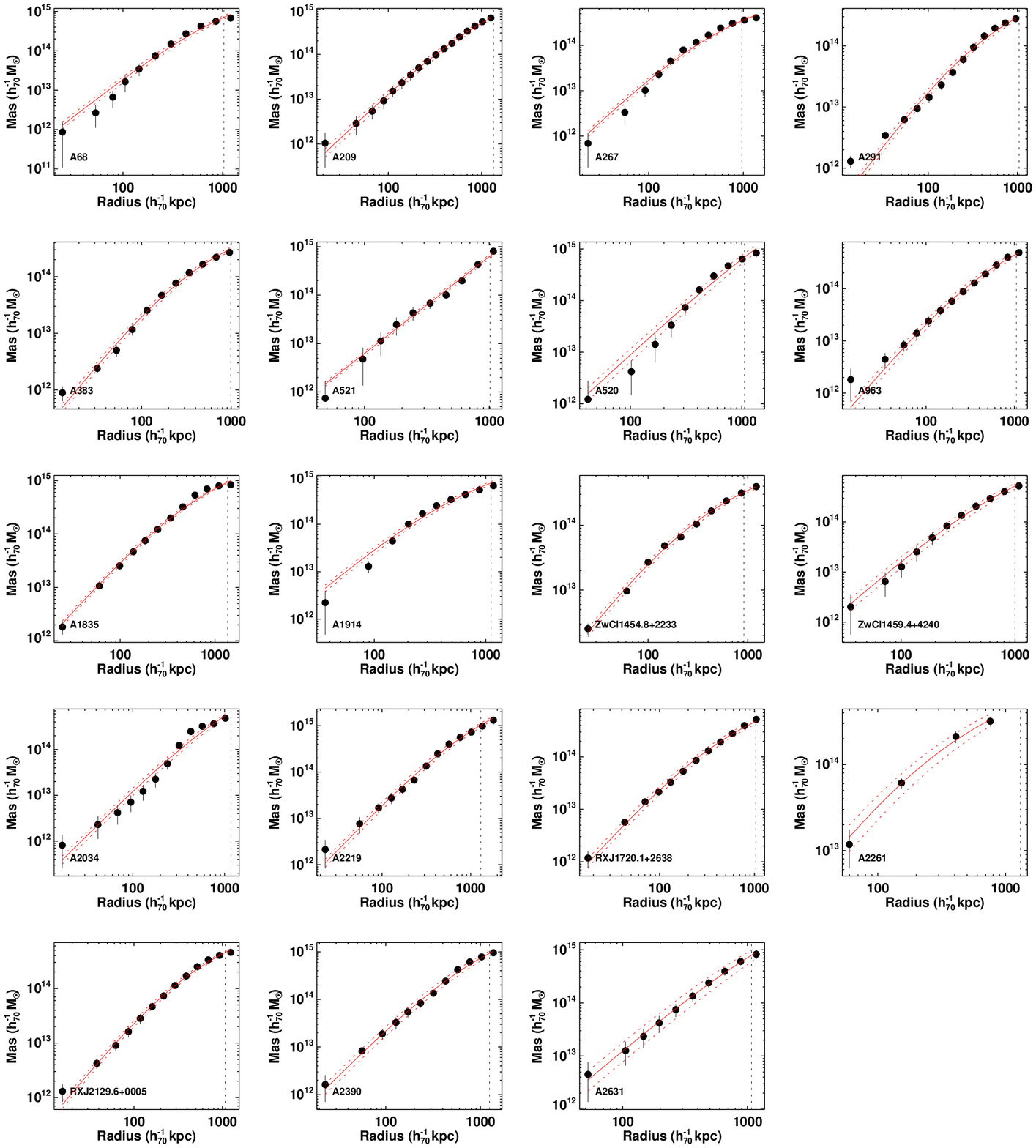}
\end{centering}
\caption{{\footnotesize  X-ray mass profiles of the sample with the best-fitting NFW model overplotted (red line). The dotted vertical line indicates $\Rv$ for each cluster, determined from the best-fitting NFW model.}}\label{fig:mprof}
\end{figure*}

The integrated mass profile of each cluster was fitted with an NFW model as described in Sect.~\ref{sec:massdiscrep}. Table~\ref{tab:mnfw} gives the corresponding best-fitting NFW model and Fig~\ref{fig:mprof} shows the mass profile of each cluster with the best-fitting NFW model overplotted.

%_______________
%% Figure: Mass profile fits
%%
%%_______________

\newpage
\begin{table}[h]
\newcolumntype{L}{>{\columncolor{white}[0pt][\tabcolsep]}l}
\newcolumntype{R}{>{\columncolor{white}[\tabcolsep][0pt]}l}
%\rowcolors{4}{light-grey}{white}
\caption[]{\footnotesize best-fitting mass profile model parameters. }
\label{tab:mnfw}
\centering
\begin{tabular}{@{}Lccccc}

\toprule
\toprule

 &      &  \multicolumn{2}{c}{X-ray} &  \multicolumn{2}{c}{Weak lensing}\\
\cmidrule[0.5pt](rl){3-4}
\cmidrule[0.5pt](rl){5-6}

Cluster &  z    & $c_{500}$ & $\Mv$ & $c_{500}$ & $\Mv$ \\
%\cmidrule[0.5pt](lr){4-4}
        &  &                & $(10^{14}$ M$_\odot$) & & $(10^{14}$ M$_\odot$) \\
 
\midrule

            A68 &  0.255  & $ 1.8 _{- 0.2}^{+ 0.2}$ & $10.0_{- 1.1}^{+ 1.2 }$ & 1.9 & $4.1_{1.0}^{1.2}$ \\
           A209 &  0.206  & $ 1.5 _{- 0.2}^{+ 0.3}$ & $ 6.3_{- 0.7}^{+ 0.7 }$ & 1.2 & $8.6_{1.2}^{1.3}$ \\
           A267 &  0.230  & $ 2.8 _{- 0.3}^{+ 0.3}$ & $ 3.6_{- 0.2}^{+ 0.3 }$ & 3.1 & $3.2_{0.6}^{0.7}$\\
           A291 &  0.196  & $ 3.4 _{- 0.3}^{+ 0.3}$ & $ 2.7_{- 0.3}^{+ 0.3 }$ & 1.0 & $4.0_{0.9}^{1.0}$ \\
           A383 &  0.188  & $ 3.8 _{- 0.4}^{+ 0.4}$ & $ 3.0_{- 0.3}^{+ 0.3 }$ & 4.6 & $3.3_{0.6}^{0.7}$ \\
           A521 &  0.248  & $ 0.0 _{- 0.0}^{+ 0.0}$ & $17.9_{- 2.3}^{+ 2.6 }$ & 1.4 & $3.9_{0.7}^{0.7}$ \\
           A520 &  0.203  & $ 0.5 _{- 0.2}^{+ 0.2}$ & $12.6_{- 2.3}^{+ 2.7 }$ & 1.4 & $4.1_{1.2}^{1.1}$ \\
           A963 &  0.206  & $ 3.1 _{- 0.4}^{+ 0.4}$ & $ 4.8_{- 0.5}^{+ 0.6 }$ & 1.2 & $4.2_{0.7}^{0.9}$ \\
          A1835 &  0.253  & $ 3.2 _{- 0.2}^{+ 0.2}$ & $ 9.4_{- 0.5}^{+ 0.5 }$ & 1.6 & $9.5_{1.5}^{1.7}$ \\
          A1914 &  0.171  & $ 3.4 _{- 0.3}^{+ 0.3}$ & $ 9.0_{- 0.7}^{+ 0.8 }$ & 2.0 & $4.7_{1.9}^{1.6}$ \\
ZwCl1454.8+2233 &  0.258  & $ 4.6 _{- 0.3}^{+ 0.3}$ & $ 3.4_{- 0.3}^{+ 0.3 }$ & 2.0 & $2.6_{0.8}^{1.0}$ \\
ZwCl1459.4+4240 &  0.290  & $ 1.8 _{- 0.3}^{+ 0.3}$ & $ 6.2_{- 0.8}^{+ 1.0 }$ & 3.5 & $3.9_{0.9}^{1.0}$ \\
          A2034 &  0.113  & $ 1.3 _{- 0.2}^{+ 0.2}$ & $ 7.8_{- 0.9}^{+ 1.0 }$ & 1.8 & $5.1_{2.4}^{2.1}$ \\
          A2219 &  0.228  & $ 1.7 _{- 0.2}^{+ 0.2}$ & $12.1_{- 1.3}^{+ 1.5 }$ & 3.5 & $8.0_{1.3}^{1.5}$ \\
 RXJ1720.1+2638 &  0.164  & $ 3.5 _{- 0.4}^{+ 0.4}$ & $ 5.3_{- 0.6}^{+ 0.6 }$ & 4.5 & $3.7_{0.9}^{1.1}$ \\
          A2261 &  0.224  & $ 5.7 _{- 1.3}^{+ 1.3}$ & $ 4.3_{- 0.7}^{+ 0.8 }$ & 3.1 & $8.0_{1.1}^{1.2}$ \\
 RXJ2129.6+0005 &  0.235  & $ 3.5 _{- 0.3}^{+ 0.3}$ & $ 4.6_{- 0.3}^{+ 0.3 }$ & 1.6 & $4.6_{1.0}^{1.1}$ \\
          A2390 &  0.231  & $ 2.2 _{- 0.3}^{+ 0.3}$ & $10.8_{- 1.3}^{+ 1.5 }$ & 3.2 & $7.0_{1.2}^{1.3}$ \\
          A2631 &  0.278  & $ 0.9 _{- 0.3}^{+ 0.3}$ & $12.4_{- 3.0}^{+ 4.0 }$ & 4.1 & $4.8_{0.7}^{0.7}$ \\

\bottomrule
\end{tabular}
\tablefoot{Columns (3,5): concentration parameter. Columns (4,6) $\Mv$ from the best-fitting NFW model. The profiles and best-fitting X-ray NFW models are illustrated in Fig.~\ref{fig:mprof}.}
\end{table}

%%%%%%%%%%%%%%%%%%%%%%%%%%%%%%%%%%%%%

\raggedright
\end{document}